\documentclass[a4paper,11pt]{article}
\usepackage[colorlinks,citecolor=red,urlcolor=blue,bookmarks=false,hypertexnames=true]{hyperref} 

\usepackage{cite}
\usepackage{bm}

\usepackage{diagbox}

\usepackage{chngcntr}

\usepackage{amssymb,amsmath,amsfonts}
\usepackage{float}
\usepackage{relsize}
\usepackage{graphicx}
\usepackage{tikz}
\usetikzlibrary{patterns}
\usetikzlibrary{shapes, arrows, shadows}

\tikzstyle{vertex} = [circle, draw, fill=blue!20, scale=1,auto=left]
\tikzstyle{vert} = [circle, draw, fill=blue!20, scale=.8,auto=left]
\tikzstyle{line} = [draw]

\usepackage{upgreek}

\usepackage{pgfplots}

\pgfplotsset{compat=newest}

\pgfdeclarepatternformonly[\LineSpace]{my north east lines}{\pgfqpoint{-1pt}{-1pt}}{\pgfqpoint{\LineSpace}{\LineSpace}}{\pgfqpoint{\LineSpace}{\LineSpace}}%
{
    \pgfsetlinewidth{0.4pt}
    \pgfpathmoveto{\pgfqpoint{0pt}{0pt}}
    \pgfpathlineto{\pgfqpoint{\LineSpace + 0.1pt}{\LineSpace + 0.1pt}}
    \pgfusepath{stroke}
}

\newdimen\LineSpace
\tikzset{
    line space/.code={\LineSpace=#1},
    line space=3pt
}

\def\be{\begin{equation}}
\def\ee{\end{equation}}
\def\bea{\begin{eqnarray}}
\def\eea{\end{eqnarray}}

\begin{document}

\begin{titlepage}
\date{\today}       \hfill

\begin{center}

\vskip .2in
{\LARGE \bf  RG boundaries and  Cardy's variational ansatz for multiple perturbations }\\
\vspace{5mm}

\today
 
\vskip .250in

\vskip .3in
{\large Anatoly Konechny}

\vskip 0.5cm
{\it Department of Mathematics,  Heriot-Watt University\\
Edinburgh EH14 4AS, United Kingdom\\[10pt]
and \\[10pt]
Maxwell Institute for Mathematical Sciences\\
Edinburgh, United Kingdom\\[10pt]
}
E-mail: A.Konechny@hw.ac.uk
\end{center}

\vskip .5in
\begin{abstract}
We consider perturbations of 2D CFTs by multiple relevant operators. The massive phases 
of such perturbations can be labeled by conformal boundary conditions. Cardy's variational ansatz 
approximates the vacuum state of the perturbed theory by a smeared conformal boundary state.
In this paper we study the limitations and propose
generalisations of this ansatz using both analytic and numerical insights based on TCSA. 
In particular we analyse the stability of Cardy's ansatz states with respect to boundary relevant perturbations 
using  bulk-boundary OPE coefficients. We show that certain transitions  between the massive phases arise 
from a pair of boundary RG flows.  The RG flows start from the conformal boundary on the transition surface and end on those that lie on the two sides of it. As an example we work out the details of the phase diagram for the Ising field theory 
and for the tricritical Ising model  perturbed by the leading thermal and magnetic fields. 
For the latter we find  a pair of novel 
 transition lines  that correspond to  pairs of RG flows. Although the mass gap remains finite at the transition lines, several one-point functions change their behaviour. We discuss how these  lines fit into 
 the standard phase diagram of the tricritical Ising model. 
We show that each line extends to a two-dimensional surface $\xi_{\sigma,c}$ in a three coupling space when we  add  perturbations by  the subleading magnetic field. 
Close to this surface we locate   symmetry breaking critical lines leading to the critical Ising model. 
Near the critical lines we find first order phase transition lines describing two-phase coexistence regions
 as predicted in Landau theory. The surface $\xi_{\sigma,c}$  is determined from   
 the CFT data using Cardy's ansatz and its properties are checked using TCSA numerics.

\end{abstract}

\end{titlepage}

\renewcommand{\thepage}{\arabic{page}}
\setcounter{page}{1}
\large

{\hypersetup{linkcolor=black,citecolor=black}
\noindent\hrulefill
\tableofcontents 
\noindent\hrulefill
}

\section{Introduction }
\renewcommand{\theequation}{\arabic{section}.\arabic{equation}}
\setcounter{equation}{0}
\large 

In this paper we consider renormalisation group (RG) flows in perturbed two-dimensional conformal field theories (2D CFTs). 
A useful tool to study such flows is an RG interface (or domain wall) \cite{BR} (see also \cite{FQ}) which is obtained by starting with a UV CFT 
on the plane and perturbing it by a relevant operator on a half-plane. Renormalising the perturbed theory and allowing the 
RG flow to arrive to the IR fixed point we obtain a conformal interface separating the UV and IR CFTs. If the latter is trivial, which happens 
if the perturbed theory develops a mass gap, we obtain a conformal boundary which we call an RG boundary. If the theory is trivially gapped 
we have an elementary (or irreducible) boundary condition and if there is a vacuum degeneracy we have a superposition of conformal boundaries. 
The RG boundaries can be used as labels for IR phases of the perturbed theory. In this paper we are particularly interested in mapping out 
phase diagrams for perturbed CFTs in terms of RG boundaries and understanding their general structure. 

Constructing an RG interface between non-trivial CFTs is quite difficult and so far it has been done non-perturbatively only for certain classes of 
integrable perturbations \cite{Gaiotto,Pog1,Stanishkov,Pog2,Pog3}. In general there are few known constructions of conformal interfaces. 
We are on a much better ground when it comes to conformal boundary conditions. For many CFTs the complete set of conformal boundary conditions is known. 
Perhaps the simplest example of this situation  is the critical Ising model. It is described by a unitary CFT which has a central charge $c=1/2$ and
 three conformal families: $1$, $\epsilon$, 
$\sigma$.  The theory has three elementary conformal boundary conditions with boundary states \cite{Cardy}
\be\label{conf_bcs}
|\pm\rangle\!\rangle = \frac{1}{\sqrt{2}} \Bigl[ |1\rangle\!\rangle + |\epsilon\rangle\!\rangle  \pm 2^{1/4}|\sigma\rangle\!\rangle \Bigr]\, , 
\qquad |F\rangle\!\rangle = |1\rangle\!\rangle - |\epsilon\rangle\!\rangle \, .
\ee
Here $| 1\rangle\!\rangle $, $|\epsilon \rangle\!\rangle $, $|\sigma\rangle\!\rangle $ stand for the Ishibashi states.  The boundary states $|\pm\rangle\!\rangle$ describe the fixed spin while $ |F\rangle\!\rangle $ describes the free  spin boundary condition of the underlying lattice model. 
The theory has two relevant perturbations - thermal and magnetic fields that couple to temperature  $t$ and the magnetic field  coupling $h$ in the perturbed critical Ising theory: 
\be\label{IFT}
H=  H_{0} + t \int\limits_{0}^{R}\! \phi_{\epsilon} (0,y)dy + h \int\limits_{0}^{R}\! \phi_{\sigma} (0,y)dy \, .
\ee
For $h=0$ the region with $t>0$ is the high temperature disordered region of the Ising model and $t<0$ corresponds to the 
low temperature symmetry broken region which has a doubly degenerate vacuum. 

For all real values of the couplings  the flows are massive and the corresponding RG boundaries were obtained in \cite{AK_Ising} 
analytically for $h=0$ and using the truncated conformal space approach (TCSA) and the truncated free fermion approach of \cite{FZ} for $h\ne 0$. The mapping of regions in terms of the conformal boundary conditions is summarised on 
  Figure \ref{Ising_diagram1}.  In that diagram for all points in the 
lower half-plane except for the points on the $h=0$ axis the RG boundary condition is $|+\rangle\!\rangle$ 
and similarly for all points in the upper half-plane we have $|-\rangle\!\rangle$. For $h=0, t<0$ (the blue line) we have a superposition $|+\rangle\!\rangle \oplus |-\rangle\!\rangle$ that reflects the 
spontaneous symmetry breaking 
while for $h=0,t>0$ (the red line) we have $|F\rangle\!\rangle$.
 \begin{center}
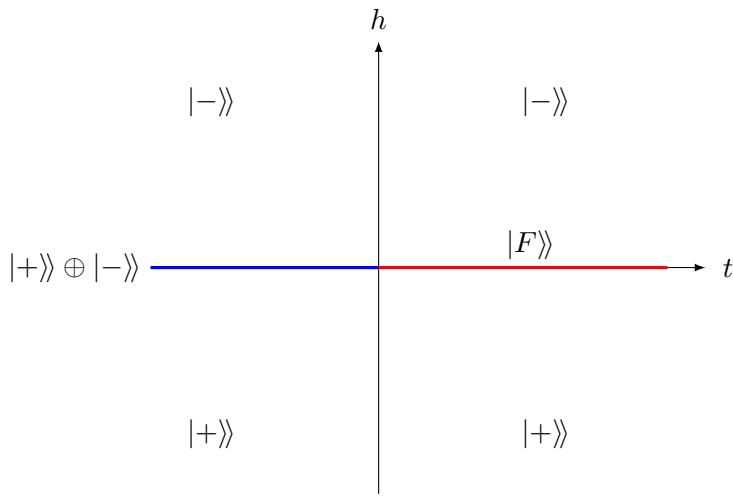
\begin{figure}[H]  
\centering
\begin{tikzpicture}[>=latex]
\draw[->] (-3,0) --(4.3,0);
\draw[->] (0,-3)--(0,3);
\draw[red,very thick] (0,0)--(3.8,0);
\draw[blue, very thick] (-3,0)--(0,0);
\draw (4.6,0  ) node {$t$};
\draw (0,3.3) node {$h$};
\draw (2.2,2.2) node {$|-\rangle\!\rangle $};
\draw (-2.2,2.2) node {$|-\rangle\!\rangle $};
\draw (-2.2,-2.2) node {$|+\rangle\!\rangle$};
\draw (2.2,-2.2) node {$|+\rangle\!\rangle$};
\draw (-4,0) node {$|+\rangle\!\rangle \oplus |-\rangle\!\rangle$};
\draw (2,0.3) node {$|F\rangle\!\rangle$};
\end{tikzpicture}
\caption{The unstable manifold of IFT and the associated conformal boundary states.}
\label{Ising_diagram1}
\end{figure}
\end{center}


A general analytic method of finding the assignment of RG boundaries to perturbations was put forward by J.~Cardy \cite{Cardy_var} 
in the form of a variational ansatz. To build the ansatz we consider a perturbed CFT on a cylinder of circumference R with coordinates $x \in {\mathbb R}$ 
along the cylinder and $y \in [0,R)$ across. The perturbed Hamiltonian can in general be written as 
\be \label{perturbed_CFT}
H=  H_{0} + \sum_{i} \lambda_{i}\!\! \int\limits_{0}^{R} \! \phi_{i}(0, y) dy 
\ee
where 
\be
H_{0} = \frac{2\pi}{R}(L_{0} + \bar L_{0} -\frac{c}{12})
\ee
is the CFT Hamiltonian, $c$ is the central charge, and 
$\phi_{i}$ are relevant primary operators with scaling dimensions $\Delta_{i}$ and (dimensionful) couplings $\lambda_{i}$. 

Following \cite{Cardy_var} we approximate the ground state (or ground states in case of their degeneracy) of (\ref{perturbed_CFT}) using smeared 
 boundary states of conformal boundary conditions. For each conformal boundary state $|\tilde a\rangle\!\rangle$ 
 we form a trial vacuum state 
 \be \label{trial_vac0}
 |\tau, a \rangle = e^{-\tau H_{0}}|\tilde a\rangle\!\rangle 
 \ee
where $\tau$ is a parameter to be determined by minimising the average energy
\be \label{var_en0}
E_{a}(\tau)= \frac{\langle \tau, a |H|\tau, a \rangle}{\langle \tau, a |\tau, a \rangle} \ . 
\ee
The numerator in this expression can be expressed in terms of  amplitudes on a cylinder of length $2\tau$ and circumference $R$ with the boundary condition $a$ 
put on each end and 
operator insertions  in the middle. An analytic expression for $E_{a}(\tau)$ can then be obtained in the limit $\tau\ll R$. Minimising all $E_{a}(\tau)$ 
over $\tau$ and choosing the smallest value we obtain the RG boundary (or boudaries) for the given perturbation.

The method can be used for any CFT with known conformal boundary conditions. Not only it selects the 
 RG boundary, but it also gives the leading finite volume correction to the vacuum state. In the rest of this paper we refer to the ansatz 
(\ref{trial_vac0}) as Cardy's ansatz. In  \cite{Cardy_var} as an example the method was applied to RG flows that start from  diagonal series minimal models.  In particular that allowed to reproduce the diagram on Figure \ref{Ising_diagram1} modulo a caveat. 
For a  small enough magnetic perturbation in the disordered region the variational energy   for the free spin  boundary condition is lower than that of the fixed spins. This is puzzling because the free spin boundary state has zero magnetisation. A similar problem 
arises  whenever we add a small symmetry breaking perturbation in the disordered region of any thermally perturbed  minimal model. 
One of the motivations for this paper is to address this problem. We are going to discuss it in detail for two models -- the Ising field theory (\ref{IFT})  and the tricritical Ising model (TIM) perturbed by the leading thermal and magnetic perturbations. 

While \cite{Cardy_var} mainly focussed on qualitative implications of the ansatz it was later analysed in \cite{LVT} with  precision by comparing 
the ansatz with numerical results  obtained using TCSA. The latter is a general method applicable to any perturbed CFT. It was invented in \cite{YZ1}, 
\cite{YZ2} and has been since applied to a wide variety of models and phenomena (see \cite{Cardy_etal,FZ,GW,HRvR,RV,LT_RG,LT_Potts,Takacs_quenches} for some examples of using TCSA and discussion of truncation errors).
A detailed review of TCSA can be found in \cite{Koniketal} and \cite{Gaboretal}. In TCSA one considers the perturbed Hamiltonian on the cylinder 
(\ref{perturbed_CFT}). This Hamiltonian is restricted to a finite dimensional space which is constructed as follows. Let $|i\rangle$ be 
a spin zero primary state. Consider Virasoro descendants of the form 
\be \label{tvec1}
L_{-n_{1}}L_{-n_{2}} \dots L_{-n_{p}} \bar L_{-m_{1}}  \bar L_{-m_{2}}  \dots  \bar L_{-m_{q}} |i \rangle \, , \quad n_{i},m_{j} \in {\mathbb N} 
\ee
and satisfying 
\be \label{tvec2} 
\sum_{i} n_{i} = \sum_{j} m_{j} \le n_{c} 
\ee
where $n_{c}$ is a fixed  integer which we call a truncation parameter. The truncated  space 
is spanned by vectors of the form (\ref{tvec1}) satisfying (\ref{tvec2}). It contains spin zero vectors whose total  descendant level\footnote{One can also truncate the space using the total level that includes the primary weights. In this paper we use the descendant level truncation, except for the Ising model where we use the free fermion oscillators and truncate in the sum of the oscillator modes. } is less or equal 
than $2n_{c}$. The Hamiltonian restricted to the truncated space is then diagonalised numerically. 

In \cite{LVT} TCSA was used to compare various predictions of Cardy's ansatz, such as the vacuum energy density, the components of the vacuum vector and 
the chiral entanglement entropy with TCSA numerical answers for single field perturbations of the Ising and TI models. 
In the absence of UV divergences a very good quantitative agreement was found which is the better the smaller the dimension of the perturbing operator is. 

The focus of the present paper is on studying Cardy's ansatz for multiple perturbations, that is when two or more perturbing operators are simultaneously switched on. We propose to supplement the use of the ansatz with analysis of stability of the ansatz states under deformations of the boundary conditions by relevant boundary operators. The leading change of the variational energy  under such deformation can be easily expressed in terms of 
bulk-boundary OPE coefficients (we do this in detail in section \ref{Cardy_gen_section}). The trial state is typically stable at the leading order when either the boundary condition does not have any relevant operators or when the relevant bulk-boundary OPE coefficients all vanish. It may happen however 
that it is stable only due to mutual cancellations between contributions from different perturbing bulk operators. This can happen only when the perturbing couplings are fine-tuned to lie in a certain submanifold of the total coupling space. 
 The transition submanifold  then itself is associated with the stable boundary condition at hand. Perturbing away from the submanifold  the boundary condition becomes unstable and a boundary RG flow is generated on it. We propose to associate the end point of the flow with the phase away from the submanifold. In the simplest situation the submanifold is a surface separating two phases  each of which is the end point of a boundary RG flow 
 that starts in the UV from the boundary condition associated with the submanifold. 
 
 We look in detail at two models -- the  Ising field theory (\ref{IFT})  and doubly perturbed TIM. The TIM is  the second  model after Ising in the diagonal 
 minimal model sequence.  It has central charge $c=7/10$, six primary states and six associated irreducible conformal boundary conditions. 
 This model gives a universality class of critical  phenomena in two dimensions in the presence of a tricritical point. 
 The TIM and its perturbations  has been extensively studied over the years (some rather incomplete list of references of field theory studies of TIM is \cite{Cardy_etal, Zam_TBA, Zam_TBA2,Friedan_etal,Qiu, TIM0, TIM1, TIM2, TIM3, TIM4, multicrit1, multicrit2}).  
 
 The underlying lattice model which describes the same universality class as perturbed TIM is the spin-1 Ising model which was originally introduced by 
Blume and Capel \cite{Blume}, \cite{Capel} and later generalised by Blume, Emery and Griffiths \cite{BEG}. For the latter (BEG) model
the lattice Hamiltonian can be written as  
\be
{\cal H}=-H\sum_{j=1}^{N} s_{j} + \Delta \sum_{j=1}^{N} s_{j}^2 -J\sum_{\langle i,j\rangle} s_{i}s_{j}  - H_{3}\sum_{\langle i,j\rangle} s_{i}s_{j}(s_{i}+s_{j})  -K\sum_{\langle i,j\rangle} s_{i}^2 s_{j}^2 \, . 
\ee
Here the indices $i,j=1, \dots, N$ label  two-dimensional lattice sites with $\langle i,j\rangle$ denoting neighbouring sites. 
On each lattice site the variable $s_{i}$ takes three possible values: $0,1,-1$.  The model has two magnetic couplings: $H$, $H_{3}$ 
and three energy couplings:$ \Delta$, $J$, $K$. The five-coupling space has one irrelevant direction and four relevant directions which in the continuum limit give rise to 
the four relevant couplings associated with the CFT scaling operators. A review of the general theory of  tricritical points as well as results for particular models, including the BEG model, can be found    in \cite{LS}. We review some aspects of Landau theory of tricritical points in section \ref{Landau_sec}.


 In this paper we focus on  perturbations of TIM by the two primary fields of lowest dimension: $\phi_{\epsilon}$ -- the leading thermal field with dimension 
 $\Delta_{\epsilon} = 1/5$ and $\phi_{\sigma}$ -- 
 the leading magnetic field which has dimension  $\Delta_{\sigma} = 3/40$. As a quick preview of our results regarding this model we offer the reader 
 its phase diagram   on Figure \ref{TIM_diagram1}.


\begin{center}
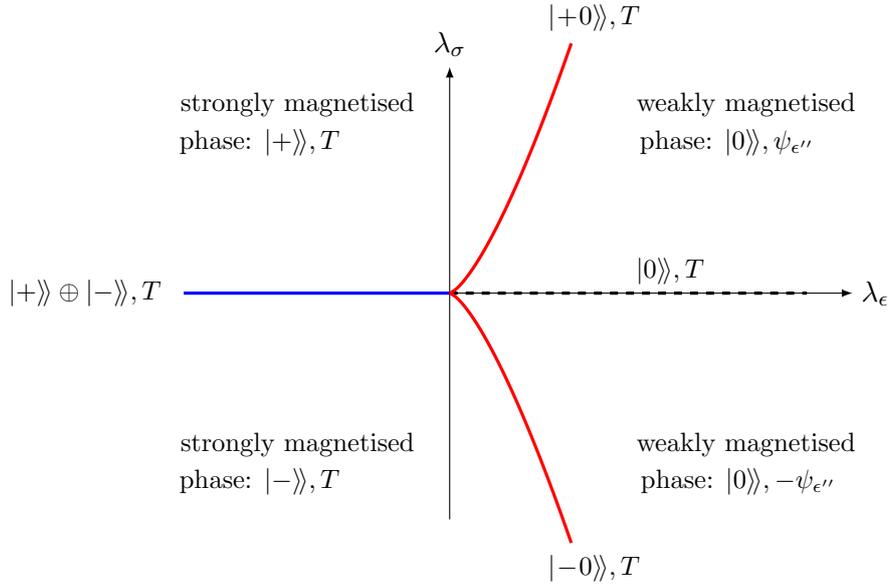
\begin{figure}[H]  
\centering
\begin{tikzpicture}[>=latex]
\draw[->] (-3,0) --(5.3,0);
\draw[very thick, dashed] (0,0)--(4.7,0);
\draw[->] (0,-3)--(0,3);
\draw[blue,very thick] (-3.5,0) --(0,0);
\draw (5.6,0  ) node {$\lambda_{\epsilon}$};
\draw (0,3.3) node {$\lambda_{\sigma}$};
\draw (-2, 2.5) node {\small strongly magnetised};
\draw (-2.5,2) node {\small  phase:      $|+\rangle\!\rangle , T$};
\draw (-2, -2) node {\small strongly magnetised};
\draw (-2.5,-2.5) node {\small  phase:      $|-\rangle\!\rangle , T$};
\draw (3.9, 2.5) node {\small weakly magnetised};
\draw (3.65,2) node {\small  phase:   $|0\rangle\!\rangle, \psi_{\epsilon''}$  };
\draw (3.9, -2) node {\small weakly magnetised};
\draw (3.8,-2.5) node {\small  phase:   $|0\rangle\!\rangle, -\psi_{\epsilon''}$  };
\draw (-4.8,0) node {\small $|+\rangle\!\rangle \oplus |-\rangle\!\rangle, T$};
\draw (2.9,0.3) node {\small $|0\rangle\!\rangle, T$}; 
\draw[scale=0.5, domain=0:3.2, smooth, variable=\x, red,very thick] plot ({\x}, {1.3*(\x)^(1.4)});
\draw[scale=0.5, domain=0:3.2, smooth, variable=\x, red,very thick] plot ({\x}, {-1.3*(\x)^(1.4)});
\draw (1.9,3.65) node {\small $|\!+\!0\rangle\!\rangle , T$} ;
\draw (1.9,-3.65) node {\small $|\!-\!0\rangle\!\rangle , T$} ;
\end{tikzpicture}
\caption{The phase diagram  of TIM perturbed by $\phi_{\sigma}$ and $\phi_{\epsilon}$. The phases are labeled 
 by conformal boundary states together with the leading irrelevant operator defining their perturbation at finite volume. 
 } 
\label{TIM_diagram1}
\end{figure}
\end{center}

Only five of the six elementary boundary conditions appear on the diagram: $|0\rangle\!\rangle$,  $|\pm\rangle\!\rangle$,  $|\!\pm\!0\rangle\!\rangle$. 
The notation for these boundary states was introduced in \cite{Chim} and reflects the 
boundary conditions imposed on the spin variables $s_{i}$ in the underlying BEG model. 
We also put on the diagram the physical interpretation of each phase as "strongly magnetised" and "weakly magnetised" which we 
explain in sections \ref{Landau_sec}, \ref{transition_subsec}.
 
In addition to the first order phase transition line (in blue) the diagram also has two   novel transition lines drawn  in red. On such a  line one of the boundary conditions 
$|\!+\!0\rangle\!\rangle$, $|\!-\!0\rangle\!\rangle$ becomes stable while moving away from the line induces a boundary RG flow towards either 
one of $|\pm\rangle\!\rangle$ boundary states or a deformed $|0\rangle\!\rangle$ boundary state. The latter deformation at the leading order  is given by the boundary irrelevant field 
$\psi_{\epsilon''}$ that replaces the stress energy tensor $T$ that (being integrated) gives the standard Cardy ansatz (\ref{trial_vac0}). 
We find that on the red lines the mass gap is finite and moving across them there are no thermodynamic singularities. However 
several expectation values change their shape as we move across. In particular the magnetisation profile looks like a kink. Inspired by that 
we call the  transition the theory experiences at the red lines a ``kink transition''. 
This  transition is  investigated in detail  in section \ref{transition_subsec}. 
We further find that the red lines  on Figure \ref{TIM_diagram1} extend into a two-dimensional surface in a three-coupling space when we switch on the  subleading magnetisation operator $\phi_{\sigma'}$.  
Close to this surface we find the symmetry breaking critical lines and more first order transition lines. These features are discussed in section \ref{sigmaprime_section}. The kink transition lines are each similar to the disorder line of the Ising model. The magnetisation  dependence on the external field has similar profiles in the vicinity of the transition with an inflection point located close or possibly exactly on the lines. Also, similarly to the Ising disorder line that is joined to the first order line at the Ising critical point, the surface of kink transitions is joined to a surface of first order transitions  at a critical line. We discuss the significance of kink transitions in phase diagrams in section  \ref{summary2_sec}.

The layout of the main part of the paper is as follows. In section \ref{Cardy_gen_section} we discuss various generalities of Cardy's ansatz including its difficulties and possible fixes. We also do the general stability analysis in that section. Section \ref{Ising_sec}  is dedicated to a detailed study of the Ising field theory. It ends with 
a quick summary of the results in subsection \ref{summary1_sec}. In section \ref{tricrit_section}  we study  perturbations of the TIM. This section also finishes with a summary and some open questions in subsection \ref{summary2_sec}. 
In the final section \ref{outlook_sec} we attempt at a general outlook on the idea of labelling phases with RG boundaries comparing  Cardy's ansatz with the classical Landau theory.  Four appendices contain some additional plots and more technical information.

\section{Cardy's variational ansatz } \label{Cardy_gen_section} \setcounter{equation}{0}
In this section we give a general discussion of Cardy's ansatz. As stated in the introduction we look at 
a perturbed CFT on a cylinder with Hamiltonian (\ref{perturbed_CFT}). 
Given a conformal boundary state $|\tilde a\rangle\!\rangle$ 
  a trial vacuum state has  the form
 \be \label{trial_vac}
 |\tau, a \rangle = e^{-\tau H_{0}}|\tilde a\rangle\!\rangle 
 \ee
 which can be described as a smeared boundary state. 
Unlike the conformal boundary state the smeared state (\ref{trial_vac}) has a finite norm
\footnote{The minimum requirement to apply 
the variational method should be that the perturbed vacuum has a finite norm when considered as a state inside the unperturbed theory state space. We know that this is not always the case in  quantum field theory, due to Haag's theorem,  but it is believed to be true in two dimensions for a sufficiently relevant perturbation. }.
The smearing parameter $\tau$ is determined by minimising the variational energy 
\be \label{var_en1}
E_{a}(\tau)= \frac{\langle \tau, a |H|\tau, a \rangle}{\langle \tau, a |\tau, a \rangle} \ . 
\ee
An explicit analytic expression for this quantity can be obtained \cite{Cardy_var} by noting that 
it  is a normalised  amplitude on a cylinder of length $2\tau$ and circumference $R$ with the boundary condition $a$ 
put on each end and 
operator insertions  in the middle. 
As  we will be interested in generalising the ansatz it will be instructive to go over the derivation of \cite{Cardy_var} in some detail.

In the limit $\tau \ll R$ we can use the ``open string channel'' quantisation in which $y$ is the Euclidean time variable. 
For $R\to \infty$ the leading contribution to the cylinder partition function comes from the vacuum state so that 
\be \label{cas_en}
\langle \tau, a |\tau, a \rangle = \langle\!\langle \tilde a|  e^{-2\tau H_{0}}    |\tilde a\rangle\!\rangle = e^{-{\cal E}_{\rm Cas} R } + \dots 
\ee
where 
\be
{\cal E}_{\rm Cas} = -\left(\frac{\pi c}{24}\right)  \frac{1}{2\tau}
\ee 
is the Casimir energy on the interval $x \in [-\tau, \tau] $ which is expressed in terms of the central charge $c$. 
The omitted terms in (\ref{cas_en})  come from excited states and are suppressed by factors $e^{-\frac{\Delta E_{i}R}{2\tau}}$, $\Delta E_{i}>0$.
Thus, we obtain 
\be \label{cas_cont}
\langle \tau, a |H_{0}| \tau, a \rangle = - \frac{1}{2} \frac{\partial}{\partial \tau}\langle\!\langle \tilde a|  e^{-2\tau H_{0}}    |\tilde a\rangle\!\rangle = R\frac{\pi c}{24(2\tau)^2} e^{-{\cal E}_{\rm Cas} R }  + \dots 
\ee
and the term in the average energy coming from $H_0$ is 
\be 
 \frac{\langle \tau, a |H_{0}|\tau, a \rangle}{\langle \tau, a |\tau, a \rangle} =  R\frac{\pi c}{24(2\tau)^2}
\ee
up to exponentially suppressed corrections. 

The perturbing terms give a factor of $R$, which comes from integrating over $y$, times a one-point function on the cylinder. 
At the leading order we can replace the cylinder by an infinite strip of width $2\tau$. The perturbing operators $\phi_{i}$ are 
inserted in the centre of the strip. Since these operators are conformal primaries and the boundary condition is conformal 
these one point functions are easily obtained by using a conformal mapping from the strip to the unit disc. It gives 
\be
 \frac{\langle \tau, a |\phi_{i}(0,0)|\tau, a \rangle}{\langle \tau, a |\tau, a \rangle} = A_{i}^{a} \left( \frac{\pi}{4\tau}\right)^{\Delta_{i}}
\ee
where 
\be
A_{i}^{a} = \frac{\langle \phi_{i}|\tilde a \rangle\!\rangle}{\langle 0|\tilde a \rangle\!\rangle}
\ee
are the disc one-point functions.

Collecting all terms together we obtain the variational energy formula from \cite{Cardy_var} 
\be \label{var_en2}
E_{a}(\tau)= R\Bigl[  \frac{\pi c}{24(2\tau)^2} + \sum_{i} \lambda_{i} A^{a}_{i} \left(\frac{\pi}{4\tau}\right)^{\Delta_{i}}    \Bigr]\, .
\ee
By minimising $E_{a}(\tau)$ (\ref{var_en2}) over $\tau$ we obtain a minimum at $\tau=\tau^{*}_{a}$. Comparing the minimal 
values $E_{a}=E_{a}(\tau_{a}^{*})$ and assuming for now that there is a single conformal boundary condition $a=\bar a$ with the 
smallest $E_{\bar a}$ we obtain an approximation to the vacuum state of perturbed theory:
\be \label{approx_vac}
|\tau_{\bar a}^{*}, \bar a \rangle = e^{-\tau^{*}_{\bar a} H_{0}}|\tilde {\bar a}\rangle\!\rangle  \, .
\ee
Note that  (\ref{approx_vac}) comes with 
a mass scale set by $\tau=\tau^{*}_{\bar a}$ which in principle is determined by the 
scales set by the couplings $\lambda_{i}$.

For a single perturbation with a coupling $\lambda_{i}$ (i.e. no summation in \ref{var_en2}) a minimum is achieved at a finite value of $\tau$ provided $\lambda_{i}A^{a}_{i} < 0 $ in which case we have 
\be \label{tau_min}
\tau = \tau^{*}_{a} =  \left(\frac{c4^{\Delta_{i}}\pi^{1-\Delta_{i}}}{48\Delta_{i} |A_{i}^{a}|}  \right)^{\frac{1}{2-\Delta_{i}}}|\lambda_{i}|^{-\frac{1}{2-\Delta_{i}}}
\ee
and the minimal value of the energy is 
\be \label{E_min}
E_{a}(\tau_{a}^{*}) = \frac{\pi c(\Delta_i -2)}{96\Delta_i} \frac{R}{(\tau^{*}_{a})^2}  \, . 
\ee
Note that the last expression is negative for relevant perturbations. 
From (\ref{tau_min}) we see that, assuming $\Delta_{i}<2$, in the far infrared where $\lambda_{i}$ tends to infinity $ \tau^{*}_{a}$ tends to zero. The approximate vacuum (\ref{approx_vac}) then tends to a conformal boundary state in accord with the general expectations 
we discussed in the introduction. If our CFT is a diagonal Virasoro minimal model the one-point function coefficients can be 
expressed in terms of the modular $S$-matrix coefficients as 
\be \label{ai_1pt}
A_{i}^{a}= \frac{S_{ia}}{S_{1a}}\left( \frac{S_{11}}{S_{i1}}\right)^{1/2} \, . 
\ee
As was noted in \cite{Cardy_var} it follows from the fact that $S_{ij}$ is a symmetric orthogonal matrix that 
for each $i$ no matter what the sign of $\lambda_{i}$ is there is always a value of $a$ such that $\lambda_{i}A_{i}^{a}<0$ 
and thus the variational minimum is always given by (\ref{tau_min}), (\ref{E_min}). 
In the presence of several couplings the large $\tau$ region of the energy function (\ref{var_en2}) is dominated by the term 
corresponding to the most relevant coupling labelled by $i=i^{*}$. Since we can choose $a$ such that $\lambda_{i^{*}}A_{i^{*}}^{a}<0$ 
the function $E_{a}(\tau)$ asymptotically approaches zero from the negative values. This means that the minimal 
variational energy is always negative. It should be noted that this does not necessarily mean that the minimal energy is
always given by $a$ chosen as above. Even if asymptotically $E_{a}(\tau)$  approaches zero from the positive values 
there can be a negative energy region at smaller values of $\tau$  where other, less relevant couplings dominate.

In passing from (\ref{var_en1}) to (\ref{var_en2}) one drops terms which contain positive powers of $e^{-R/\tau}$. 
Such terms can be neglected unless there is a degeneracy, that is several boundary conditions $\bar a$  yielding  the same 
minimal value of the energy average. This happens when we are in a symmetry breaking region of the perturbed theory. 
 In this case, following \cite{LVT}, we need to diagonalise the matrix 
\be \label{deg_matrix}
{\cal M}_{\bar a \bar b} = \frac{\langle \tau, \bar a |H|\tau, \bar b \rangle}{\sqrt{\langle \tau, \bar a |\tau, \bar a \rangle \langle \tau, \bar b |\tau, \bar b \rangle}} \, .
\ee
The off-diagonal elements in this matrix involve an amplitude on a strip with different boundary conditions on the two ends and an insertion 
of a bulk operator in the middle. Such an amplitude  mapped on a disc  involves the bulk operator and two boundary-condition changing boundary operators. 
For the interaction terms this involves conformal blocks. For the insertion of $H_{0}$ the calculation is model independent  and was done in \cite{LVT}: 
\be \label{deg_matrix0} 
{\cal M}_{\bar a \bar b}^{0} = \frac{\langle \tau, \bar a |H_{0}|\tau, \bar b \rangle}{\sqrt{\langle \tau, \bar a |\tau, \bar a \rangle \langle \tau, \bar b |\tau, \bar b \rangle}} = R \frac{\pi(c-24h_{\bar a\bar b})}{24(2\tau)^2} e^{-R\pi h_{\bar a\bar b}/(2\tau)} 
\ee
where $ h_{\bar a\bar b}$ is the smallest weight of boundary condition changing fields linking $\bar a$ with $\bar b$. 
Since the off-diagonal terms are exponentially suppressed at the leading order one can set $\tau = \tau^{*}_{\bar a}$ in their expressions. 
Diagonalising the matrix (\ref{deg_matrix})  yields a unique vacuum together with approximately degenerate states whose energies differ by exponentially small 
terms determined by $ h_{\bar a\bar b}$.
This is what is physically expected as at finite volume there is always a unique vacuum plus approximately degenerate 
states with energy gaps exponentially suppressed with  the volume.

The state space for diagonal minimal models with $P$ primary fields decomposes as 
\be
{\cal H} = \bigoplus_{i=1}^{P}  {\cal H}_{i}^{L} \otimes {\cal H}_{i}^{R}
\ee
 where $ {\cal H}_{i}^{L}$ and ${\cal H}_{i}^{R}$ are the irreducible representation spaces for the holomorphic and anti-holomorphic copies of the Virasoro algebra respectively. Furthermore, each irreducible subspace $ {\cal H}_{i}^{L,R}$ is further graded by the eigenvalues of $L_{0} - h_{i}$ and $\bar L_{0} - h_{i}$, i.e. by the descendant level $N$,   where $h_{i}$ is the conformal weight of the corresponding primary. 
 We can thus write  
 \be
 {\cal H}_{i}^{L} = \bigoplus_{N=0}^{\infty} {\cal H}_{i,N}^{L}\, , \qquad  {\cal H}_{i}^{R} = \bigoplus_{N=0}^{\infty} {\cal H}_{i,N}^{R} \, .
 \ee
 Let $|k_{i,N}\rangle_{L} $, $k_{i,N}=1, \dots, {\rm dim} {\cal H}_{i,N}^{L}$ be the basis vectors of an orthonormal basis in 
 ${\cal H}_{i,N}^{L}$ and let $|k_{i,N}\rangle_{R} $ denote the conjugate   basis vectors\footnote{ If the basis states $|k_{i,N}\rangle_{L} $ 
 are written as linear combination of Virasoro monomials built by a successive action of  $L_{-n}$, $n>0$ operators on the primary states  then to obtain the conjugate vectors 
 $|k_{i,N}\rangle_{R} $ we replace each $L_{-n}$ by $\bar L_{-n}$ and each numerical coefficient by its complex conjugate. } in  ${\cal H}_{i,N}^{R}$.
 The Ishibashi states then can be written as 
 \be \label{ishibashi}
 |i\rangle\!\rangle = \sum_{N} \sum_{k_{i,N}} |k_{i,N}\rangle_{L} \otimes |k_{i,N}\rangle_{R} \, .
 \ee
 With these notations the ground state candidate obtained using Cardy's ansatz can be written as 
\be \label{vac_form}
|{\rm var}\rangle = \sum_{i} \frac{a_{i}}{\sqrt{\cal N}}  \sum_{N,k_{i,N}} e^{-\frac{2\pi \tau^{*}}{R}(2h_{i} + 2N)} |k_{i,N}\rangle_{L} \otimes |k_{i,N} \rangle_{R} 
\ee
where the coefficients $a_{i}$ depend only on the $S$-matrix elements and ${\cal N}$ is the normalisation factor. 
This formula applies in either a degenerate or a non-degenerate case. In the non-degenerate case 
\be
a_{i} = \frac{S_{\bar a i}}{\sqrt{S_{1i}}} 
\ee
while in the degenerate case each $a_{i}$ is a particular linear combination of such quantities. 
Any state of the form (\ref{vac_form}) has several characteristic features. For each primary sector all components have 
the same sign -- that of $a_{i}$, and exponentially decrease in magnitude with their weight.  Moreover it contains only states 
for which the descendant content is the same in the holomorphic and anti-holomorphic sectors with equal coefficients. 
Note that for each $i, N$ the vector 
\be 
 \sum_{k_{i,N}} |k_{i,N}\rangle_{L} \otimes |k_{i,N}\rangle_{R} 
\ee
is independent of the particular choice of an orthonormal basis $|k_{i,N}\rangle_{L}$. 

We would like now to discuss some limitations of the ansatz. Firstly, we note that the expression (\ref{perturbed_CFT})
used in the derivation of \ref{var_en2} is a bare Hamiltonian. When the dimension of  one of the   perturbing operators 
is greater than 1 the vacuum energy diverges and one needs to introduce a regularisation and add  counterterms  
to (\ref{perturbed_CFT}). A logarithmic divergence appears at the second order for the thermal perturbation of the critical Ising model and is a universal term which must be present in the infinite space energy density. 
It was suggested in \cite{Cardy_var} to handle this divergence by adding  a term proportional to $\ln(\tau) $ to (\ref{perturbed_CFT}). We will discuss this 
proposal in more detail in section \ref{Ising_sec}. Another example of divergence in the vacuum energy arises for 
the subleading energy perturbation of the tricritical Ising model. In this case this is a perturbative power divergence and one may think 
it should not  affect the infinite space energy density. It should be noted however that for the negative sign of the coupling the theory is massive and has an unbroken supersymmetry \cite{Friedan_etal}. The infinite space energy density then should be zero (as confirmed by the TBA solution \cite{Zam_TBA}). On the other hand, as we noted above,  Cardy's ansatz always gives a negative energy density and thus in this case  we do not even seem to get a valid bound on the vacuum energy.  
It seems to us that the root of the problem is in the fact that the $N\to \infty$ tail of the variational vacuum (\ref{vac_form}) is different 
from the one in the actual vacuum. If we add an infinite counterterm to (\ref{perturbed_CFT}) then Cardy's asatz formally gives us a positive infinite energy density which at least is above the true vacuum energy. 
It is not clear to us how to improve  the ansatz for this theory to obtain a finite positive or zero energy density. 
It may happen though that the ansatz correctly identifies the underlying conformal boundary states (the RG boundary) but needs additional ingredients to give a useful bound on the energy. In this paper we consider two models. While the Ising field theory with $t\ne 0$ has a logarithmic divergence of the vacuum energy
the doubly perturbed TIM is free of divergences.

Another interesting problem emerges in the presence of several perturbations in relation to symmetry breaking. 
When our CFT has a symmetry e.g. the ${\mathbb Z}_{2}$ symmetry of the Virasoro minimal models, all bulk operators 
as well as all conformal boundary conditions either preserve the symmetry or break it. If we perturb this CFT by a symmetry 
preserving operator  then in the disordered region we expect Cardy's ansatz to give us a unique vacuum based on a symmetry preserving boundary condition. The variational states built on the symmetry breaking states will then have 
finite energy gaps above the vacuum. If we then switch on an additional  small symmetry breaking perturbation then by continuity 
the previously chosen symmetric boundary state will still have the minimal variational energy for sufficiently small 
symmetry breaking couplings. This seems unphysical as the vacuum must develop a non-vanishing expectation value of the order parameter no matter how small the external  symmetry breaking (magnetic) fields are. 
The $e^{-\tau H_{0}}$ term in (\ref{trial_vac}) expanded to the first order in $\tau$ and applied to the boundary condition coincides with the first order deformation 
of the boundary by an irrelevant local operator $T$ -- the stress-energy tensor. The whole exponential can be considered as 
some truncated version of the finite deformation. The fact that $T$ is irrelevant on the boundary ensures that in the far infrared 
we approach a conformal boundary. The $T$-perturbation is special in that the  operator $T$ is always available and due to its relatively small dimension is often the leading irrelevant operator for more complicated perturbations. Also the perturbation in (\ref{trial_vac}) has a simple geometric interpretation of putting the theory on a finite length cylinder making the ansatz analytically tractable.  
However $T$ is symmetry preserving and thus cannot give a 
non-vanishing magnetisation in the cases at hand. 
It is natural then to enlarge the ansatz to consider other perturbations by symmetry breaking irrelevant boundary operators. 
 On the other hand a symmetry breaking relevant boundary operator may be also available and may be induced on the boundary 
 by some of the bulk perturbations. In the case of the perturbed Ising model it was suggested in \cite{Cardy_var} to include 
 the boundary magnetic field perturbation into an enlarged variational ansatz based on the free conformal boundary state. 
 In this paper we are going to discuss both situations -- enlarging the ansatz by switching on a boundary relevant or irrelevant operator. 
It is hard to say something general about perturbing the boundary by another irrelevant operator so we will postpone the discussion of this case  until section \ref{small_sigma_sec}  where we consider a concrete example. 
 Perturbing by a relevant operator on  the other hand, when done together with the exponential in (\ref{trial_vac}), can be considered as 
 a local deformed boundary condition put on a strip of finite width so that it is tractable at least perturbatively. 
 The deformed boundary condition for a large circle (i.e. for $R\to \infty$) will approach a new conformal boundary state. Thus, for 
 a large but finite $R$ the corresponding vacuum can be described as this new conformal boundary state perturbed by the leading irrelevant operator along which the RG flow arrives to this IR fixed point. This means that we can treat perturbations by relevant boundary operators as a stability analysis of the solutions given by    (\ref{trial_vac}), (\ref{var_en2}). We can do this at the leading order keeping the discussion general. 
 
 Let us consider a conformal boundary condition with the boundary state $|\tilde a\rangle\!\rangle$  deformed by a relevant boundary operator $\psi$ of dimension $\Delta_{\psi}<1$ with a dimensionful coupling $\alpha$. The cylinder amplitude (\ref{var_en1})
 is deformed  by inserting  
 \be
e^{-\alpha \int\limits_{0}^{R}\psi(y)dy }
 \ee
 on both ends. %
 For the  $H_{0}$ average one can still use the quantisation in the $y$-direction and the cylinder amplitude at the leading order 
 can be expressed in terms of the Casimir energy. The latter receives the first contribution from the perturbation at the quadratic order 
 in $\alpha$. The linear order in $\alpha$ thus can only come from the bulk perturbing operators and is given by an integrated two-point function 
 \be
 \langle \phi_{i}(0,0) \rangle_{\alpha, a} = \langle \phi_{i} \rangle_{\alpha=0} - 2\alpha \int\limits_{-\infty}^{\infty} dy \langle \phi_{i}(0,0) \psi(y)\rangle_{\rm strip} + {\cal O}(\alpha^2) 
 \ee
 where $\psi$ is inserted on one end of the strip and the factor of 2 takes care of the insertion on the other end. 
 Mapping the correlator from the strip to the upper half plane and taking the integral we obtain  
 \be
  \frac{\langle\!\langle  a, \alpha |e^{-\tau H_{0}} \phi_{i}(0,0)e^{-\tau H_{0}} | a, \alpha \rangle\!\rangle}{\langle\!\langle  a, \alpha |e^{-2\tau H_{0}} | a, \alpha \rangle\!\rangle} = A_{i}^{a} \left( \frac{\pi}{4\tau}\right)^{\Delta_{i}}
  - {}^{(a)}\!B_{i}^{\psi}\alpha \left(\frac{\pi}{4\tau} \right)^{\Delta_{i}+\Delta_{\psi}-1} f(\Delta_{\psi}) +  {\cal O}(\alpha^2) 
 \ee
 where $|a,\alpha\rangle\!\rangle$  denotes the deformed boundary state, $ {}^{(a)}\!B_{i}^{\psi}$ is the bulk-boundary OPE coefficient (see Appendix \ref{appendix_OPE} for our conventions) and 
 \be
 f(\Delta_{\psi}) = \sqrt{\pi}\frac{\Gamma\left(\frac{\Delta_{\psi}}{2}\right)}{\Gamma\left(\frac{1+\Delta_{\psi}}{2}\right)} \, . 
 \ee
 Since the sign of $\alpha$ is not fixed we see that the variational energy will always go down at the linear order in $\alpha$ unless 
 a special condition takes place:
 \be \label{equilibrium_gen}
 \sum_{i}   {}^{(a)}\!B_{i}^{\psi}  \lambda_{i}  \left(\frac{\pi}{4\tau^{*}_{a}} \right)^{\Delta_{i}} = 0
 \ee 
 where $\tau_{a}^{*}$ gives the minimum over $\tau$ of the unperturbed variational energy $E_{a}(\tau)$. 
 We will encounter a situation when this equilibrium condition holds in a concrete example of perturbations 
 of the tricritical Ising model which we discuss in detail in section \ref{tricrit_section} (see subsection \ref{transition_subsec} in particular). The simplest way in which a variational solution can be stable at the leading order with respect to  relevant boundary perturbations is 
 when all relevant bulk-boundary coefficients ${}^{(a)}\!B_{i}^{\psi}$ vanish.

At the second order in $\alpha$ both the average of $H_{0}$ and the average of the perturbing operators receive a correction. 
  The cylinder partition function receives a 
leading contribution of the form 
\be
\langle\!\langle a, \alpha|e^{-2\tau H_{0}}|a,\alpha\rangle\!\rangle = e^{-{\cal E}_{\rm Cas}(\alpha)  R } + \dots
\ee
 where the Casimir energy has the following scaling form 
 \be
 {\cal E}_{\rm Cas}(\alpha)  =b \alpha^{\frac{1}{1-\Delta_{\psi}}}  -\frac{1}{2\tau} C(\alpha (2\tau)^{1-\Delta_{\psi}}) 
 \ee
where the first term is the non-perturbative infinite volume boundary energy while the function $C=C(x)$ sums up the perturbative corrections. Generalising formula (\ref{cas_cont}) we obtain 
\be \label{gen_f_H0}
\frac{\langle\!\langle  a, \alpha |H_{0}e^{-2\tau H_{0}}|a,\alpha \rangle\!\rangle }{\langle\!\langle a, \alpha|e^{-2\tau H_{0}}|a,\alpha\rangle\!\rangle} = \frac{R}{{(2\tau)^2}}\Bigl[ C(\tilde \alpha) -\tilde \alpha(1-\Delta_{\psi})C'(\tilde \alpha) \Bigr] \, 
\ee
where $\tilde \alpha = \alpha (2\tau)^{1-\Delta_{\psi}}$. 
The function $C$ is in general a complicated function which  is  known analytically in an integral form for integrable boundary flows. 
If one knows where the boundary flow arrives in the infrared   one may hope to analyse the small and large $\alpha$ asymptotics of the above general expression. The leading order is quadratic and can be easily calculated to give 
\be \label{cas_en_b}
{\cal E}_{\rm Cas}(\alpha)  = - \frac{1}{2\tau} \Bigl[  \left(\frac{\pi c}{24}\right)  + \tilde \alpha^2 \left( \frac{\pi}{2}\right)^{2\Delta_{\psi}-1} 
\frac{\Gamma(\frac{1}{2} - \Delta_{\psi})\Gamma(\Delta_{\psi})}{2\sqrt{\pi}} + {\cal O}(\tilde \alpha^3) \Bigr] \, . 
\ee
 This expression is valid for all $\Delta_{\psi} \ne 1/2$. For  $\Delta_{\psi} = 1/2$  we have a logarithmic divergence due to a resonance. 
 We will assume here this is not the case. Note that for $1/2<\Delta_{\psi}<1$ the integral at hand has a power divergence and 
  (\ref{cas_en_b}) is obtained by dropping this divergence using analytic continuation. Substituting the last expression into 
  (\ref{gen_f_H0}) we obtain the leading order contribution 
  \be
  \frac{\langle\!\langle  a, \alpha |H_{0}e^{-2\tau H_{0}}|a,\alpha \rangle\!\rangle }{\langle\!\langle a, \alpha|e^{-2\tau H_{0}}|a,\alpha\rangle\!\rangle} = \frac{R}{(2\tau)^2}\Bigl[  \frac{\pi c}{24} - \frac{\tilde \alpha^2}{\sqrt{\pi}}  \left( \frac{\pi}{2}\right)^{2\Delta_{\psi}-1} \Gamma(3/2 - \Delta_{\psi})\Gamma(\Delta_{\psi}) \Bigr]
  \ee
 up to higher order terms in $\tilde \alpha$. It is interesting to note that the correction term is negative for all $0<\Delta_{\psi}<1$, $\Delta_{\psi} \ne 1/2$. Thus, even if the linear order contribution vanishes, say due to all relevant bulk-boundary OPE coefficients 
 being zero, at the second order the perturbation may be unstable. The only way the second order correction to the complete variational energy can be positive is if there is a positive contribution coming from the perturbing terms. Those corrections can be expressed via 
the integrated three point function with one bulk and two copies of the boundary operator $\psi$ inserted. Such a correlator depends on conformal blocks. It would be interesting to calculate this correction for particular models. We leave this question to future work. 
 In the rest of the paper we focus on the linear stability. 
 
 As a final remark in this section we would like to mention that boundary RG flows can be triggered by bulk deformations \cite{FGK}. 
 If one starts with a CFT on a half plane (or half-cylinder) with a conformal boundary condition then deforming the bulk by an operator $\phi_{i}$ 
 may induce a beta function for a boundary operator $\psi_{j}$ via the bulk-boundary OPE coefficient $B_{i}^{j}$. We do not see an immediate connection between our analysis of the variational method and  this mechanism. In the picture of RG boundaries we start with an interface between the UV CFT and the perturbed theory. This interface then flows to a boundary condition in the infrared. Thus, when we look at any additional perturbation we add it 
 on the massive side of the interface rather than on the CFT side as one does in the bulk induced boundary RG flows.


\section{Ising field theory} \label{Ising_sec}
\setcounter{equation}{0}
\subsection{Some generalities}
In this section we consider perturbations of the critical Ising model (\ref{IFT}). 
 The mapping of regions in terms of the conformal boundary conditions was summarised in the introduction. For the sake of reader's convenience we present the same diagram here on 
  Figure \ref{Ising_diagram}. 


\begin{center}
\begin{figure}[H]  
\centering
\begin{tikzpicture}[>=latex]
\draw[->] (-3,0) --(4.3,0);
\draw[->] (0,-3)--(0,3);
\draw[red,very thick] (0,0)--(3.8,0);
\draw[blue, very thick] (-3,0)--(0,0);
\draw (4.6,0  ) node {$t$};
\draw (0,3.3) node {$h$};
\draw (2.2,2.2) node {$|-\rangle\!\rangle $};
\draw (-2.2,2.2) node {$|-\rangle\!\rangle $};
\draw (-2.2,-2.2) node {$|+\rangle\!\rangle$};
\draw (2.2,-2.2) node {$|+\rangle\!\rangle$};
\draw (-4,0) node {$|+\rangle\!\rangle \oplus |-\rangle\!\rangle$};
\draw (2,0.3) node {$|F\rangle\!\rangle$};
\end{tikzpicture}
\caption{The unstable manifold of IFT and the associated conformal boundary states.}
\label{Ising_diagram}
\end{figure}
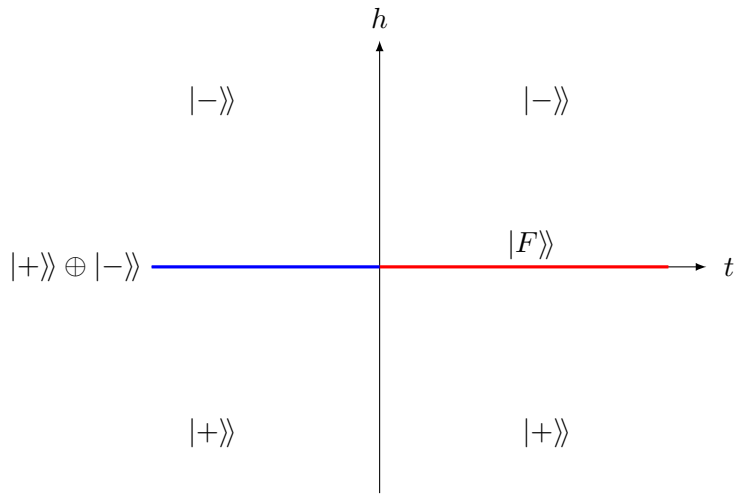
\end{center}


The numerical identification done in \cite{AK_Ising} that lead to the phase diagram on Figure \ref{Ising_diagram} relied primarily on 
calculating the ratios of the low-lying components of  the vacuum vector and comparing them to those in 
the conformal boundary states. The two simplest ratios are 
\be \label{Ising_overlaps}
\Gamma_{\epsilon} = \frac{\langle \epsilon| 0\rangle_{t,h}}{\langle  0| 0\rangle_{t,h}} \, , \quad 
\Gamma_{\sigma} = \frac{\langle \sigma| 0\rangle_{t,h}}{\langle  0| 0\rangle_{t,h}}
\ee
where $| 0\rangle_{t,h}$ denotes the vacuum of the perturbed theory (\ref{IFT}) obtained as a vector in the 
CFT state space. Also $|0\rangle$ denotes the conformal vacuum. It was found that these ratios for large $R$ approach the asymptotic values 
$\Gamma_{\epsilon} \to \pm 1$, $\Gamma_{\sigma}\to \pm 2^{1/4}$ that correspond to those in the conformal 
boundary states (\ref{conf_bcs}). Obtaining  the asymptotics of such ratios 
from TCSA numerics in general is 
a tricky business.  TCSA is an approximation scheme and the truncation errors become large for large $R$ so one searches 
for a physical window where $R$ is large enough so that the finite size corrections are small but at the same time $R$ is small enough so that the truncation errors are small. For the vacuum component ratios this is particularly relevant as they approach their asymptotic values slowly, as inverse powers of $R$. 
The reader can find a more detailed discussion of   TCSA computational strategies
in appendix \ref{appendix_TCSA}. Here it suffices to note that  the identification of conformal boundary states underlying the vacuum becomes more reliable if we are able to control the leading corrections to the vacuum state at finite $R$. The Cardy's ansatz 
proposes a concrete form of these corrections. 
The  variational energies (\ref{var_en2})
 for the Ising field theory (\ref{IFT}) have the following form 
\bea \label{Ising_varens}
E_{+} &= &R\Bigl[ \frac{\pi}{48 (2\tau)^2 } + \frac{t\pi}{4\tau} + h 2^{1/4} \left(\frac{\pi}{4\tau}\right)^{1/8} \Bigr] \, , \nonumber \\
E_{-} &= &R\Bigl[ \frac{\pi}{48 (2\tau)^2 } + \frac{t\pi}{4\tau} - h 2^{1/4} \left(\frac{\pi}{4\tau}\right)^{1/8} \Bigr]\, , \nonumber \\
E_{ F}&=& \Bigl[\frac{\pi}{48 (2\tau)^2 } - \frac{t\pi}{4\tau} \Bigr] \, . 
\eea
Before we look in detail at the ansatz predictions  it is instructive to 
look at the exact solution one has for $h=0$.   

\subsection{Pure thermal perturbation} \label{Ising_thermal}

For $h=0$ the theory (\ref{IFT}) describes the  perturbation of massless fermions by a mass term with mass\footnote{The mass coupling is usually defined as $m=-2\pi t$ hence the absolute value to get the physical mass.}  $|m|=2\pi |t|$. 
The exact vacuum state of the perturbed theory written in terms of the massless fermion creation operators is\footnote{At finite size the true vacuum is unique and is a vector in the NS sector. For $t<0$ there is a vector in the Ramond sector whose energy is above the vacuum energy by exponentially small term in $R$. For $Rt\to -\infty$ it becomes after rescaling the $|\sigma\rangle\!\rangle$ Ishibashi state.}
\be \label{exact_vac}
|0\rangle_{m} = {\cal N} {\rm exp}\left(i{\rm sign}(t)\sum_{n=0}^{\infty}f_{n} \bar a_{n+1/2}^{\dagger} a_{n+1/2}^{\dagger}\right)  |0\rangle
\ee
where 
\be
f_{n} = \sqrt{1 + \left(\frac{n+1/2}{\tilde t}\right)^2 } - \frac{n+1/2}{|\tilde t|} \,  ,
\ee
$ \tilde t = t R$ is the dimensionless coupling and 
${\cal N}={\cal N}(\tilde t)$ is the normalisation factor. 
Noting that when $n\ll |\tilde t| $
\be \label{f_approx}
f_{n} = e^{-(n+1/2)/{|\tilde t|}} + {\cal O}\left(      \left(\frac{n+1/2}{\tilde t} \right)^3 \right) 
\ee
  the corresponding components in $|0\rangle_{m}$ can be approximated by 
  \bea \label{low_tail}
&&  |0\rangle_{m} \sim {\cal N} {\rm exp}\left(i{\rm sign}(t)\sum_{n\ll \tilde t} e^{-(n+1/2)/{\tilde t}} \bar a_{n+1/2}^{\dagger} a_{n+1/2}^{\dagger}\right)  |0\rangle
\nonumber \\
&& = {\cal N'} e^{-\tau_{e} H_{0}}  {\rm exp}\left(i{\rm sign}(t) \sum_{n\ll \tilde t }\bar a_{n+1/2}^{\dagger} a_{n+1/2}^{\dagger}\right)  |0\rangle 
\eea
where 
\be \label{taue}
\tau_{e} = \frac{1}{4\pi |t|} = \frac{1}{2|m|}\, , \qquad {\cal N'}= {\cal N}e^{c/(24 |\tilde t|)} \, .
\ee
To compare the last expression in (\ref{low_tail})  with (\ref{ishibashi}), (\ref{vac_form}) we introduce 
 the states 
\be \label{mode_basis} 
 \Bigl|\frac{\bf k}{2}\Bigr\rangle = \eta_{n}a^{\dagger}_{k_{1}/2} \dots  a^{\dagger}_{k_{n}/2} \bar a^{\dagger}_{k_{1}/2} \dots \bar a^{\dagger}_{k_{n}/2}  |0\rangle \, ,\quad k_{1} > k_{2} > \dots > k_{n}
\ee
where 
\be
\eta_{n} = \left\{ \begin{array}{l@{\qquad}l} 
1\, , & n \mbox{  even} \\[1ex] 
i\, , & n \mbox{  odd  }\, ,
\end{array} \right.
\ee
 $k_{i}$ are positive odd integers and $ {\bf k}=(k_{1}, k_{2}, \dots , k_{n}) $ which labels such a state is an integer partition of the state's total conformal weight 
 $\sum_{j} k_{j} $ into odd numbers without repetition. For $n$ even these states are of the  form $|{\bf k}\rangle_{L} \otimes |{\bf k}\rangle_{R}$ where 
\be \label{chiral_b}
|{\bf k}\rangle_{L} = a^{\dagger}_{k_{1}} \dots  a^{\dagger}_{k_{n}} |0\rangle 
\ee
form an orthonormal basis in ${\cal H}_{{\bf 1},N}^{L}$ with  $N=\sum k_{i} $ and $ |{\bf k}\rangle_{R}$ denote the conjugate basis as in (\ref{ishibashi}).
Furthermore, in  our conventions 
\be \label{epsilon_convention}
|\epsilon\rangle = i a^{\dagger}_{1/2}\bar a^{\dagger}_{1/2} |0\rangle\, 
\ee 
and for $n$ odd we can write 
 \be
  \Bigl|\frac{\bf k}{2}\Bigr\rangle = |{\bf k}, \epsilon \rangle_{L} \otimes |{\bf k}, \epsilon\rangle_{R}
 \ee
where the states 
\be
|{\bf k}, \epsilon \rangle_{L} = \left(\frac{2}{k_{n}}\right)^2 a^{\dagger}_{k_{1}/2} \dots  a^{\dagger}_{k_{n-1}/2} L_{(1-k_{n})/2}|\epsilon\rangle 
\ee
form an orthonormal basis in  ${\cal H}_{{\bf \epsilon} ,N}^{L}$, $N=-1/2+ \sum k_{i}$ and the states $|{\bf k}, \epsilon\rangle_{R}$ 
are conjugate. 

Using this and formula (\ref{ishibashi})  we see that  when $t>0$ the vector on the right hand side of (\ref{low_tail}) can be rewritten as a Cardy's ansatz state
\be
 |0\rangle_{m} \sim {\cal N'} e^{-\tau_{e} H_{0}} (|1\rangle\!\rangle - |\epsilon\rangle\!\rangle) = {\cal N'} e^{-\tau_{e}H_{0}}   |F\rangle\!\rangle \, . 
\ee 
As can be seen from (\ref{f_approx}) when $n$ is comparable or larger than $|\tilde t|$ we should expect discrepancies with the ansatz. 
In particular the UV tail of the exact vacuum state involves states with $n\gg |\tilde t|$ for which the corresponding components\footnote{Of course a generic state in the tail will contain both types of modes -- with $n> |\tilde t|$ 
and with $n<|\tilde t|$. For simplicity here we look only at contributions of large mode numbers.} look like 
\be
|0\rangle_{m} \sim {\cal N} {\rm exp}\left(i{\rm sign}(t) \sum_{n\gg \tilde t} \frac{\tilde t }{2 (n+1/2)} \bar a_{n+1/2}^{\dagger} a_{n+1/2}^{\dagger}\right)  |0\rangle 
\ee
that is  the coupling and the weight dependences are essentially interchanged. Qualitatively the components in the UV tail are suppressed much stronger than in 
the Cardy's ansatz. 

\begin{center}
\begin{figure}[h!]

\begin{minipage}[b]{0.5\linewidth}
\centering
\includegraphics[scale=0.8]{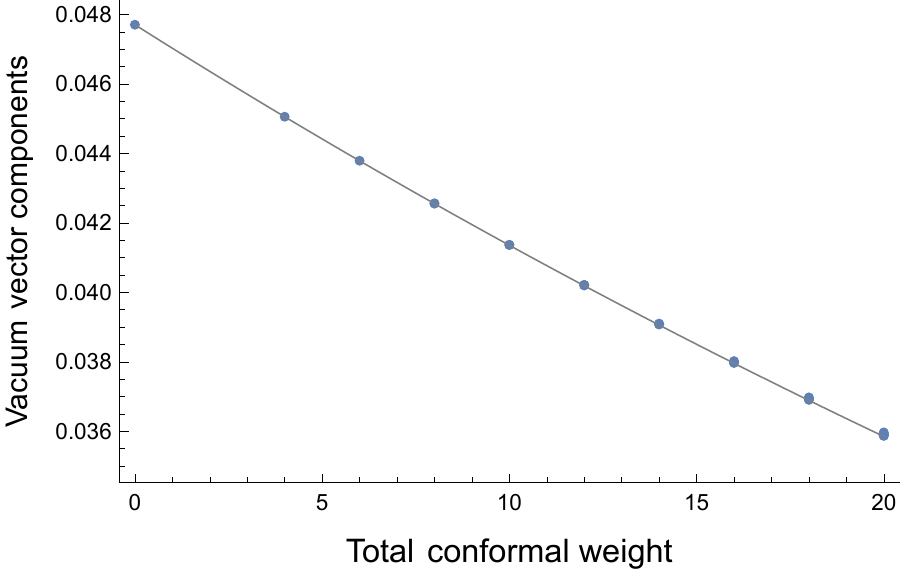}
\end{minipage}%
\begin{minipage}[b]{0.5\linewidth}
\centering
\includegraphics[scale=0.8]{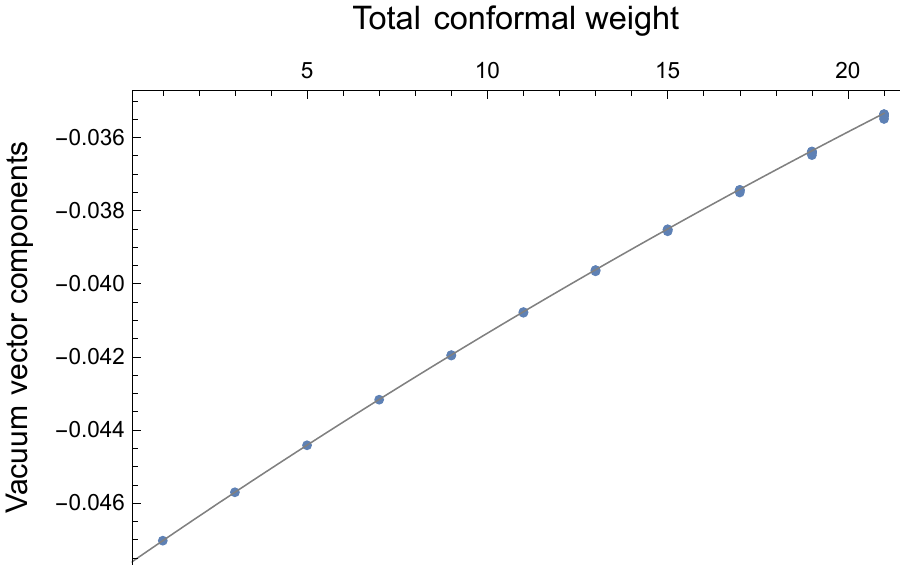}
\end{minipage}
\caption{The exact vacuum vector components $C_{i}$ against the total conformal weight of the basis vectors for $\tilde t=35$. The left plot 
represents the vacuum sector and the right plot -- the $\epsilon$-sector. The solid grey line is the exponential function $\pm\, {\cal N} e^{-{\rm weight}/(2\tilde t)}$.}
\label{fig_spec1}
\end{figure}
\end{center}

To get some intuition on how the corrections look for the  intermediate components it is instructive to look at a plot of the vacuum vector components 
written in an orthonormal factorised basis. Such a basis is provided by the states  (\ref{mode_basis}). If we denote them as $|i\rangle$ where $i$ is some 
label then we can write 
\be
|0\rangle_{m} = \sum_{i} C_{i} |i\rangle \, .
\ee
On figure \ref{fig_spec1} we see the coefficients $C_{i}$ plotted against the total conformal weight of $|i\rangle$ for $\tilde t=35$. The components 
shown go up to the weight $21$ and thus involve the modes $n\le n_{\rm max}=10$. 

\begin{center}
\begin{figure}[h!]

\begin{minipage}[b]{0.5\linewidth}
\centering
\includegraphics[scale=0.8]{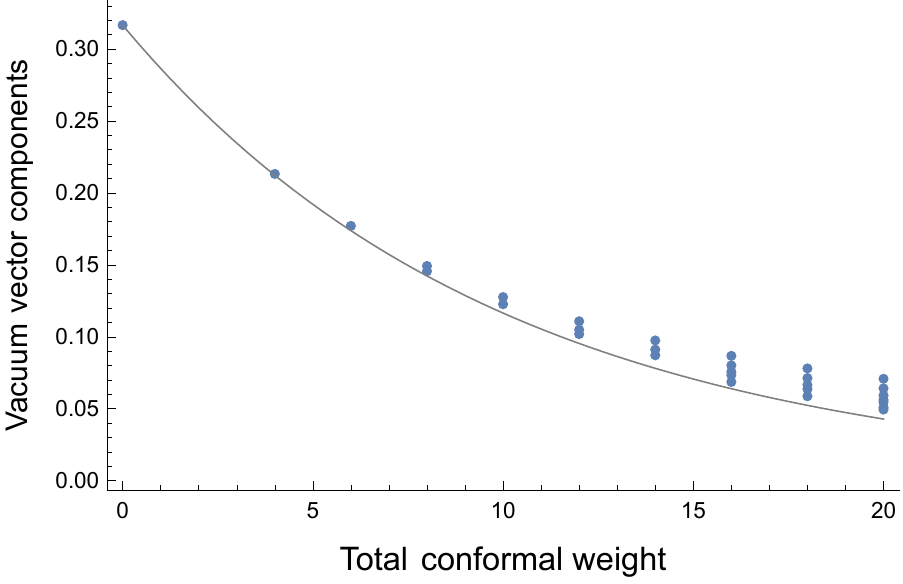}
\end{minipage}%
\begin{minipage}[b]{0.5\linewidth}
\centering
\includegraphics[scale=0.8]{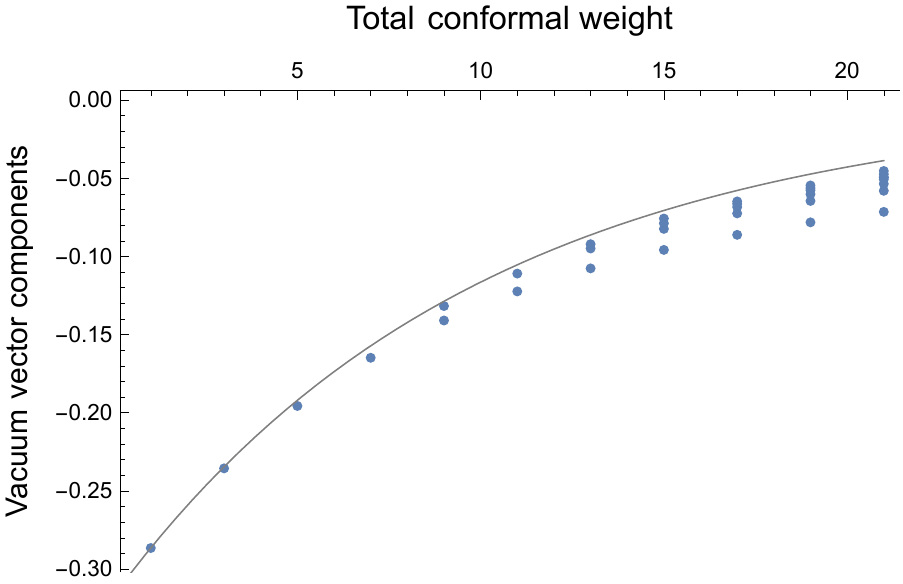}
\end{minipage}
\caption{The exact vacuum vector components $C_{i}$ against the total conformal weight of the basis vectors for $\tilde t=5$. The left plot 
represents the vacuum sector and the right plot -- the $\epsilon$-sector. The solid grey line is the exponential function $\pm\, {\cal N} e^{-{\rm weight}/(2\tilde t)}$.}
\label{fig_spec2}
\end{figure}
\end{center}

We see that the components shown all fall on the Cardy's ansatz trajectory. On Figure \ref{fig_spec2} the same components are shown for a 
smaller value $\tilde t=5$. In that plot we observe that for high enough weights the components deviate from the Cardy's ansatz exponential. Moreover 
the high weight components of the same weight split into different values that shows deviation from Ishibashi states. One can account for  these 
differences by adding  to the exponential in the ansatz   higher dimension fermion bilinears, that is the KdV charges:
\be \label{kdv}
I_{2n+1} = \int\limits_{0}^{R}\! dy :\! \psi \partial^{2n+1} \psi \!:\!\! (y) 
\ee
with $n\ge 1$. 

Let us next look at the Cardy's ansatz prediction for the pure thermal perturbation. 
From (\ref{Ising_varens})  we find that when $h=0$ and $t>0$ (the disordered region) the minimum is obtained for $E_{ F}$ at 
\be \label{Ising_taustar}
\tau =\tau^{*}= \frac{1}{24|t|} = \frac{\pi}{12 |m|} \, . 
\ee
When $h=0$ and $t<0$ (the symmetry breaking region) the minimum is degenerate and is obtained by $E_{+}$ and $E_{-}$ 
for the same value of $\tau$ given 
in (\ref{Ising_taustar}). To find the true vacuum we need to diagonalise the appropriate $2\times 2$ matrix (\ref{deg_matrix}). 
Since  $h_{+-}=1/2$ we obtain 
\be
{\cal M} = \left( \begin{array}{cc} 
E & \epsilon_{+-}Re^{-3|m| R} \\
\epsilon_{+-}Re^{-2|m|R}& E
\end{array}
\right) 
\ee
where 
\be
E = E_{+} (\tau^{*})=E_{-}(\tau^{*}) = \frac{3|m|^2R}{4\pi} 
\ee
and $\epsilon_{+-}$ is a dimension $mass^2$ quantity which depends only on the couplings and, as suggested in \cite{LVT},  can be approximated by 
(\ref{deg_matrix0}).  If $\epsilon_{+-}<0$ (that is true if we use  (\ref{deg_matrix0})) then the lowest eigenvector is 
\be
e^{-\tau^{*}H_{0}}(|+\rangle\!\rangle + |-\rangle\!\rangle) \qquad \mbox{ with eigenvalue } E_{0}=E + \epsilon_{+-}Re^{-3|m| R}
\ee
and the first excited eigenvector is 
\be
e^{-\tau^{*}H_{0}}(|+\rangle\!\rangle - |-\rangle\!\rangle) \qquad \mbox{ with eigenvalue } E_{0}=E - \epsilon_{+-}Re^{-3|m| R} \, .
\ee
From (\ref{conf_bcs}) we see that the true vacuum is contained entirely in the NS sector of the theory while the first excited state lies in the Ramond 
sector and has an exponentially small gap. 

It is interesting to observe that $\tau^{*}$ given in (\ref{Ising_taustar}) is different from the value $\tau_{e}$ given in (\ref{taue}) which we obtained by analysing 
the exact solution. This fits with  the discrepancy observed in \cite{LVT} in a plot of vacuum components obtained via TCSA (Figure 3 on page 16). 
The difference with Cardy's ansatz is visible on that plot even for lowest conformal weights which seem to lie on an exponent with a larger coefficient. 
As rightly noted in \cite{LVT} the discrepancy could be related to the UV logarithmic divergence present in the model. It was suggested in 
\cite{Cardy_var} to  account for this divergence by adding the appropriate term proportional to $\ln \tau$. Fixing the coefficient\footnote{The sign in front of the logarithmic term is chosen so that at the minimum we obtain a contribution  of $+(m^2/4\pi)\ln |m|$ giving the standard Onsager's singularity term in free energy. } we obtain  the following 
modified variational energy for the free boundary condition
\be \label{log_term}
E_{F}'= \Bigl[\frac{\pi}{48 (2\tau)^2 } - \frac{t\pi}{4\tau}  -\pi t^2 \ln \tau \Bigr] 
\ee
 Finding the extrema of this function for $t>0$ amounts to a quadratic equation for which the root corresponding to a local minimum is 
 \be
 \tau^{**} = \frac{1}{8t}\left( 1- \frac{1}{\sqrt{3}} \right) = \frac{\pi }{4|m|}\left( 1- \frac{1}{\sqrt{3}} \right) \approx \frac{0.33}{|m|} \, . 
 \ee
This gets us closer to the value $\tau_{e} = 0.5/|m|$ than $\tau^{*}\approx 0.26/|m|$ but is still sufficiently different to make us think that some other modifications of the ansatz are needed, for example adding the KdV charges we mentioned before.   
The situation with the value of $\tau$ is even more puzzling given a good match between the slope of the chiral (left-right) entanglement entropy 
obtained for this model via the Cardy's ansatz without the logarithmic term (that is using $\tau^{*}$) and TCSA\footnote{We have calculated the chiral entanglement entropy for the exact 
vacuum vector \ref{exact_vac} and found exactly  the same coefficient predicted by the Cardy's ansatz without the logarithmic term. We plan to report on this calculation and discuss the chiral entanglement in more detail elsewhere \cite{Kon_wip} } \cite{LVT}. 

Notwithstanding  the above quantitative discrepancy we note that the case of the thermal coupling in the Ising model is clearly special due to the divergence. 
TCSA computations presented in \cite{LVT} for more relevant perturbations such as pure magnetic perturbation in the Ising model and pure thermal and magnetic perturbation in the TIM show excellent agreement between the ansatz and the low lying components of the vacuum vector obtained numerically.

\subsection{Mixed thermal and magnetic perturbations for $t<0$} \label{Ising_SSB}
We assume now that $h>0$ and $t\ne 0$. We will label the RG trajectory by a scaling parameter\footnote{Note that in some papers, e.g. in \cite{FZ}, 
one defines a similar parameter using $|m|$  rather than $|t|$. }
\be
\xi = \frac{h}{|t|^{15/8} }
\ee
which is convenient for small values of $h$. The variational energies $E_{\pm}$ can then be rewritten as 
\be \label{Ising_varens2}
E_{\pm }(\tilde \tau)= Rt^2 \Bigl[ \frac{1}{12 \pi \tilde \tau^2 } +  \frac{ {\rm sign}(t)}{\tilde \tau} \pm  \frac{ \xi  2^{1/4}}{\tilde \tau^{1/8}} \Bigr] 
\ee
where we introduced a dimensionless variable 
\be
\tilde \tau = \frac{4|t| \tau}{\pi}\, . 
\ee
We will first consider the low temperature region:   $t<0$.  As the free boundary condition does not couple to the magnetic field in this region  we have 
\be
E_{-}(\tilde \tau) <E_{+}(\tilde \tau)  \, , \qquad E_{-}(\tilde \tau)<E_{ F} (\tilde \tau)
\ee
for all $h>0$ and $t<0$. In fact the minimum of $E_{ F}$ is at $\tau \to \infty$ and thus can be discarded for all $t<0$. 
For sufficiently small $\xi$ the variational energy $E_{+}(\tilde \tau)$ has a local minimum at a finite value of $\tilde \tau$. The corresponding smeared 
state thus represents a metastable state.  For a certain threshold value of $\xi$ the minimum disappears just like it does in the classical Landau or mean field 
theory.

For $h=0$ we discussed the near degeneracy of the vacuum in the previous subsection. The true vacuum is given by the superposition 
 $e^{-\tau^{*}H_{0}}(|+\rangle\!\rangle + |-\rangle\!\rangle)$. Switching on $h>0$ replaces this by $e^{-\tau H_{0}} |-\rangle\!\rangle$ 
 for an appropriate $\tau$. It is tempting to connect this change with an RG flow triggered by switching on the boundary identity field on 
 the $ |-\rangle\!\rangle$ component with a positive coupling and on the  $ |+\rangle\!\rangle$ component with a negative coupling. We can indeed observe a connection 
 between the flow and the variational method if we analyse the leading finite size corrections. To that end consider a trial state given by a 
 superposition 
 \be
|\tau, \beta_{\pm} \rangle = e^{-\tau H_{0}}  (e^{\beta_{+}}  |+\rangle\!\rangle + e^{\beta_{-}}  |-\rangle\!\rangle ) 
 \ee
where $\beta_{\pm}$ are additional variational parameters. Such a superposition describes a local boundary condition if each $\beta_{\pm}$ 
depends linearly on $R$ when it gives a superposition of the two conformal boundary conditions  deformed 
by the boundary identity fields. We will see how this will come out from the variational method. For the average of the variational energy we obtain 
\be
\frac{\langle \tau, \beta_{\pm}|H |\tau, \beta_{\pm}  \rangle}{\langle \tau, \beta_{\pm} |\tau, \beta_{\pm}  \rangle }
= Rt^2 \Bigl[ \frac{1}{12 \pi \tilde \tau^2 } - \frac{ 1}{\tilde \tau} +  \frac{\epsilon_{+-}e^{-3|m| R}}{t^2 \cosh(\beta_{+}-\beta_{-})} +  \frac{ \xi  2^{1/4}}{\tilde \tau^{1/8}} {\rm tanh}(\beta_{+}-\beta_{-})  \Bigr] \, .
\ee
Considering $\xi$ to be small we can take $\tau=\tau_{*}$  so that 
\be
\frac{\langle \tau^{*}, \beta_{\pm}|H |\tau^{*}, \beta_{\pm}  \rangle}{\langle \tau^{*}, \beta_{\pm} |\tau^{*}, \beta_{\pm}  \rangle }  = R \Bigl[ 
\frac{3|m|^2}{4\pi}  + \frac{\epsilon_{+-} e^{-3|m| R}}{ \cosh(\beta_{+}-\beta_{-})} +   \xi t^2 2^{1/4}(6\pi)^{1/8} {\rm tanh}(\beta_{+}-\beta_{-})  \Bigr]\, .
\ee 
When $\epsilon_{+-}<0$ this expression considered as a function of $\beta_{+}-\beta_{-}$ has a local minimum when
\be
\sinh( \beta_{+} - \beta_{-} ) = C e^{3|m|R}\, , \qquad C = \xi  2^{1/4} (6\pi)^{1/8} \left(\frac{t^2}{\epsilon_{+-}}\right) \, .
\ee
For large $R$ this gives 
\be
 \beta_{+} - \beta_{-} \sim  - {\rm sign}(h) 3|m|R  -  {\rm sign}(h)\ln (2|C|) \, .
\ee
So indeed the leading finite size corrections give $\beta_{\pm}$ that depend linearly on $R$ and  describe the RG flow 
of the superposition of the two fixed spin boundary conditions triggered  by the boundary identity operators. 
It is curious that the leading value of the boundary identity coupling we obtained is independent of $h$. This may be due to
the fact that the small $h$ and large $R$ limits do not commute. 

\begin{center}
\begin{figure}[H]
\begin{minipage}[b]{0.5\linewidth}
\centering
\includegraphics[scale=0.82]{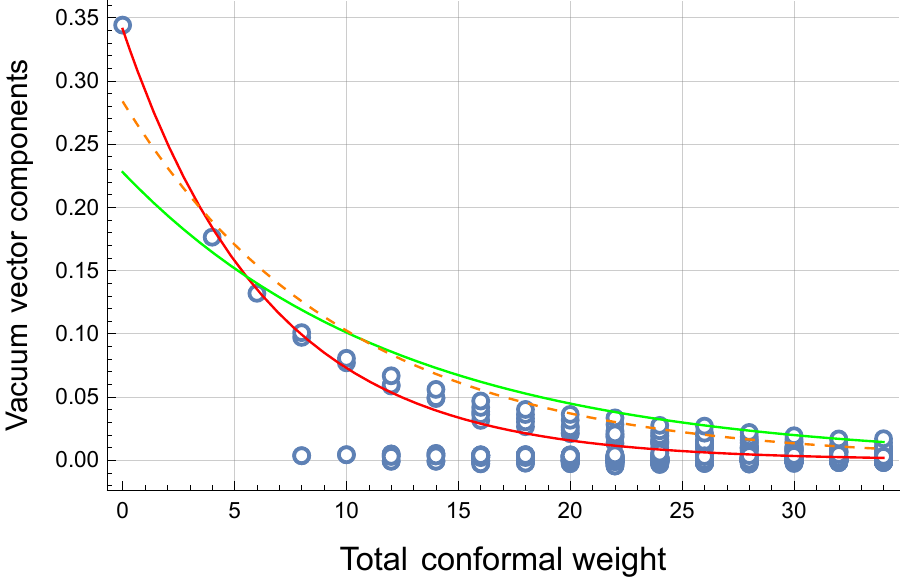}
\end{minipage}%
\begin{minipage}[b]{0.5\linewidth}
\centering
\includegraphics[scale=0.82]{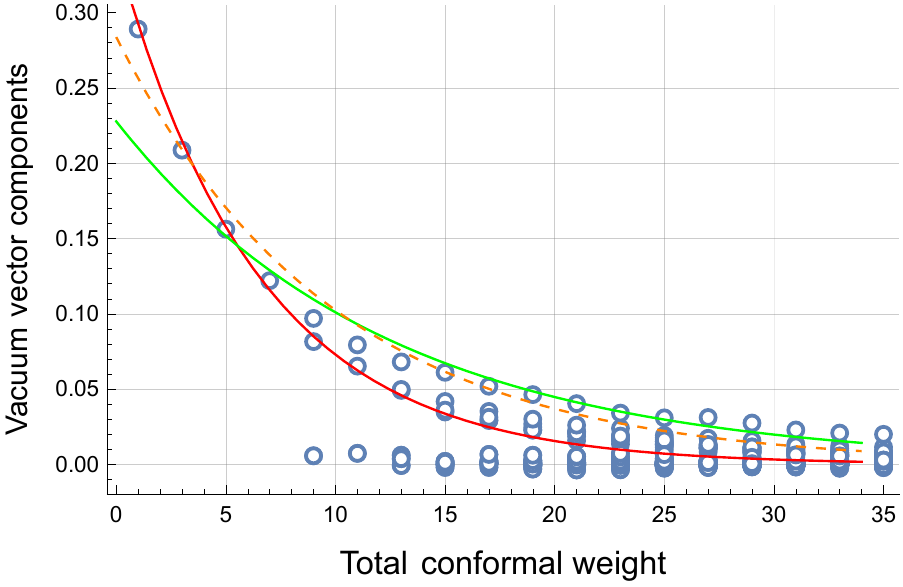}
\end{minipage}
\caption{The TCSA vacuum vector components $C_{i}$ against the total conformal weight of the basis vectors taken for the mixed perturbation with $\xi=1$, $t<0$ and at  $|m|R=20$. The left plot 
represents the vacuum sector and the right plot -- the $\epsilon$-sector. The green line corresponds to the Cardy ansatz with $\tau^{*}_{1}$, 
the orange dashed line  to the Cardy ansatz with $\tau^{*}_{2}$ and the red line  to the best fit exponential with $\tau_{\rm fit}^{*}$. }
\label{fig_spec3}
\end{figure}

\end{center}

\begin{center}
\begin{figure}[H]
\centering
\includegraphics[scale=0.82]{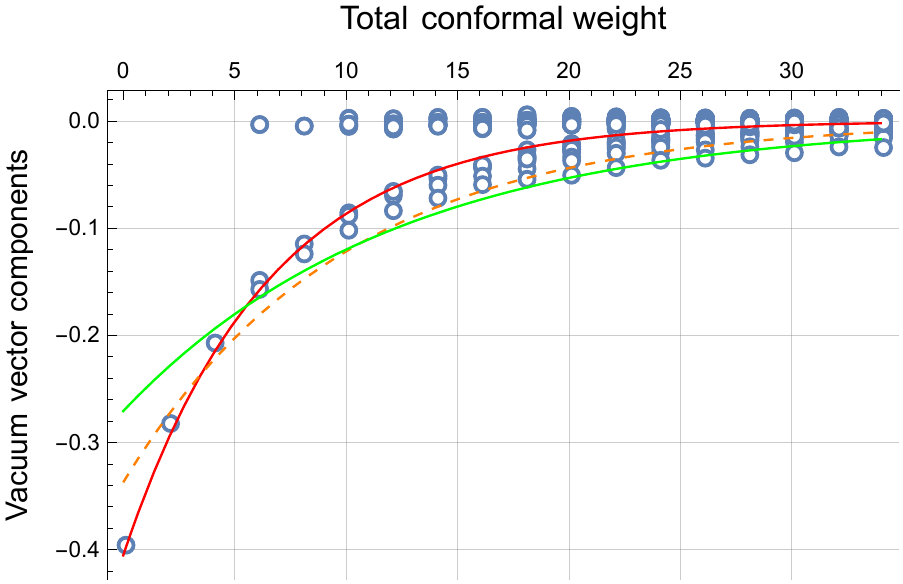}
\caption{The TCSA vacuum vector components $C_{i}$ in the $\sigma$-sector plotted against the total conformal weight. 
See the caption to Figure \ref{fig_spec3} for more detail.}
\label{fig_spec4}
\end{figure}

\end{center}

Cardy's ansatz predicts that for all $t<0$  the underlying conformal boundary condition (RG boundary) is $|-\rangle\!\rangle$ for $h>0$ 
and $|+\rangle\!\rangle$ for $h<0$. This was also observed in \cite{AK_Ising} based on the overlaps (\ref{Ising_overlaps}). We can get 
a more detailed confirmation and also a comparison with Cardy's ansatz by looking at the weight spectrum of the vacuum state similar to 
those depicted on Figure \ref{fig_spec2}. As an illustration on Figures \ref{fig_spec3} and \ref{fig_spec4} we show  TCSA numerical results\footnote{The  numerics on figures \ref{fig_spec3} and \ref{fig_spec4} is obtained for descendants level cut off with $n_{c}= 17$ that corresponds to the truncated state space of dimension 9615.} for the weight spectrum of the vacuum at  $\xi=1$ and taken at $R|m|=20$.  
We show on the same plots the exponentials corresponding to a state of the form (\ref{vac_form}) with three different values of $\tau^{*}$. 
Using the variational energy $E_{-}$ given in (\ref{Ising_varens}), that is without the logarithmic term, we obtain numerically 
$\tau^{*}_{1}|m|\approx 0.2588$. Adding the logarithmic term we obtain $\tau^{*}_{2}|m|\approx 0.3243$. Numerically fitting the lowest lying components 
to the same ansatz we obtain $\tau^{*}_{\rm fit}|m| \approx 0.4911$. All three values are quite close to the three values we considered in section 
\ref{Ising_thermal} where we compared the exact solution at $h=0$ to Cardy's ansatz. This is because $\xi=1$ corresponds to a relatively small 
magnetic field. Changing the radius $R$ one sees qualitatively similar weight distributions: with all components of the same sign, the lowest components on an exponential curve and the higher weight components split over a range of values. Increasing the radius the spectrum becomes 
closer to  (\ref{vac_form}). We also checked a range of values for $\xi$ finding that larger  $\xi$   are better described by the Cardy's ansatz with
the logarithmic term becoming negligible for $\xi$ very large. This fits with the results of \cite{LVT} that demonstrate that for the pure magnetic field perturbation 
the asymptotic vacuum state is well approximated by  Cardy's ansatz.

\subsection{Mixed thermal and magnetic perturbations for $t>0$} \label{Ising_sec_disorder}
 For $h>0$ and $t>0$ we have negative local minima for $E_{ F}$ and $E_{-}$. We find numerically that for  $\xi<\xi_{1}=8.7633...$ 
 the minimum of $E_{ F}$ is lower than that of $E_{-}$. This is not  physical as the  magnetisation is zero in the 
 vacuum corresponding to $|F\rangle\!\rangle$. This problem was noted in \cite{Cardy_var} where it was suggested that the ansatz should 
 be modified for the free boundary condition to include a perturbation by the boundary magnetic field. It was proposed to write a trial state 
 \be
 e^{-\tau H_{0}}|h_{b}\rangle\!\rangle
 \ee
 that depends on two couplings: $\tau $ and $h_{b}$. The latter is the value of the boundary magnetic field. 
  The boundary magnetic field in the Ising model was considered 
in the context of QFT in \cite{GZ}. It is obtained by taking the critical Ising model on a half plane with a free spin boundary condition 
and perturbing it by a boundary action 
\be
S_{\rm boundary} = h_{b} \int\! \epsilon(y) dy
\ee
where $\epsilon(y)$ is the scaling boundary  field of dimension 1/2 representing the boundary spin and $h$ is the coupling constant. Here we assume that 
$\epsilon(y)$ is canonically normalised so that it has a two-point function $\langle \epsilon(y) \epsilon(0)\rangle = 1/|y|$. 
 The boundary state $|h_{b}\rangle\!\rangle$ for this model was obtained in \cite{Chatterjee_crit}. 
For the smeared state we have the following expression\footnote{In our conventions  (\ref{epsilon_convention}) has the opposite sign from that in   \cite{Chatterjee_crit}  that 
results in our boundary state having  the opposite sign in the exponents. } 
\be \label{bmf_bs}
 e^{-\tau_{\rm uv} H_{0}}|h_{b}\rangle\!\rangle =  g \exp\Bigl( 
-i\sum_{n=0}^{\infty} f_{n+1/2} \,  a^{\dagger}_{n+1/2} \bar a^{\dagger}_{n+1/2} \Bigr) |0\rangle 
+ \tilde g \exp\Bigl( 
-i\sum_{n=1}^{\infty}  f_{n}  a^{\dagger}_{n}  \bar  a^{\dagger}_{n} \, \Bigr) |\sigma \rangle 
\ee
where 
\be \label{bmf_bs2}
f_{k}=f_{k}(\alpha,\tau_{\rm uv}/R) =  \frac{k -\alpha}{k + \alpha} e^{-\frac{4\pi \tau_{\rm uv}}{R}k} \, , 
\ee
and $k$ is an integer or a half-integer. Furthermore, in (\ref{bmf_bs})
\be \label{bmf_bs3}
   g=  \frac{\sqrt{\pi} (R\mu)^{-\alpha}} {\Gamma(\alpha + 1/2) }\, , 
\qquad \tilde g =  \pm \sqrt{\pi} (R\mu)^{-\alpha}e^{-\frac{\pi \tau}{4R}} \frac{2^{1/4}\sqrt{\alpha}}{\Gamma(\alpha + 1)} \, ,
\ee 
and
\be
 \alpha = 2h_{b}^2 R \, .
\ee
The  sign in front of $\tilde g$   in (\ref{bmf_bs3}) is given by the sign of $h_{b}$. We have added a subindex "uv" to the smearing parameter $\tau$ 
as it will be clear from the forthcoming discussion that the ansatz (\ref{bmf_bs2}) gives a good description for sufficiently small values of $R$ while for 
the larger values of $R$  a different  smearing parameter $\tau$ will emerge which we will denote as $\tau_{\rm ir}$. 

 Although solving the variational problem with this ansatz is a complicated problem we can compare it qualitatively with the TCSA data using 
 the dependence  of components $C_{i}$ on the weight as we did before. 
 We expect the values of $h_{b}$ and $\tau_{\rm uv}$ to depend only on the  couplings: $t$ and $h$. 
 The $R$-dependence of the trial state thus only comes from $\alpha$ and the exponentials containing $\tau_{\rm uv}/R$. 
 To see how the weight spectrum of (\ref{bmf_bs}) behaves when we change $R$ we use the 
 basis introduced in (\ref{mode_basis}) in the NS sector and an analogous basis in the Ramond sector where the half-integers $k_{i}$ 
 should be replaced by integers $n_{i}$. For each level $N$ we introduce four different sets of integer partitions: 
 $$
 P^{\rm NS, e}_{2n}\, , \quad  P^{\rm NS, o}_{2n+1}\, , \quad P^{\rm R, e}_{n} \, , \quad P^{\rm R, o}_{n} \, , \quad n\in {\mathbb Z}\, , \enspace n\ge 0\, .
 $$ 
 The elements  in the first  set are ordered  partitions without repetitions of $2n$ into odd integers containing an even number of integers. 
 The elements  in the second  set are ordered  partitions without repetitions of $2n+1$ into odd integers containing an odd number of integers. 
Similarly  the elements in the last two sets are ordered partitions without repetitions of $n$ into integers (of any parity) containing an even or an odd number 
 of integers as indicated. For the first two sets we associate with each partition $(k)$  the state $|{\bf k} /2\rangle$ introduced in  
 (\ref{mode_basis}). For ${\bf n}=(n_{1},n_{2}, \dots , n_{2s}) \in  P^{\rm R, even}_{N}$ we introduce 
 \be \label{mode_basisReven}
 |{\bf n}\rangle_{\rm even} = a^{\dagger}_{n_{1}}a^{\dagger}_{n_2}\dots a^{\dagger}_{2s} \bar a^{\dagger}_{n_{1}}\bar a^{\dagger}_{n_2}\dots \bar a^{\dagger}_{2s} |\sigma\rangle 
 \ee
 while to ${\bf n}=(n_1, n_2, \dots , n_{2s+1})\in  P^{\rm R, odd}_{N}$ we associate 
 \be \label{mode_basisRodd}
 |{\bf n}\rangle_{\rm odd} = a^{\dagger}_{n_{1}}a^{\dagger}_{n_2}\dots a^{\dagger}_{2s+1}a_{0} 
 \bar a^{\dagger}_{n_{1}}\bar a^{\dagger}_{n_2}\dots \bar a^{\dagger}_{2s+1}\bar a_{0} |\sigma\rangle 
 \ee
 where $a_{0}$ and $\bar a_{0}$ are the Ramond zero modes normalised so that $a_{0}^2=\bar a_{0}^2=1/2$. We note that each of these vectors has 
 a factorised form $ |k_{i,N}\rangle_{L} \otimes |k_{i,N} \rangle_{R} $
 as appears in (\ref{ishibashi}). 
 With these notations it is straightforward to decomposethe boundary state (\ref{bmf_bs}) as\footnote{Note that equation  (\ref{epsilon_convention}) and 
 $C_{\sigma \sigma \epsilon} =1/2$ imply that $2ia_{0}\bar a_{0}|\sigma\rangle = |\sigma\rangle$. } 
 \be \label{bmf_sectors}
 e^{-\tau_{\rm uv} H_{0}}|h_{b}\rangle\!\rangle = |{\rm NS, even}\rangle_{\alpha,\tau} +  |{\rm NS, odd}\rangle_{\alpha,\tau} +  |{\rm R, even}\rangle_{\alpha,\tau} 
 +  |{\rm R, odd}\rangle_{\alpha,\tau} 
 \ee
 where 
 \be \label{NS_components}
 |{\rm NS, even}\rangle_{\alpha,\tau} = \sum_{n=0}^{\infty} \sum_{{\bf k} \in P^{\rm NS, e}_{2n}}  C({\bf k}) 
 \Bigl| \frac{\bf k}{2}\Bigr\rangle  \, , \quad  |{\rm NS, odd}\rangle_{\alpha,\tau} = \sum_{n=0}^{\infty} \sum_{{\bf k} \in P^{\rm NS, o}_{2n+1}} C({\bf k})  \Bigl| \frac{\bf k}{2}\Bigr\rangle \, , 
  \ee
  \be \label{R_components}
 |{\rm R, even}\rangle_{\alpha,\tau} = \sum_{n=0}^{\infty} \sum_{{\bf n} \in P^{\rm R, e}_{n}}\tilde C({\bf n})
|{\bf n} \rangle_{\rm even}  \, , \quad 
 |{\rm R, odd}\rangle_{\alpha,\tau} = \sum_{n=1}^{\infty} \sum_{{\bf n} \in P^{\rm R, o}_{n}} \tilde C({\bf n}) 
|{\bf n} \rangle_{\rm odd}  \, , 
  \ee
  \be \label{factorised_coefs1}
  C({\bf k}) = g \prod_{k_{i}\in {\bf k}} f_{k_{i}/2}  \, , \enspace {\bf k} \in P^{\rm NS, e}_{2n}\, ,  \qquad  C({\bf k}) = - g \prod_{k_{i}\in {\bf k}} f_{k_{i}/2}\,  
   \enspace {\bf k} \in P^{\rm NS, o}_{2n+1}
  \ee
  \be \label{factorised_coefs2}
   \tilde C({\bf n}) = \tilde g \prod_{n_{i}\in {\bf n}}  f_{n_{i}} \, , \enspace {\bf n} \in P^{\rm R, e}_{n}\, ,  \qquad  \tilde C({\bf n}) = -\tilde g \prod_{n_{i}\in {\bf n}}  f_{n_{i}} \, , \enspace {\bf n} \in P^{\rm R, o}_{n} \, .
  \ee
  Let us assume from now on that $h_{b}<0$ and thus $\tilde g<0$.  
  We see that if $\alpha<1/2$ then all values  $f(k_{i}/2)$ and $\tilde f(n_{j})$ are positive and the coefficients 
  appearing in  $|{\rm NS, even}\rangle_{\alpha,\tau}$ and  $ |{\rm R, odd}\rangle_{\alpha,\tau}$   are all positive while those 
  appearing in $|{\rm NS, odd}\rangle_{\alpha,\tau}$ and $ |{\rm R, even}\rangle_{\alpha,\tau}$   are all negative. 
  Moreover, for $\alpha \ll 1/2$ we have 
  \be
  f_{k} \approx e^{-\frac{4\pi \tau_{\rm uv}}{R}k} \, , \qquad  g\approx 1\, , \quad \tilde g\approx -2^{1/4} \sqrt{\alpha \pi} 
  \ee
  and, using (\ref{epsilon_convention}),  we can approximate our state as 
  \be \label{beginning_RG}
   e^{-\tau_{\rm uv} H_{0}}|{\bf 1}\rangle\!\rangle -  e^{-\tau_{\rm uv} H_{0}}|\epsilon\rangle\!\rangle + 2^{1/4} \sqrt{\alpha \pi}  e^{-\tau_{\rm uv} H_{0}} \Bigl[
\sum_{n=1}^{\infty} \sum_{{\bf n} \in P^{\rm R, o}_{n}} 
|{\bf n} \rangle_{\rm odd}   -\sum_{n=0}^{\infty} \sum_{{\bf n} \in P^{\rm R, e}_{n}} 
|{\bf n} \rangle_{\rm even} \Bigr] \, . 
  \ee
 
\begin{center}
\begin{figure}[H]
\begin{minipage}[b]{0.5\linewidth}
\centering
\includegraphics[scale=0.82]{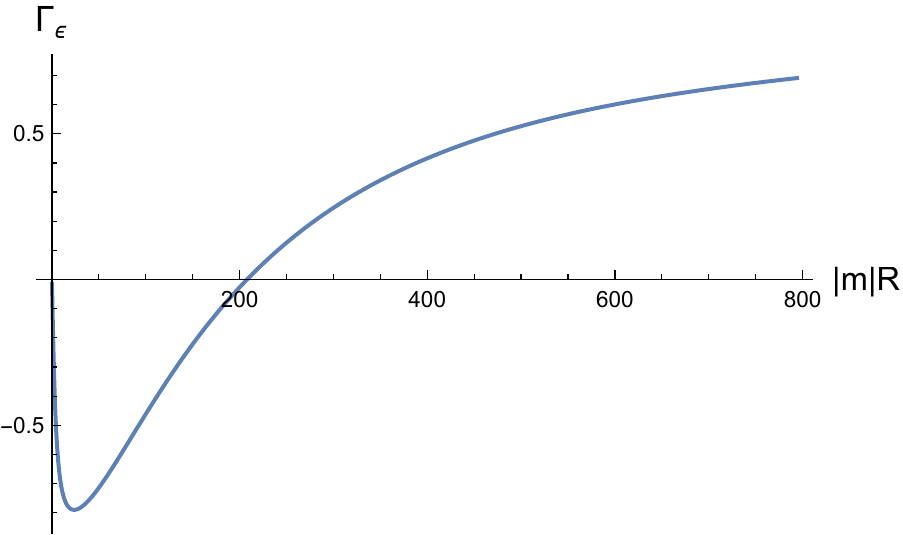}
\end{minipage}%
\begin{minipage}[b]{0.5\linewidth}
\centering
\includegraphics[scale=0.82]{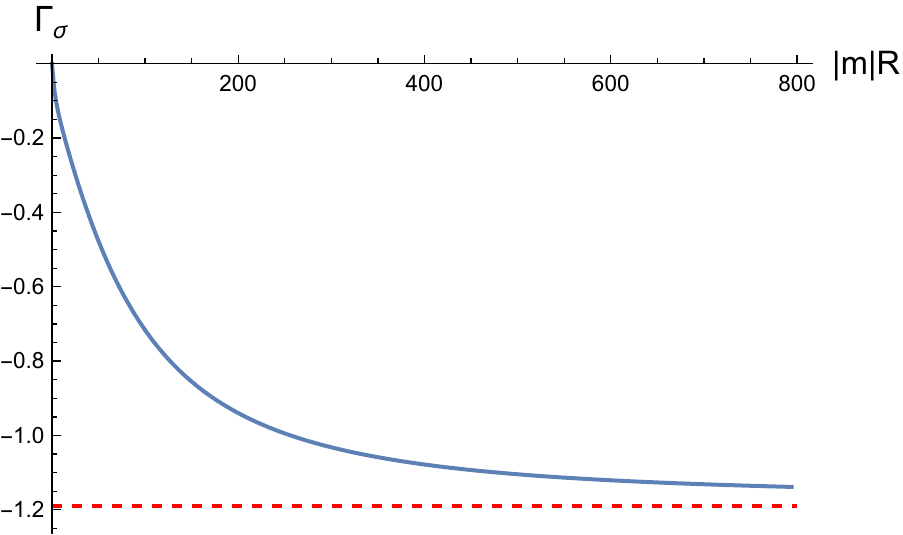}
\end{minipage}
\caption{Ratios of vacuum components obtained using TCSA with oscillator basis truncated at $n_{c}=17$ for 
$\xi=1$, $t>0$.  }
\label{overlap_plots}
\end{figure}

\end{center}

 We investigated the region $t>0$ using  TCSA numerics done with the basis built via the creation operators. This basis includes diagonal states 
 of the form  (\ref{mode_basis}),  (\ref{mode_basisReven}),  (\ref{mode_basisRodd}) for which the oscillator content in the holomorphic part is the same, including the order, as in the anti-holomorphic part. The truncation is done imposing\footnote{This truncation differs  in the $\epsilon$-sector  from the truncation in Virasoro descendant level,  but the choice is only a matter of convenience.  For all numerical data presented in  this subsection $n_{c}=17$ that gives a truncated space of dimension 8774.} 
 \be
 \sum_{i} n_{i} \le n_{c} \, , \qquad \sum_{j} \frac{k_{j}}{2} \le n_{c}  \, .
 \ee
  There are also non-diagonal basis vectors for which the two partitions specifying the oscillator content in the holomorphic and anti-holomorphic part are different. We would expect the vacuum vector  to be approximated by (\ref{bmf_bs}) for sufficiently small $h$  when it can be treated as a perturbation of the smeared boundary state $e^{-\tau_{\rm uv}H_{0}}|F\rangle\!\rangle$ and 
 for some range of  $R|m|$. At the beginning of that range, where the vector  settles at the beginning of the boundary RG flow, its components should 
 look like the ones in (\ref{beginning_RG}). A good guide to finding this point is the plot of the vacuum component $\Gamma_{\epsilon}$ 
 defined in  (\ref{Ising_overlaps}). We use for illustration the $\xi=1$ trajectory. The TCSA plots for both $\Gamma_{\epsilon}$ and $\Gamma_{\sigma}$ are presented on Figure 
 \ref{overlap_plots}. We see that while $\Gamma_{\sigma}$ monotonically approaches its asymptotic value of $-2^{1/4}$ that corresponds to $|-\rangle\!\rangle$, 
 the ratio $\Gamma_{\epsilon}$ is initially negative with a minimum value close to $R|m|=25$ then increases crossing zero and continues a monotonic increase. 
 While the TCSA data at such large values of $R$ becomes unreliable we can conjecture the correct asymptotic value is  $\Gamma_{\epsilon}=1$ as in $|-\rangle\!\rangle$. 
 
\begin{center}
\begin{figure}[H]
\begin{minipage}[b]{0.5\linewidth}
\centering
\includegraphics[scale=0.82]{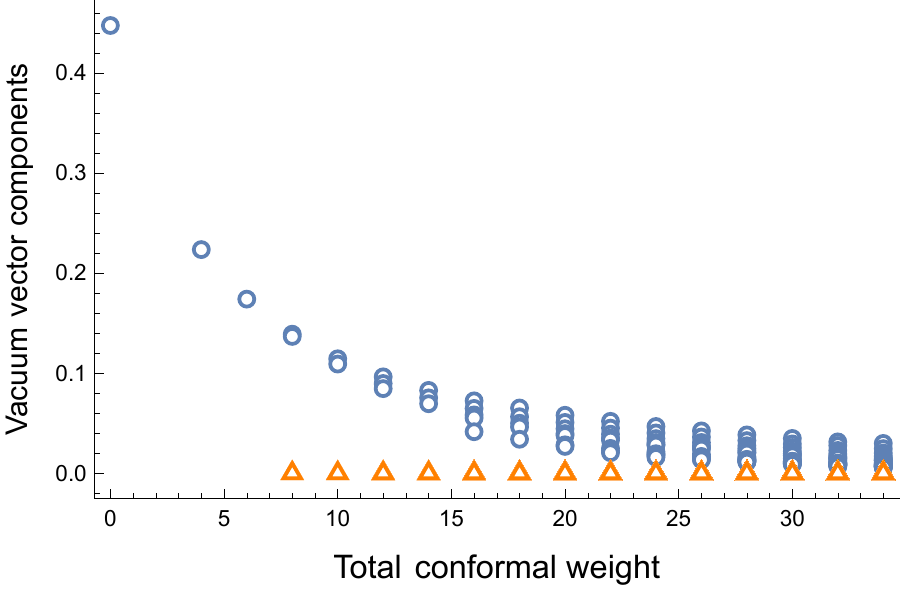}
\end{minipage}%
\begin{minipage}[b]{0.5\linewidth}
\centering
\includegraphics[scale=0.82]{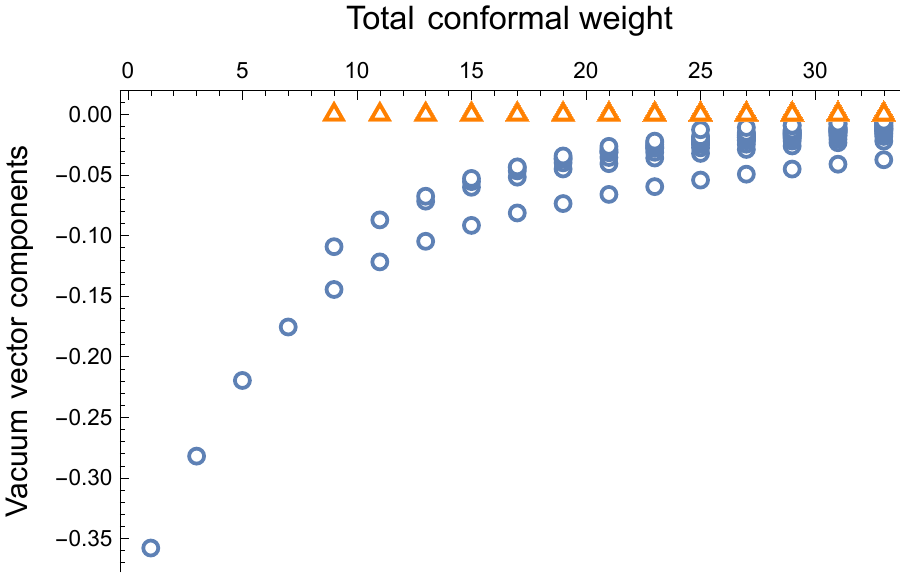}
\end{minipage}
\caption{The TCSA vacuum vector components $C_{i}$ against the total conformal weight of the basis vectors taken for the mixed perturbation with $\xi=1$, $t>0$ and at  $|m|R=25$. The left plot 
represents the vacuum sector and the right plot -- the $\epsilon$-sector. The blue circles mark the diagonal components and the orange triangles -- the non-diagonal ones.}
\label{fig_bmfspec1}
\end{figure}

\end{center}

\begin{center}
\begin{figure}[H]
\centering
\includegraphics[scale=0.85]{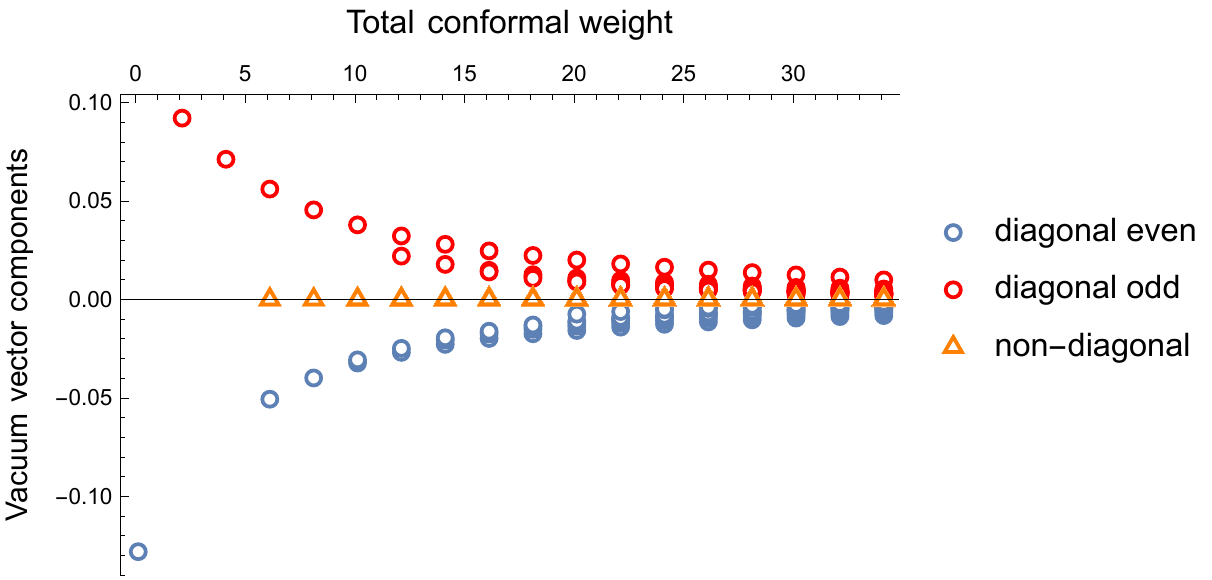}
\caption{The TCSA vacuum vector components $C_{i}$ in the $\sigma$-sector plotted against the total conformal weight of the basis vectors taken for the mixed perturbation with $\xi=1$, $t>0$ and at  $|m|R=25$.  }
\label{fig_bmfspec2}
\end{figure}

\end{center}

Looking at  the numerics describing the vacuum vector  at the minimum of $\Gamma_{\epsilon}$ we find that its spectrum of weights is qualitatively 
the same as that of (\ref{beginning_RG}). The plots of the vacuum components $C_{i}$ in the oscillator basis are presented on Figures \ref{fig_bmfspec1}
and \ref{fig_bmfspec2}.  

   While the vacuum vector contains components corresponding to the non-diagonal basis vector they are 
small (on the plots they are marked by orange triangles which cluster on the horizontal axis). The exponential decay is clearly visible and the signs 
are distributed according to the parity of the partitions' length. We can numerically fit the low weight components to 
exponential functions. Although they differ slightly in each sector the average value is around $\tau_{\rm uv}|m| \approx 0.457$.
Thus at the value  $R|m|=25$ the vacuum vector is well-approximated by the smeared boundary magnetic field boundary state with  particular values of 
$\tau_{\rm uv}$ and  $\alpha\ll 1/2$. It is interesting to note that the value of $\tau_{\rm uv}|m|$ we got is not too far from $\tau_{e}|m|=0.5$ 
that is the exact value for $\xi=0$.

\begin{center}
\begin{figure}[H]
\centering
\includegraphics[scale=1]{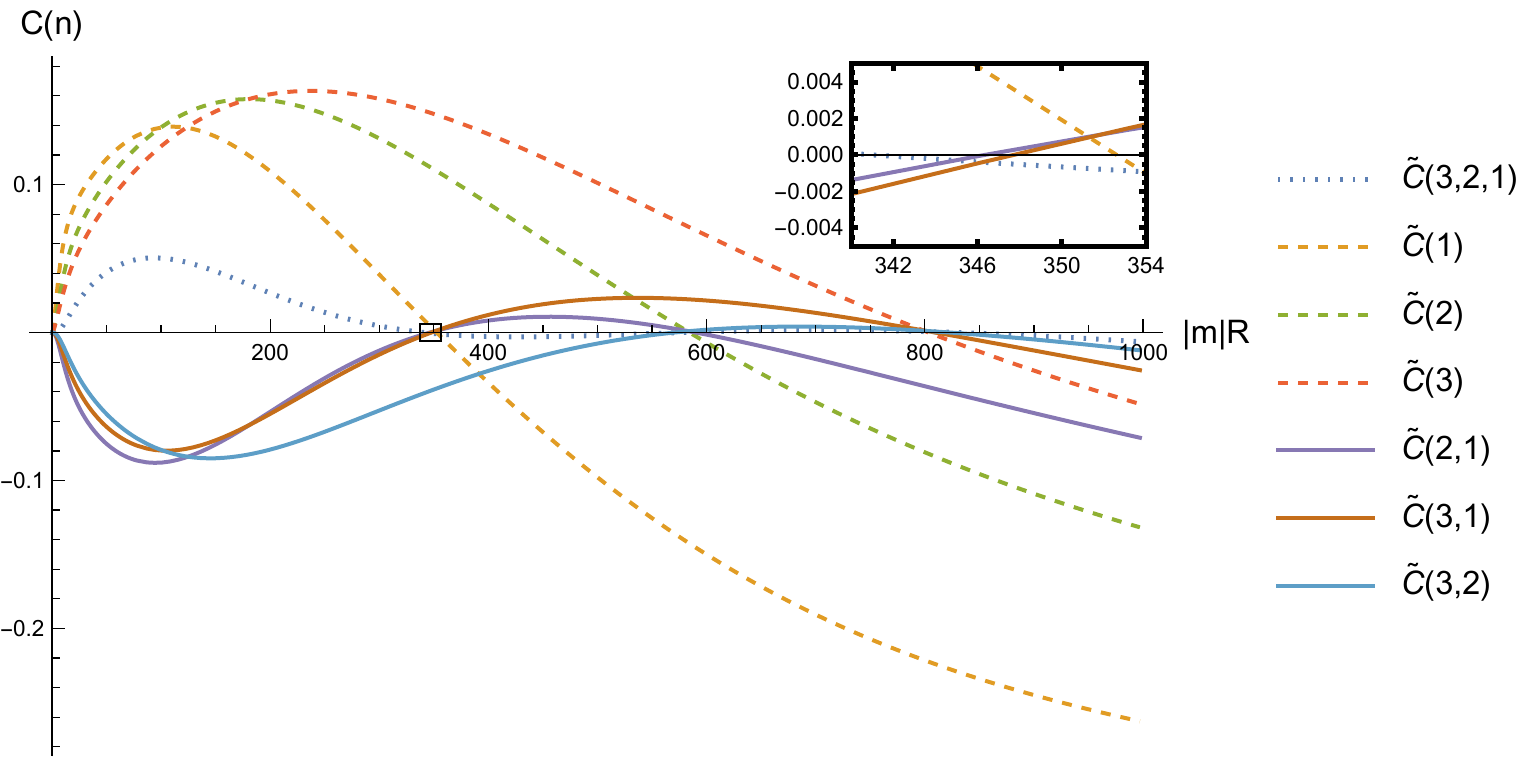}
\caption{The first seven components $\tilde C({\bf n})$ in the Ramond sector of the perturbed Ising model with $\xi=1$, $t>0$. These are obtained 
for TCSA using the oscillator basis and level truncation at $n_{c}=17$.  The insert plot shows a magnified region near the first crossings of the horizontal axis.  }
\label{fig_oscillations}
\end{figure}

\end{center}

Increasing $R$ we expect that the effective value of $\alpha$ will start changing and the vector components start moving. 
In our theoretical vector (\ref{bmf_sectors}) $\alpha$ depends linearly on $R$. When,   starting from the region where $\alpha \ll 1/2$, we increase $R$   the functions 
$f_{k}(\alpha, \tau/R)$ each decreases from the initial values $f_{k}(\alpha, \tau/R) \approx e^{-\frac{4\pi \tau_{\rm uv}}{R}k}$, 
 pass through zero at $\alpha=k$  and then asymptote to $-1$. For the $N$-particle components $C({\bf k})$, $\tilde C({\bf n})$  in (\ref{NS_components}), (\ref{R_components}) 
 will initially oscillate passing through zero $N$ times when $\alpha$ crosses the values $k_{i}/2 \in {\bf k}/2$ and $n_{i}\in {\bf n}$ respectively, before 
 reaching the asymptotic value which is $1$ in the NS-sector and $-1$ in the Ramond sector. 
 We find such oscillations using TCSA numerical calculations. The plots for the first seven components $\tilde C({\bf n})$  in the Ramond sector 
 as a function of $R|m|$ are presented on Figure \ref{fig_oscillations}. The points where the components vanish are slightly off-set with respect to 
 the ones we expect from the factorisation (\ref{factorised_coefs1}),  (\ref{factorised_coefs2}) in the theoretical model but increasing the TCSA cutoff brings them closer so the 
 mismatch may be attributed to the truncation errors.

  We find the same oscillations in the NS sector as well. We also checked the factorisation relations like 
  \be
  \frac{\tilde C(2,1) \tilde C(\mathsmaller{ \O})}{\tilde C(2) \tilde C(1) } = 1
  \ee
 where  $\tilde C(\mathsmaller{ \O})=\tilde g$ is the coefficient of the 
  Ramond lowest weight vector $|\sigma\rangle$ that is formally labelled by the empty partition.
  We found such relations to hold over a large range of $R$  with good accuracy except for small regions near the points where some components vanish. The mismatch between 
  the position of zeroes mentioned above causes the relation to break down in those regions.

\begin{center}
\begin{figure}[H]
\centering
\includegraphics[scale=1]{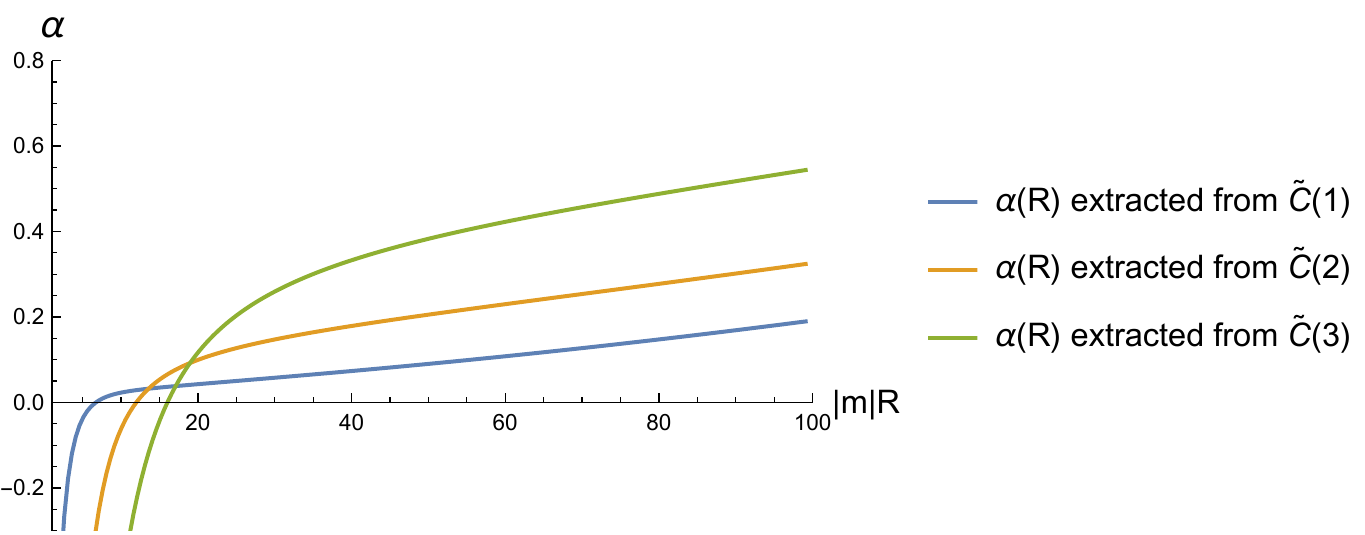}
\caption{The running boundary coupling $\alpha(R)$ for $\xi=1$, $t>0$ extracted from the lowest 3 one-particle states in the Ramond sector using the value   $\tau_{\rm uv}|m|\approx 0.457$. 
The plots are obtained using TCSA  with oscillator basis truncated at level 17. }
\label{fig_alpha}
\end{figure}

\end{center}

  We can further test the theoretical model (\ref{NS_components}), (\ref{R_components}) 
  more quantitatively by extracting the value of $\alpha$ from the TCSA data. We use the value of $\tau |m|$ we found from the vacuum vector components in the region of small $\alpha$ and the one-particle components to express 
  \be \label{runalphaNS}
  \alpha \equiv \alpha_{k/2}= \frac{k}{2}\left(\frac{1+ \frac{C(k)}{C(\mathsmaller{ \O})}
  e^{2\pi k \tau_{\rm uv}/R}}{1- \frac{C(k)}{C(\mathsmaller{ \O})}e^{2\pi k \tau_{\rm uv/R}}} 
  \right)
  \ee
  in the NS sector or 
  \be \label{runalphaR}
  \alpha \equiv \alpha_{n}= n\left(\frac{1+ \frac{\tilde C(n)}{\tilde C(\mathsmaller{ \O})}e^{4\pi n \tau_{\rm uv}/R}}{1- \frac{\tilde C(n)}{\tilde C(\mathsmaller{ \O})}e^{4\pi n \tau_{\rm uv}/R}} \right)
  \ee
  in the Ramond sector. Here $C(\mathsmaller{ \O})=g$ is the component of the UV vacuum $|0\rangle$. On Figure \ref{fig_alpha} we show a plot of $\alpha_{1}$, $\alpha_{2}$ and $\alpha_{3}$ as functions 
  of $R|m|$ obtained numerically. We see that they all increase and for sufficiently large $R$ become approximately linear with the same slope but 
  different intercepts. This fits with our general expectation that a massive vacuum can be described by a local boundary condition (in this case 
  by the boundary magnetic field) only asymptotically for large $R$. The intercepts then give a finite size correction of order $1/R$  to the 
  boundary magnetic field coupling $h_{b}^2$. This is similar to our analysis of finite size corrections for the boundary identity field flows we discussed in section 
  \ref{Ising_SSB}. We also mention that we find similar plots in the NS sector.

  If we focus our attention on a group of low weight components in the theoretical vacuum vector (\ref{bmf_bs}) then for sufficiently large  $R$ after a series of oscillations these components settle into the values corresponding to $|-\rangle\!\rangle$. The effect of the exponentials containing $\tau$ 
  goes away way before the oscillations finish and thus such exponentials can be neglected.
  Retaining the leading $1/R$ correction 
  we have\footnote{The approximation (\ref{f_exp}) resulting in (\ref{tau_ir}) is another way to obtain the leading term in the infrared effective action 
  for the boundary magnetic field model. The leading term was previously obtained in \cite{Toth}, \cite{me_bmf}. The complete IR effective action was obtained in \cite{me_random}.  }
  \be \label{f_exp}
  f_{k} \approx \frac{k-\alpha}{k+\alpha} \approx -1 + \frac{2k}{\alpha} \approx -e^{-\frac{2\pi}{R} \tau_{\rm ir}2k } 
  \ee
  where 
  \be \label{tau_ir}
  \tau_{\rm ir} = \frac{R}{2\pi \alpha} = \frac{1}{4\pi h_{b}^2} \, .
  \ee
  Indeed in TCSA numerics we find that for sufficiently large $R$ the low weight components of the vacuum take the form of a smeared boundary state 
  \be \label{ir_smeared}
  e^{-\tau_{\rm ir}H_{0}}|-\rangle\!\rangle
  \ee
   with plots similar to those on Figures \ref{fig_spec3} and \ref{fig_spec4}. 
  
 The above example of $\xi=1$ numerics demonstrates that for sufficiently small $\xi$ we find a full RG flow triggered by the boundary magnetic field 
 and modified by the smearing factor $e^{-\tau_{\rm uv}H_{0}}$ that for the low weight components quickly becomes 1 while a new smearing factor 
emerges from $|h_{b}\rangle\!\rangle$ and asymptotically the vacuum approaches  (\ref{ir_smeared}).  
For larger values of $\xi$ this picture is modified in certain parts.   
We find that for small values of $R$ we can no longer find a vacuum vector with a weight spectrum 
similar to the one on Figures \ref{fig_bmfspec1} and \ref{fig_bmfspec2} (see Figures \ref{fig_bmfspec1b} and \ref{fig_bmfspec2b} in Appendix 
\ref{xi10_appendix} for a sample plot). 
However the oscillations similar to the ones shown on Figure \ref{fig_oscillations} are still in place (see Figure \ref{fig_oscNS_xi10} in Appendix 
\ref{xi10_appendix}) and asymptotically after the oscillations stop we get (\ref{ir_smeared}). 
Moreover the running of $\alpha$ can be estimated using (\ref{runalphaNS}), (\ref{runalphaR}) with $\tau_{\rm uv}$ set to zero and one gets plots with a linear asymptote (see plots on Figure \ref{fig_alpha_NSR_xi10} in Appendix \ref{xi10_appendix}). Our interpretation of what is going on is that at the onset of the boundary 
magnetic field flow for larger $\xi$ we have  the boundary state $|h_{b}\rangle\!\rangle$ modified by a more complicated combination of irrelevant operators, possibly involving the higher KdV charges (\ref{kdv}). As $R$ increases these irrelevant operators die out and we settle into the boundary magnetic field flow 
whose end point is (\ref{ir_smeared}). The larger magnetic field we take the shorter becomes the region where we can discern the boundary RG flow patterns. 
  
  \subsection{Summary}  \label{summary1_sec}
  
  In the preceeding subsections we have looked at the case of mixed thermal and magnetic field perturbations of the critical Ising model both in the symmetry breaking ($t<0$) and in the disorder ($t>0$) regions. For small magnetic fields (small $\xi$) we demonstrated that the evolution of the vacuum vector with the system size $R$ can be described by a boundary RG flow. The associated RG flows start in the UV from the RG boundary associated with the zero magnetic field: $|F\rangle\!\rangle$ for $t>0$ and $|+\rangle\!\rangle\oplus |-\rangle\!\rangle$  for $t<0$. They are triggered by the boundary magnetic field 
  in the disordered region and by the boundary identity fields in the symmetry breaking region. In both cases for sufficiently large $R$ the vacuum is described by the IR fixed point of these boundary  flows. For the low weight components of the vacuum vector the leading finite $R$ correction is described in all cases by  Cardy's ansatz. This brings us back to the puzzle noted in \cite{Cardy_var}: for sufficiently small $\xi$ the variational ansatz 
  gives lower energy to the smeared free boundary condition that has zero magnetisation. While, following the suggestion of \cite{Cardy_var}, 
  we modified the ansatz to $e^{-\tau_{\rm uv}H_{0}}|h_{b}\rangle\!\rangle$ and then got numerical confirmations for it, at the end of the RG flow the low weight components 
   assume the form  $e^{-\tau_{\rm ir}H_{0}}|-\rangle\!\rangle$ which was in the original ansatz. The resolution of this seems to be in the fact that the 
  tail of  components with high conformal  weights in the modified ansatz is different from the original anastz. 
  The tail is essentially that of $|F\rangle\!\rangle$.   With both ranges of the weights contributing to the vacuum energy the latter should be below the  variational energy $E_{F}$ due to the low weight components. 
   We would like to stress that even though we don't know how to calculate the contribution  of the UV tail analytically we know qualitatively that once the magnetic field is switched on the phase corresponds to the  $|\pm\rangle\!\rangle$ state which is the end point of the boundary magnetic field flow.

  A similar reasoning may explain the discrepancy of the value of $\tau^{*}$ 
  one obtains from Cardy's ansatz for $h=0$ with the one that follows from the exact solution.  Although in the $h=0$ example the discrepancy gets smaller when one includes the logarithmic term (see formula (\ref{log_term})) in general it is not clear to us how to systematically improve the contribution of the high weight tail of the vacuum in the ansatz. While this sensitivity to the shape of the tail can be attributed to the UV divergence present in the $t\ne 0$ Ising field theory, in the next section we will see another example of when the ansatz needs to be modified even in the absence of divergences.


\section{Tricritical Ising model perturbed by the leading thermal and magnetic  operators}\setcounter{equation}{0}
\label{tricrit_section}

\subsection{Landau theory} \label{Landau_sec}

\begin{center}
\begin{figure}[H]
\centering
\includegraphics[scale=0.8]{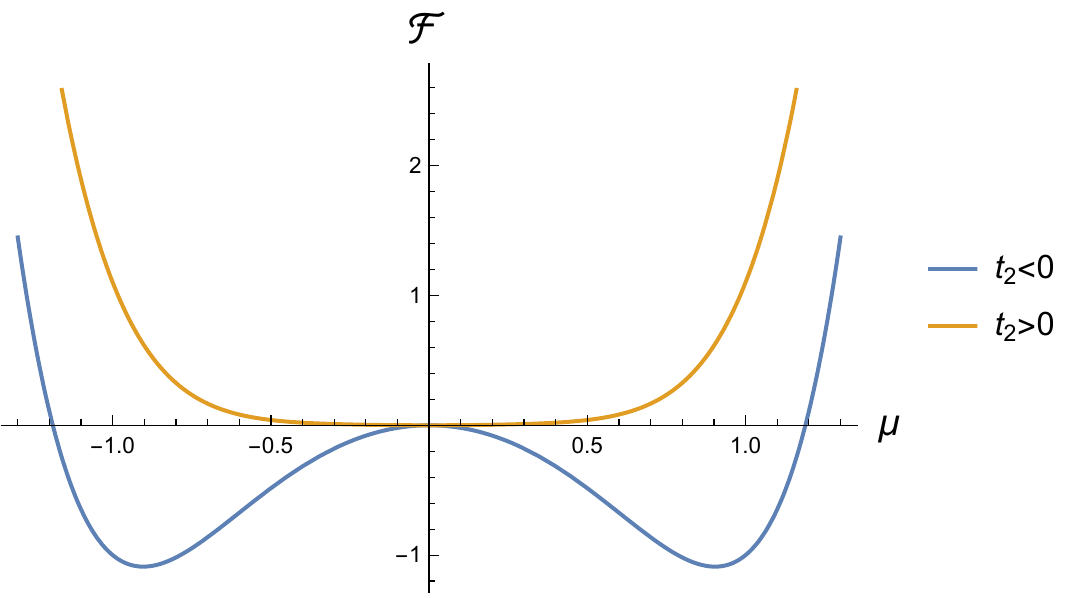}
\caption{ Shape of Landau free energy function for $h=t_{3}=0$.}
\label{potential_fig}
\end{figure}
\end{center}

\begin{center}
\begin{figure}[H]
\centering
\includegraphics[scale=0.8]{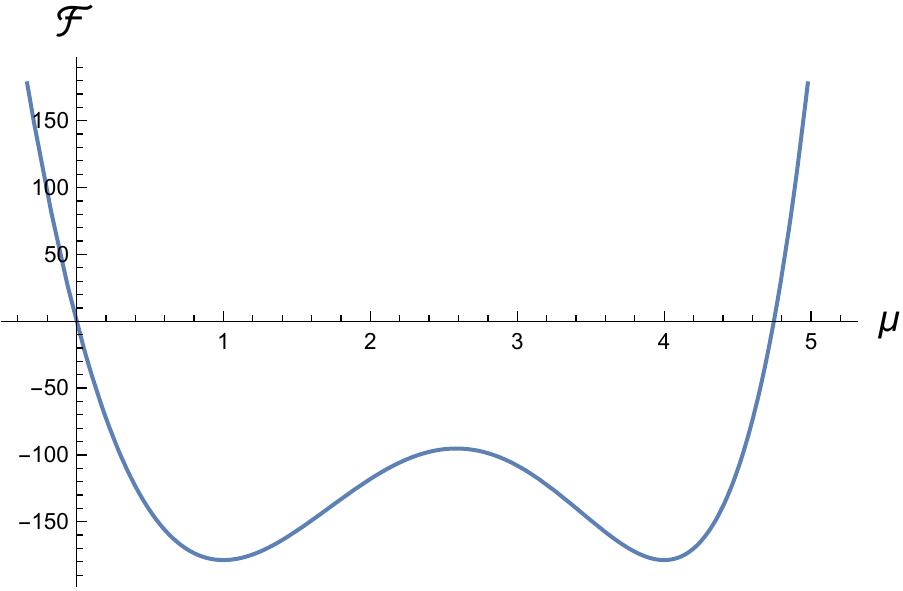}
\caption{ Landau free energy function at a sample point on the two-phase coexistence surface: $t_{3}=190,t_{2}=609,h=420$ .}
\label{potential_fig2}
\end{figure}
\end{center}

To build some intuition about the perturbed TIM we will discuss in this section the  Landau theory of a tricritical point. 
It is defined by the Landau free energy function that is a 6-th degree polynomial in the order parameter which we denote by $\mu$. 
Without loss of generality this polynomial can be chosen as 
\be \label{LF}
{\cal F}(\mu) = -h\mu + \frac{t_{2}}{2}\mu^2 -  \frac{t_3}{3}\mu^3 +  \frac{t_{4}}{2}\mu^4 +  \frac{\mu^6}{6} \, .
\ee
Here $t_{2}$ and $t_{4}$ are the two thermal and $h$ and $t_{3}$ are the two magnetic couplings (the signs are just a matter of convenience). 
The analysis of (\ref{LF}) usually starts from the symmetry plane: $h=t_{3}=0$. On the symmetry plane in the half-plane $t_{2}<0$ the theory has a doubly degenerate vacuum. 
The half-line $t_{2}=0,t_{4}>0$ is a symmetry-preserving critical line leading to the critical Ising theory in the infrared. This half-line joins at the tricritical point $t_{2}=t_{4}=0$ the triple phase coexistence line that lies in the $t_{2}>0, t_{4}<0$ quadrant. On the QFT side the symmetry plane has been studied in 
\cite{Cardy_etal, Zam_TBA, Zam_TBA2, TIM1} where the same phase structure was found. The vacuum degeneracies on the symmetry plane are supported by  integrability. The analysis usually proceeds by adding to the two thermal perturbations the leading magnetic perturbation, i.e. taking $h\ne 0$, see for example
\cite{LS} section III D for details. We are going instead to consider the three-coupling theory (\ref{LF})  with $t_{4}=0$ as this will be relevant to the  
field theory perturbations of TIM we consider  in this paper. The Landau free energy is thus 
 \be \label{LF2}
{\cal F}(\mu) = -h\mu + \frac{t_{2}}{2}\mu^2 -  \frac{t_3}{3}\mu^3  +  \frac{\mu^6}{6} \, .
\ee
At the extremal points we have $\frac{d{\cal F}}{d\mu}=0$ that gives 
\be \label{derF}
h = t_{2}\mu -t_{3}\mu^{2} + \mu^5 \, .
\ee
This equation can have at most three real solutions for $\mu$ due to the absence of the cubic and quartic terms. Thus, we can have at most two phases 
coexisting in this region. For  $h=t_{3}=0$ the free energy becomes 
\be
 \label{LF3}
{\cal F}(\mu) =  \frac{t_{2}}{2}\mu^2  +  \frac{\mu^6}{6} \, 
\ee
and it is easy to see that $t_{2}<0$ is a two-phase coexistence line. We call this a symmetry breaking or the $\alpha$-line. On Figure \ref{potential_fig} we see the shape of free energy for $t_{2}<0$ and 
$t_{2}>0$. 
The two vacua on this line are related by  the ${\mathbb Z}_{2}$-symmetry: 
$\mu \to -\mu$. 
In general, for  given values of $h, t_{2},t_{3}$, equation (\ref{derF}) has either one or three real solutions. In the latter case we can denote them as 
\be
\mu_{1}\le \mu_{2} \le  \mu_{3} \, .
\ee 
Then the phase coexistence condition is 
\be \label{coex_eq}
{\cal F}(\mu_1) = {\cal F}(\mu_3) \, , \qquad \mu_{1} \ne \mu_{3} \, , \quad \frac{d^{2} {\cal F}}{d\mu^2} (\mu_{2}) < 0  \, . 
\ee
This imposes one equation on three real parameters (plus inequalities) and thus gives a surface in the three-coupling space labeled by $t_{2}, h, t_{3}$. 
A sample shape of free energy on this surface is shown on Fugure \ref{potential_fig2}. 
The values of the order parameter in the two phases: $\mu_{1}$ and $\mu_{3}$ on the coexistence surface away from the $\alpha$-line are not related by reflection. When $\mu_{1}$ and $\mu_{3}$ are of the same sign, as on Figure \ref{potential_fig2}, the phase that corresponds to $\mu_{i}$ with a larger magnitude   can be called "strongly magnetised" and the one with the lesser magnitude can be called "weakly magnetised". 
In other regions of the coexistence surface $\mu_{1}$ and $\mu_{3}$ have different signs  and different magnitude so that we can call them positively and negatively magnetised. 

One can systematically construct the coexistence surface parameterising it by the values of $\mu_{1}$ and $\mu_{2}$. We put the relevant formulae in Appendix \ref{appendix_Landau}.
It is not hard to show analytically that the phase coexistence surface can be defined as the graph of a function $h=h_{\rm co}(t_{2}, t_{3})$ 
whose domain is given by 
\be \label{domain_coe}
t_{2} < \frac{3}{2\cdot 10^{1/3}}|t_{3}|^{4/3} \, .
\ee
On Figure \ref{Landau_domain} we show this domain together with three lines where  $h_{\rm co}=0$ and the signs of $h_{\rm co}$ in between these lines. 

\begin{center}
\begin{figure}[H]  
\centering
\begin{tikzpicture}[>=latex]

\fill [gray!20!white, domain=-5:0, variable=\x]
      (-5, -4)
      --   plot ({\x}, {0.5*(-\x)^(1.4)}) 
      --(0,-4)
      -- cycle;
\fill [gray!20!white, domain=0:5, variable=\x]
      (0, -4)
      --   plot ({\x}, {0.5*(\x)^(1.4)}) 
      --(5,-4)
      -- cycle;

\draw[->] (-5.2,0) --(5.2,0);
\draw[->] (0,-4)--(0,4);
\draw[very thick, dashed] (0,-4) --(0,0);

\draw (5.5,0  ) node {$t_{3}$};
\draw (0,4.3) node {$t_{2} $};
\draw[ domain=0:5, smooth, variable=\x, red,very thick] plot ({\x}, {0.5*(\x)^(1.4)});
\draw[ domain=-5:0, smooth, variable=\x, red,very thick] plot ({\x}, {0.5*(-\x)^(1.4)});

\draw[ domain=0:5, smooth, dashed, variable=\x, blue, very thick] plot ({\x}, {0.2*(\x)^(1.4)});
\draw[ domain=-5:0, smooth, dashed, variable=\x, blue, very thick] plot ({\x}, {0.2*(-\x)^(1.4)});

\draw (0,-4.3) node { $h_{\rm co}=0$} ;
\draw (5.7,1.9) node { $h_{\rm co}=0$} ;
\draw (-5.7,1.9) node { $h_{\rm co}=0$} ; 
\draw (3.8,2) node { $h_{\rm co}>0$} ; 
\draw (-3.7,2) node { $h_{\rm co}<0$} ; 
\draw (3,-2) node { $h_{\rm co}<0$} ; 
\draw (-3,-2) node { $h_{\rm co}>0$} ;
\end{tikzpicture}
\caption{The domain of  function $h_{\rm co}(t_{2},t_{3})$ giving the two-phase coexistence surface. 
 } 
\label{Landau_domain}
\end{figure}
\end{center}
\begin{center}
\begin{figure}[H]
\centering
\includegraphics[scale=0.55]{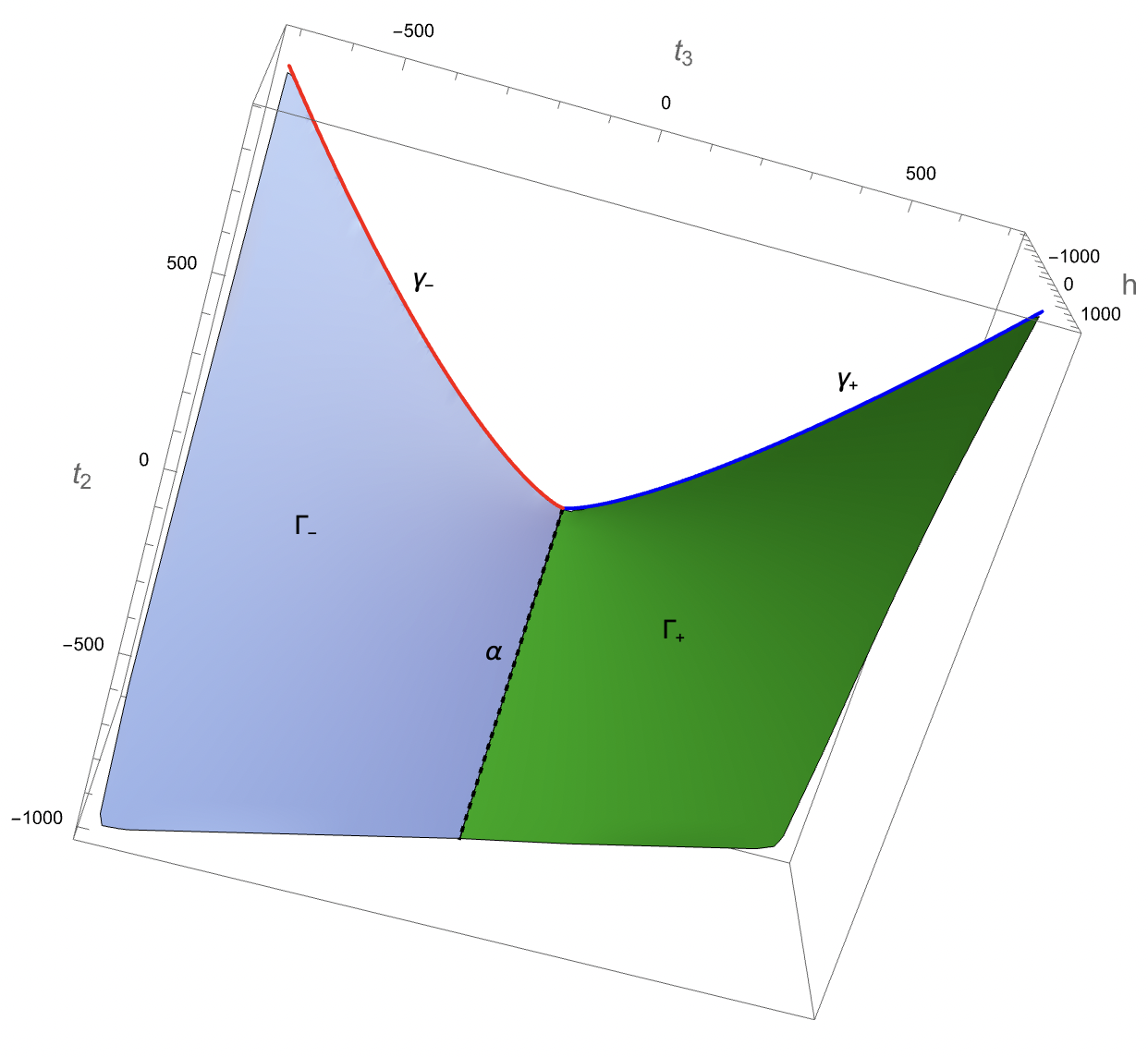}
\caption{ The two-phase coexistence surface of the three-coupling  Landau theory. It consists of two sheets: $\Gamma_{+}$ and $\Gamma_{-}$ that are joined at the symmetry breaking line $\alpha$. Each sheet ends on a critical line $\gamma_{\pm}$.}
\label{fig_Landau2}
\end{figure}
\end{center}

The red lines on Figure \ref{Landau_domain} are the graphs of the function 
\be
t_{2} = \frac{3}{2\cdot 10^{1/3}}|t_{3}|^{4/3}
\ee
and the blue dashed lines where $h_{\rm co}$ vanishes are given by 
\be
t_{2} = \frac{1}{2^{4/3}}|t_{3}|^{4/3} \, .
\ee
The regions between the red and blue dashed lines have $\mu_{1}$ and $\mu_{3}$ of the same sign thus representing the weakly and strongly magnetised 
phases. In the rest of the domain $\mu_{1}$ and $\mu_{3}$ have opposite signs describing positively and negatively magnetised phases. 
The coexistence surface consists of two sheets: $\Gamma_{+}$ where $t_{3}>0$ and $\Gamma_{-}$ where $t_{3}<0$. They are related by the reflection in the magnetic couplings: $(h,t_{3}) \to (-h,-t_{3})$. The two sheets are glued to each other at the symmetry breaking line $\alpha$. This is depicted on a three-dimensional plot on Figure \ref{fig_Landau2} which is a scaled version of the exact numerical surface. 

Each sheet $\Gamma_{\pm}$ ends on a critical line 
$\gamma_{\pm}$ on which the three extrema coincide: $\mu_{1}=\mu_{2}=\mu_{3}=0$. The equations for these lines can be found by solving 
in addition to  equation (\ref{derF}) two more equations:
\be \label{second_der}
\frac{d^2 {\cal F}}{d\mu^2} = t_{2} -2t_{3}\mu + 5\mu^{4} = 0 \, , 
\ee
\be  \label{third_der}
\frac{d^3 {\cal F}}{d\mu^3} = -2t_{3} + 20\mu^{3} = 0 \, .
\ee
Solving them we obtain the following equations for the critical lines $\gamma_{\pm}$
\be \label{crit_lines_Landau}
t_{2} = \frac{3}{2\cdot 10^{1/3}} |t_{3}|^{4/3} \, , \qquad h= {\rm sign}(t_{3}) \frac{3}{5\cdot 10^{2/3}}  |t_{3}|^{4/3}
\ee
where taking $t_{3}>0$ gives $\gamma_{+ }$ and taking $t_{3}<0$ gives $\gamma_{-}$.

Note that while  condition (\ref{second_der}) is necessary for the vanishing of the mass gap (infinite correlation length), the condition 
(\ref{third_der}) is also needed to ensure that we have a local minimum rather than an inflection point of ${\cal F}(\mu)$. 
For a later discussion it is instructive to look at a surface specified by two conditions 
\be \label{kink_Landau}
\frac{d{\cal F}}{d\mu} = 0 \, , \qquad \frac{d^3{\cal F}}{d\mu^3} = 0 \, .
\ee 
The significance of this surface can be clarified if we look at the magnetisation function $\mu = \mu(h ,t_{2}, t_{3})$ that in a single phase region is obtained 
as an inverse function to $h(\mu, t_{2}, t_{3})$ given by the right hand side of (\ref{derF}). Using rules for differentiating an inverse function we 
obtain 
\be
\frac{\partial^{2} \mu}{\partial h^{2}} = - \frac{\frac{\partial^{2} h}{\partial \mu^{2}}}{\left( \frac{\partial h}{\partial \mu}\right)^{3}} = 
- \frac{\frac{\partial^{3} {\cal F}}{\partial \mu^{3}}}{\left( \frac{\partial^{2} {\cal F}}{\partial \mu^2}\right)^{3}} \, .
\ee
We see from the last expression that  away from criticality, that is when $\frac{\partial^{2} {\cal F}}{\partial \mu^2} \ne 0$, 
on the surface  (\ref{kink_Landau})  we have an inflection point of the magnetisation $\mu$ as a function of the external field $h$. 
Moreover, since  (\ref{kink_Landau}) is part of the conditions for criticality (\ref{second_der}), (\ref{third_der}), the surface  (\ref{kink_Landau}) 
should pass through the critical region. 
We can make these observations more explicit by solving  (\ref{kink_Landau}) for our three coupling model. We obtain that this surface is given by 
equation
\be \label{Lkink_eq}
h= h^{*}\equiv {\rm sign} (t_{3}) (t_{2} \left| \frac{t_{3}}{10}\right|^{1/3} - 9  \left|\frac{t_{3}}{10}\right|^{5/3}) 
\ee
and the order parameter 
\be
\mu = \mu^{*} \equiv \left( \frac{t_{3}}{10}\right)^{1/3} \, .
\ee
Expanding near this surface in the order parameter $\mu=\mu^{*} + \xi$ we obtain 
\be \label{Lkink2}
h-h^{*} = |t_{3}|^{2/3} 10^{1/3}\xi^{3}  + T\xi  + \frac{5}{10^{1/3}}|t_{3}|^{1/3}\xi^4 + \xi^{5}
\ee
where 
\be
 T= t_{2} - \frac{3}{2\cdot 10^{1/3}} |t_{3}|^{4/3} \, . 
\ee
We see that $h-h^{*}$ as a function of $\xi$ has an inflection point at $\xi=0$ and its derivative at the origin becomes zero at criticality, that is when $T=0$ 
 (see (\ref{crit_lines_Landau})). When $T>0$ we are outside of the domain of the phase coexistence region   (\ref{domain_coe}). Moreover, it is 
 not hard to check that when $T>0$ we can take $|h-h^{*}|$ small enough so that (\ref{Lkink2}) has a unique inverse function $\xi(h-h^{*})$ 
 on such an interval (that means we are away from the region where there is a metastable second vacuum). The function  $\xi(h-h^{*})$ 
 has a kink-like shape with an inflection point at the origin. At criticality the magnetic susceptibility $\frac{\partial \mu}{\partial h}$ goes to infinity. 
 
\begin{center}
\begin{figure}[H]
\begin{minipage}[b]{0.5\linewidth}
\centering
\includegraphics[scale=0.82]{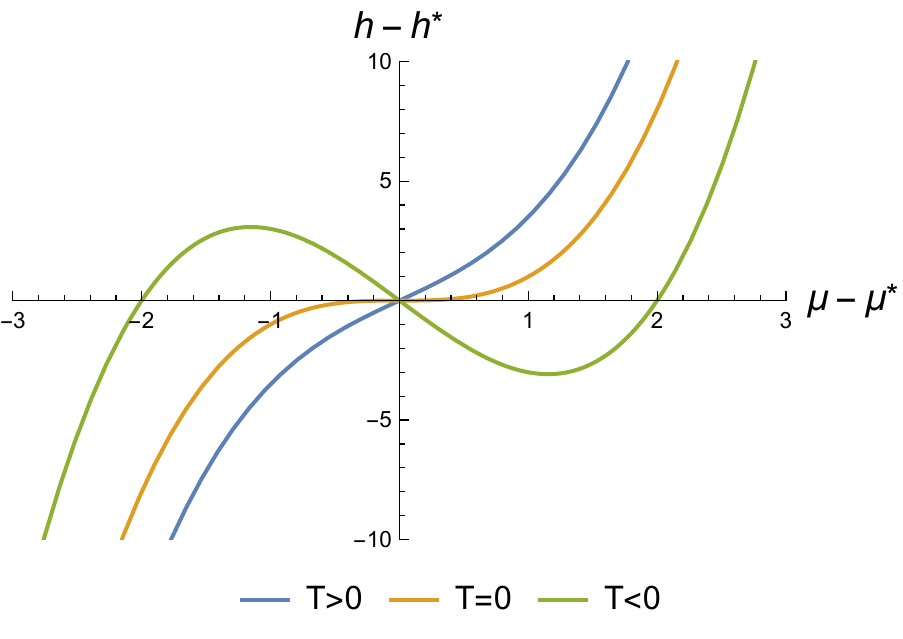}
\end{minipage}%
\begin{minipage}[b]{0.5\linewidth}
\centering
\includegraphics[scale=0.82]{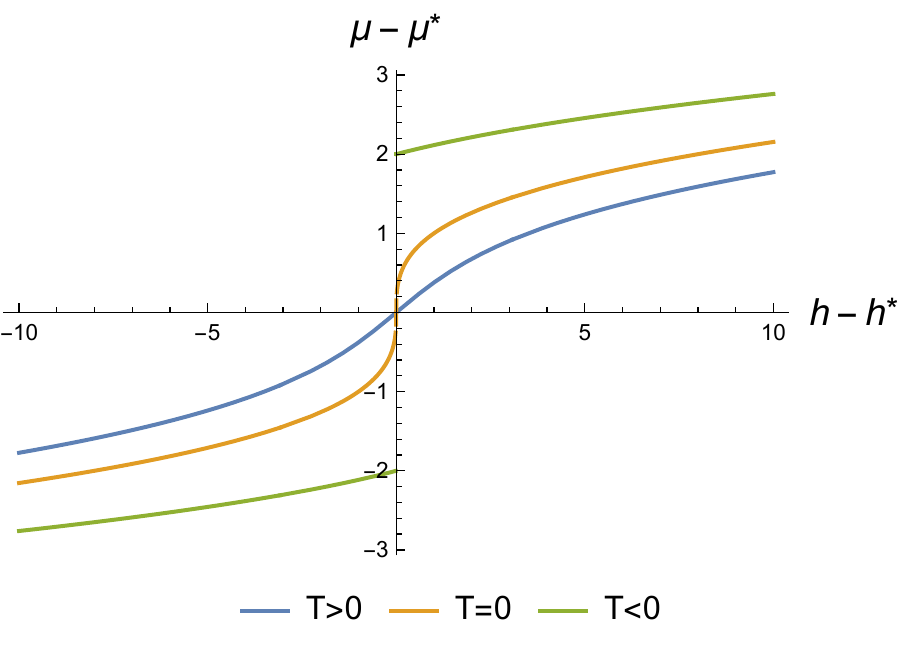}
\end{minipage}
\caption{The external field as a function of magnetisation (left) and the magnetisation as a  function of external field (right) at different values of parameter $T$.}
\label{Landau_kinks}
\end{figure}
\end{center}

 When $T<0$ there is no longer an inverse function for (\ref{Lkink2}) however small the difference $|h-h^{*}|$ is. When 
 \be \label{smallxi}
 |\xi|\ll \frac{10^{2/3}}{5} |t_{3}|^{1/3} 
 \ee
 we can neglect the quartic and quintic terms in (\ref{Lkink2}). This gives us an effective Ising-like free energy 
 \be 
 f(\mu) = \frac{T}{2} \xi^2 + \frac{10^{1/3}}{4}|t_{3}|^{2/3} \xi^{4} 
 \ee
and the magnetisation behaves like on the disorder line of the Ising model as long as (\ref{smallxi}) is satisfied. 
For $T<0$ we move into the two-phase coexistence region where there is an interval of $\xi$ where the inverse function has 3 branches 
(two of which give degenerate free energy minima) that results in a discontinuity of the magnetisation function at the origin.  
The behaviour of the magnetisation function  at different values of $T$ 
is illustrated on Figure \ref{Landau_kinks}. 
The pictures on Figure  \ref{Landau_kinks} are valid when  (\ref{smallxi}) holds  and  $|h-h^{*}|$ is small enough. 
To ensure that the magnetisation discontinuity satisfies  (\ref{smallxi}) we need 
\be
|T|\ll \frac{10^{5/3}}{25} |t_{3}|^{4/3} \, . 
\ee
When $T<0$ is too large in magnitude we can no longer neglect the quartic and quintic terms in (\ref{Lkink2}) that result in moving away from the two-phase 
coexistence surface. Thus, the surface defined by (\ref{kink_Landau}) passes through the critical region, as expected, but past the critical line it approximates the phase coexistence surface only when it is close enough to the critical line.

\subsection{Some generalities}

We next turn to the perturbed CFT analysis of TIM.  We first introduce in detail the notation for the bulk primaries and Cardy boundary conditions. These are presented 
 in Table  \ref{TIM:table}.
\begin{center}
\begin{table}[h!]
\centering
\begin{tabular}{|c|>{\centering\arraybackslash}p{1cm}|>{\centering\arraybackslash}p{1cm}|>{\centering\arraybackslash}p{1cm}|>{\centering\arraybackslash}p{1.4cm}|>{\centering\arraybackslash}p{1.4cm}|>{\centering\arraybackslash}p{2cm}|}
\hline 
\rule{0pt}{4ex}  \rule[-3ex]{0pt}{0pt}  Kac table label & (1,1)& (2,1)& (3,1)& (1,2)& (1,3) & (2,2) \\
\hline 
\rule{0pt}{4ex}  \rule[-3ex]{0pt}{0pt}  Conformal weight & 0& 7/16& 3/2& 1/10& 3/5 & 3/80 \\
\hline 
\rule{0pt}{4ex}  \rule[-3ex]{0pt}{0pt}  Conformal family notation & {\bf 1} & $\sigma'$ & $\epsilon''$ & $\epsilon$ & $\epsilon'$& $\sigma$\\
\hline 
\rule{0pt}{4ex}  \rule[-3ex]{0pt}{0pt}  Cardy b.c. notation & $(-)$ & $(0)$ & $(+)$ & $(-0)$ & $(+0)$ & $ (d)$\\

\hline
\rule{0pt}{4ex}  \rule[-3ex]{0pt}{0pt} Boundary fields & (1,1)& (1,1), (3,1)& (1,1)& (1,1), (1,3)& (1,1), (1,3) & 
(1,1), (1,3), (1,2), (3,1)\\
\hline 
\end{tabular}
\caption{Primary fields and Cardy boundary conditions  in the Tricritical Ising Model}
\label{TIM:table}
\end{table}
\end{center}

As we mentioned in the introduction the notation for boundary conditions is meant to  reflect the 
boundary conditions imposed on the spin variables $s_{i}$ in the underlying BEG model. 
The notation for $(d)$ comes from the word ``degenerate''\footnote{In some papers "d"  is interpreted as "disordered" that seems to be a better choice.} as the order parameter on the boundary is allowed to be in any of the three vacua. 
The boundary states for the Cardy boundary conditions can be written as\footnote{Note that we use here the notation introduced in \cite{Chim} in which the plus or minus sign in the notation corresponds to the opposite sign in front of $|\sigma\rangle\!\rangle$. This is the opposite rule to that used for the Ising model boundary states (\ref{conf_bcs}).}
\bea \label{b_states}
 |-\rangle\!\rangle\equiv |\tilde  1\rangle\!\rangle&= &5^{-1/4}\sqrt{s_2}\Bigl[  |1\rangle\!\rangle + |\epsilon''\rangle\!\rangle + 
\sqrt{\varphi}(|\epsilon\rangle\!\rangle + 
|\epsilon'\rangle\!\rangle) + 2^{1/4}\sqrt{\varphi}|\sigma\rangle\!\rangle + 2^{1/4}|\sigma'\rangle\!\rangle \Bigr] \, , \nonumber \\
 |+\rangle\!\rangle\equiv |\tilde \epsilon''\rangle\!\rangle & = & 5^{-1/4}\sqrt{s_2}\Bigl[  |1\rangle\!\rangle + |\epsilon''\rangle\!\rangle + \sqrt{\varphi}(|\epsilon\rangle\!\rangle + 
|\epsilon'\rangle\!\rangle) - 2^{1/4}\sqrt{\varphi}|\sigma\rangle\!\rangle - 2^{1/4}|\sigma'\rangle\!\rangle \Bigr] \, , \nonumber \\
 |\!-\!0\rangle\!\rangle \equiv |\tilde \epsilon\,\rangle\!\rangle &=& 5^{-1/4}\varphi\sqrt{s_2}\Bigl[  |1\rangle\!\rangle + |\epsilon''\rangle\!\rangle - 
\varphi^{-3/2}( |\epsilon\rangle\!\rangle + 
|\epsilon'\rangle\!\rangle) + 2^{1/4}\varphi^{-3/2}|\sigma\rangle\!\rangle - 2^{1/4}|\sigma'\rangle\!\rangle \Bigr] \, , \nonumber \\
 |\!+\!0\rangle\!\rangle \equiv |\tilde \epsilon'\,\rangle\!\rangle &=& 5^{-1/4}\varphi\sqrt{s_2}\Bigl[  |1\rangle\!\rangle + |\epsilon''\rangle\!\rangle - 
\varphi^{-3/2}( |\epsilon\rangle\!\rangle + 
|\epsilon'\rangle\!\rangle) - 2^{1/4}\varphi^{-3/2}|\sigma\rangle\!\rangle + 2^{1/4}|\sigma'\rangle\!\rangle \Bigr] \, , \nonumber \\
 |0\rangle\!\rangle \equiv |\tilde \sigma'\, \rangle\!\rangle &=& 5^{-1/4}\sqrt{2s_2}\Bigl[ |1\rangle\!\rangle - |\epsilon''\rangle\!\rangle + 
   \sqrt{\varphi}( |\epsilon'\rangle\!\rangle - 
|\epsilon\rangle\!\rangle)  \Bigr] \, , \nonumber \\
 |d\rangle\!\rangle \equiv |\tilde \sigma\, \rangle\!\rangle &=& 5^{-1/4}\sqrt{2s_2}\varphi\Bigl[ |1\rangle\!\rangle - |\epsilon''\rangle\!\rangle + 
 \varphi^{-3/2}( |\epsilon\rangle\!\rangle - 
|\epsilon'\rangle\!\rangle)  \Bigr] \, 
\eea 
where the states $|a\rangle\!\rangle$ on the right hand sides are the Ishibashi states and 
\be
 s_2=\sin(4\pi/5)=\frac{\sqrt{5-\sqrt{5}}}{\sqrt{8}} \, ,
\qquad 
\varphi=\ \frac{1+\sqrt{5}}{2} \, .
\ee
In (\ref{b_states}) we factored out the coefficient at the Ishibashi state corresponding to the identity so that the coefficients 
at the other Ishibashi states give the normalised disc one-point functions of the corresponding operators. Having in mind the Cardy's variational ansatz it is useful to note that the above coefficients 
satisfy 
\be  \label{order_of_constants}
\varphi^{-3/2}<2^{1/4}\varphi^{-3/2}<2^{1/4} <\sqrt{\varphi}<2^{1/4}\sqrt{\varphi} \, .
\ee

We are going to focus from now on on the TIM perturbed by its two most relevant operators 
\be\label{TIFT}
H=  H_{0} + \lambda_{\epsilon} \int\limits_{0}^{R} \phi_{\epsilon} (0,y)dy + \lambda_{\sigma} \int\limits_{0}^{R} \phi_{\sigma} (0,y)dy \, .
\ee
The Cardy's ansatz variational energies labelled by the boundary states (\ref{b_states}) for this theory are as follows
\bea \label{TIM_var_1}
E_{\pm}(\tau) &=& R\Bigl[ V_{0}(\tau) +\lambda_{\epsilon} \sqrt{\varphi} V_{\epsilon}(\tau) \mp \lambda_{\sigma} 2^{1/4} \sqrt{\varphi} V_{\sigma}(\tau) \Bigr] \,  , \nonumber \\
E_{\pm 0}(\tau) &=& R\Bigl[ V_{0}(\tau) - \lambda_{\epsilon} \varphi^{-3/2} V_{\epsilon}(\tau) \mp \lambda_{\sigma} 2^{1/4} \varphi^{-3/2} V_{\sigma}(\tau) \Bigr] \,  , \nonumber \\
E_{0}(\tau) &=& R\Bigl[ V_{0}(\tau) - \lambda_{\epsilon}\sqrt{ \varphi}  V_{\epsilon}(\tau) \Bigr] \,  , \nonumber \\ 
E_{d}(\tau) &=& R\Bigl[ V_{0}(\tau) + \lambda_{\epsilon} \varphi^{-3/2}  V_{\epsilon}(\tau) \Bigr]
\eea
where for brevity we used the notation 
\be
V_{0}(\tau) = \frac{7\pi}{240(2\tau)^2} \, , \quad V_{\epsilon}(\tau) = \left(\frac{\pi}{4\tau}\right)^{1/5} \, , \quad  V_{\sigma}(\tau) = \left(\frac{\pi}{4\tau}\right)^{3/40} \, .
\ee

The cases $\lambda_{\epsilon}=0$ and $\lambda_{\sigma}=0$ were studied in \cite{Cardy_var}, \cite{LVT}. 
For $\lambda_{\sigma}=0$  and  $\lambda_{\epsilon}>0$ we are in the disordered region with a unique vacuum which is approximated by a smeared $|0\rangle\!\rangle$ boundary state. When $\lambda_{\sigma}=0$  and  $\lambda_{\epsilon}<0$ we are in the broken symmetry region 
and the two degenerate vacua are described as  smeared $|+\rangle\!\rangle$ and $|-\rangle\!\rangle$. The true vacuum at finite  $R$ is 
the smeared linear combination $|+\rangle\!\rangle + |-\rangle\!\rangle$. 
When $\lambda_{\epsilon}=0$ the variational ansatz gives for the vacuum the smeared $|+\rangle\!\rangle$ boundary state for $\lambda_{\sigma}<0$ 
and the smeared $|-\rangle\!\rangle$ when $\lambda_{\sigma}>0$. 
In \cite{LVT} the Cardy's ansatz results were quantitatively compared with TCSA numerical results and a rather good  
numerical agreement was found.

\subsection{Mixed leading thermal and magnetic perturbations for $\lambda_{\epsilon}<0$} 
\begin{center}
\begin{figure}[H]
\begin{minipage}[b]{0.5\linewidth}
\centering
\includegraphics[scale=0.82]{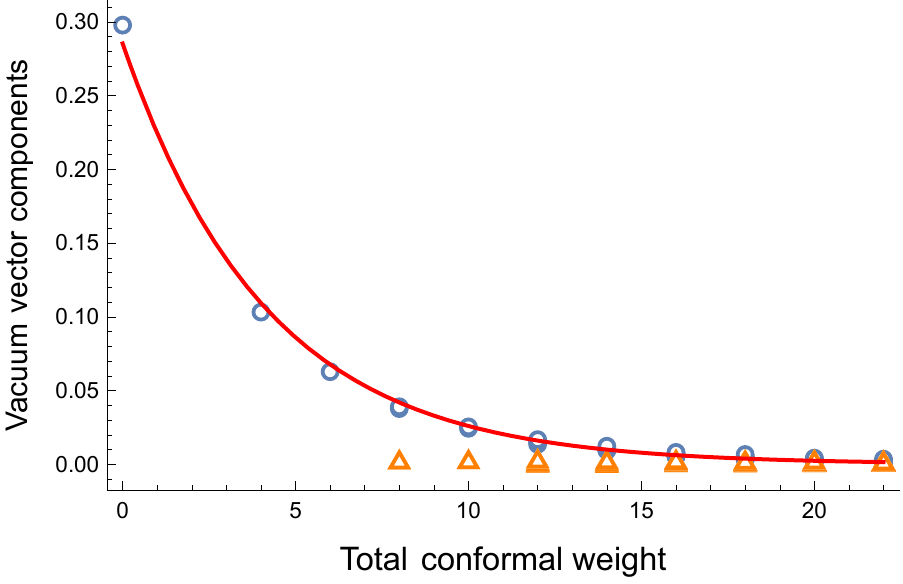}
\end{minipage}%
\begin{minipage}[b]{0.5\linewidth}
\centering
\includegraphics[scale=0.82]{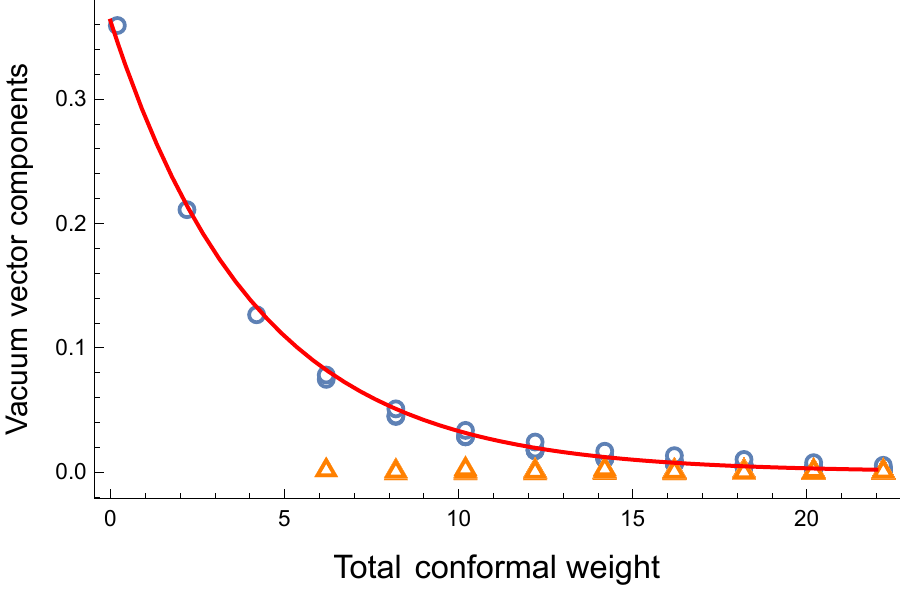}
\end{minipage}
\caption{The TCSA vacuum vector components $C_{i}$ against the total conformal weight of the basis vectors taken for the TIM with $\xi_{\sigma}=1.5$, $\lambda_{\epsilon}<0$ and at  $r=30$. The left plot 
represents the vacuum sector and the right plot -- the $\epsilon$-sector. The red line coresponds to the distribution in the Cardy's ansatz solution. 
The orange triangles mark the non-diagonal components.}
\label{fig_TIM_spec1}
\end{figure}

\end{center}

\begin{center}
\begin{figure}[H]
\begin{minipage}[b]{0.5\linewidth}
\centering
\includegraphics[scale=0.82]{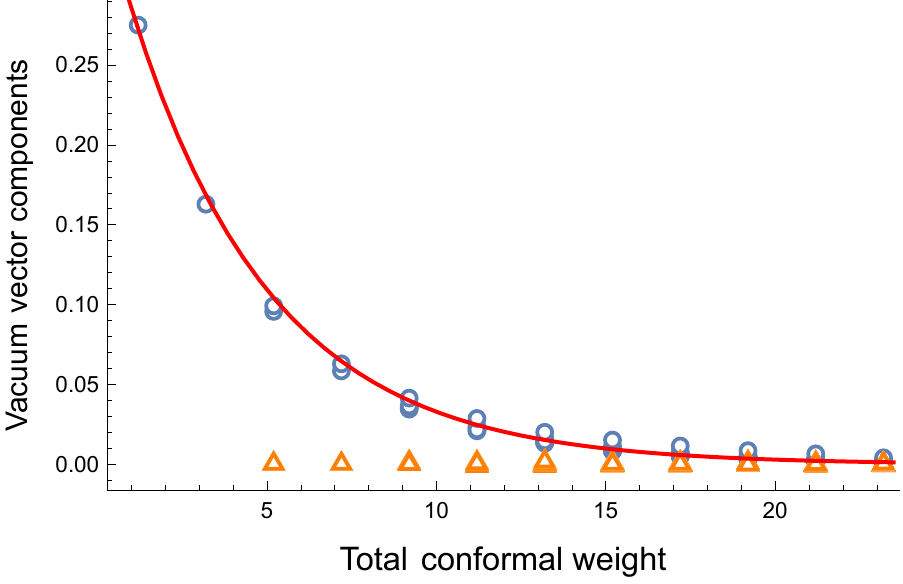}
\end{minipage}%
\begin{minipage}[b]{0.5\linewidth}
\centering
\includegraphics[scale=0.82]{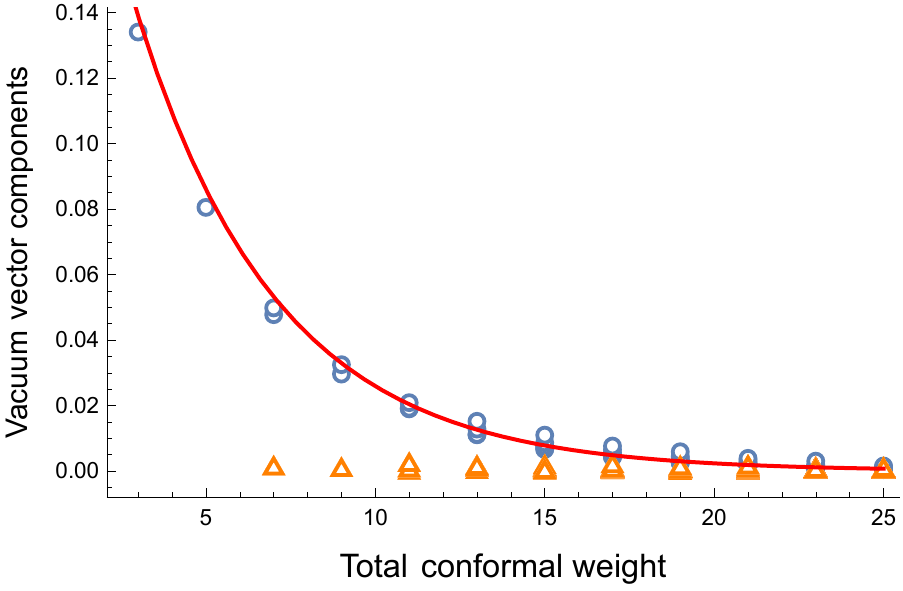}
\end{minipage}
\caption{The TCSA vacuum vector components in the $\epsilon'$-sector (left) and the  $\epsilon''$-sector (right). See the caption to Fig. \ref{fig_TIM_spec1} 
for more details.}

\label{fig_TIM_spec2}
\end{figure}

\end{center}

\begin{center}
\begin{figure}[H]
\begin{minipage}[b]{0.5\linewidth}
\centering
\includegraphics[scale=0.82]{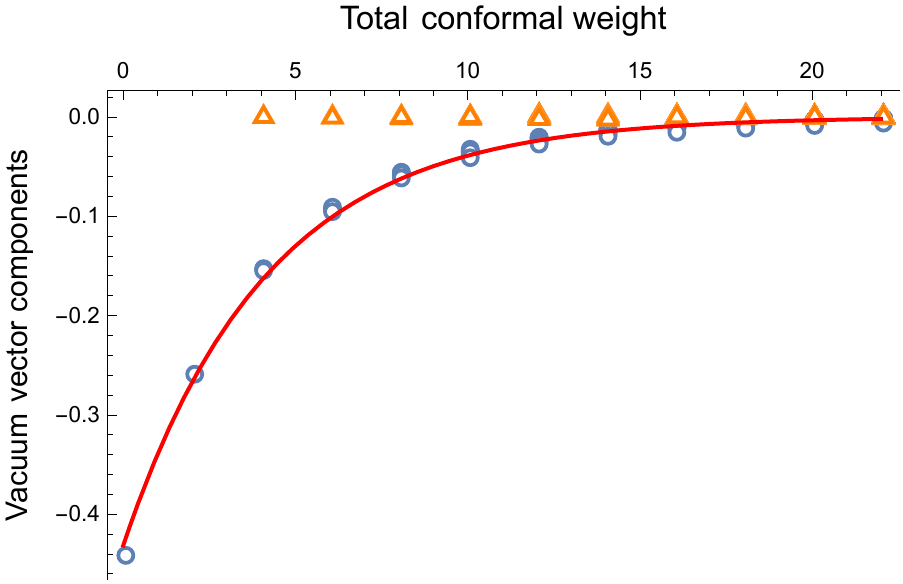}
\end{minipage}%
\begin{minipage}[b]{0.5\linewidth}
\centering
\includegraphics[scale=0.82]{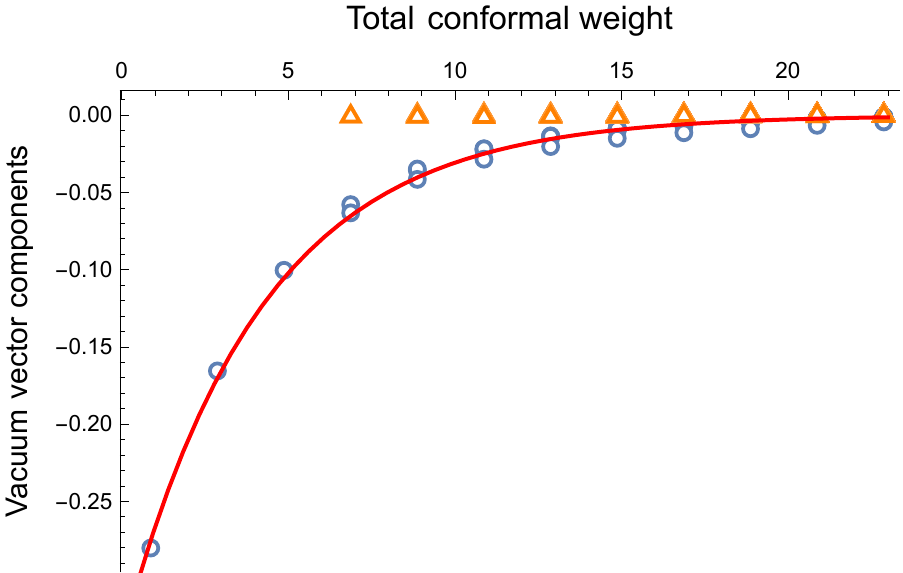}
\end{minipage}
\caption{The TCSA vacuum vector components in the $\sigma$-sector (left) and the  $\sigma'$-sector (right). See the caption to Fig. \ref{fig_TIM_spec1} 
for more details.}

\label{fig_TIM_spec3}
\end{figure}

\end{center}

When both perturbations are present in (\ref{TIFT}) but the magnetic coupling is relatively small  it is convenient to parameterise the RG trajectories by 
a parameter 
\be \label{xi_def}
\xi_{\sigma} = \frac{\lambda_{\sigma}}{|\lambda_{\epsilon}|^{77/72}}
\ee
and to measure the radius in units set by the mass gap of the pure $\lambda_{\epsilon}$ deformation of the tricritical theory. 
The letter theory is integrable and the mass--coupling relation is known exactly: 
\be\label{kappaepsilon}
\lambda_{\epsilon} = \kappa_{\epsilon} m^{2-2h}=\kappa_{\epsilon} m^{9/5}
\ee
where 
\be
\kappa_{\epsilon} = \frac{1}{\pi}\sqrt{\frac{\Gamma^3(1/5)\Gamma(3/5)}{4\Gamma^3(4/5)\Gamma(2/5)}}\left( \frac{\Gamma(2/3)\Gamma(5/9)}{2\Gamma(2/9)}\right)^{9/5} \approx  0.0928344
\ee
We introduce the  dimensionless length of the system 
\be
r = R m= (\kappa_{\epsilon})^{-5/9} R|\lambda_{\epsilon}|^{5/9} 
\ee
and the  dimensionless variational energies 
\be
e_{i}= \frac{E_{i}}{ (\kappa_{\epsilon}|\lambda_{\epsilon}|)^{5/9} } \, . 
\ee
Explicitly the variational energies are given by 
\bea \label{var_e}
e_{\pm}(\tilde \tau) &= & r \Bigl[\frac{7}{60 \pi \tilde \tau^2} + a \frac{\sqrt{\varphi}}{(\tilde \tau)^{1/5}}   \mp \xi_{\sigma} \frac{2^{1/4} \sqrt{\varphi}}{(\tilde \tau)^{3/40}} \Bigr] \, , \nonumber \\
e_{\pm 0}(\tilde \tau) &= &r \Bigl[[\frac{7}{60 \pi \tilde \tau^2} - a \frac{\varphi^{-3/2}}{(\tilde \tau)^{1/5}}  \mp \xi_{\sigma}  \frac{2^{1/4} \varphi^{-3/2}}{(\tilde \tau)^{3/40}} \Bigr] \, , \nonumber \\
e_{0}(\tilde \tau) &=& r\Bigl[\frac{7}{60 \pi \tilde \tau^2}  - a  \frac{\sqrt{\varphi}}{(\tilde \tau)^{1/5}}      \Bigr]   \, , \nonumber \\
e_{d}(\tilde \tau) &=&r\Bigl[\frac{7}{60 \pi \tilde \tau^2}  + a  \frac{\varphi^{-3/2}}{(\tilde \tau)^{1/5}}      \Bigr] 
\eea
where 
\be \label{tt_def}
a= {\rm sign}(\lambda_{\epsilon}) \, , \qquad \tilde \tau = \frac{4}{\pi} \tau |\lambda_{\epsilon}|^{5/9} \, . 
\ee
  Using (\ref{order_of_constants}) we see that in the symmetry breaking region, when $a=-1$, $e_{0}(\tilde \tau)$ and $e_{-0}(\tilde \tau)$ do not have a local minimum at finite $\tilde \tau$ 
  and thus can be discarded. For the remaining energies, for all values of $\tilde \tau$  we have 
  \be
  e_{+0}(\tilde \tau) > e_{+}(\tilde \tau) \, , \quad  e_{-}(\tilde \tau) > e_{+}(\tilde \tau) \, , \quad  e_{d}(\tilde \tau) > e_{+}(\tilde \tau) 
  \ee
  that indicates that throughout this region the asymptotic vacuum is given by a smeared $|+\rangle\!\rangle$ boundary state. 
  Using TCSA data\footnote{ Everywhere in this section the numerics was obtained with the descendant level cutoff at $n_{c}=11$ that gives a dimension 8810 truncated state space.} we probed a  range of values of $\xi_{\sigma}$ and found that at the end of the physical window (the largest value of $r$ where the vacuum energy is well approximated by a linear function)  the vacuum state qualitatively always looks like a smeared  $|+\rangle\!\rangle$. Comparing with Cardy's ansatz we find a good match for a range of values of $\xi_{\sigma}$. 
  As an illustration we present 
  on Figures \ref{fig_TIM_spec1} - \ref{fig_TIM_spec3} plots of the components' distribution for the TCSA vacuum vector taken with $\xi_{\sigma}=1.5$,
  $\lambda_{\sigma}>0$
  and at $r=30$. The diagonal states are marked by 
  blue circles and the non-diagonal ones by orange triangles. 
  On every plot we also put the 
  Cardy's ansatz solution corresponding to $e^{-\tau^{*} H_{0}}|+\rangle\!\rangle$ with numerically found minimum of $e_{+}$ at $\tau^{*} m \approx 1.1454$.

  
\subsection{Small magnetic perturbation in the presence of  $\lambda_{\epsilon}>0$}  \label{small_sigma_sec}
In the disordered region we have $\lambda_{\epsilon}>0$ and $a=1$. We see from (\ref{var_e}) that $e_{d}$ and $e_{-}$ do not have minima 
at finite values of $\tilde \tau$. Also $e_{-0}(\tilde \tau) >e_{+0}(\tilde \tau)$ for all values of $\tilde \tau$. This leaves us with three competing variational energies in this 
region: $e_{0}$, $e_{+0}$, and $e_{+}$. We find numerically that the minimal value is provided by $e_{0}$ for $\xi_{\sigma}< \xi_{c}^{(1)}$, 
by $e_{+0}$ in the range: $\xi_{c}^{(1)} < \xi_{\sigma} < \xi_{c}^{(2)} $, and by $e_{+}$ for $\xi_{\sigma}> \xi_{c}^{(2)}$ with 
\be
\xi_{c}^{(1)}= 1.4593... \, , \qquad \xi_{c}^{(2)} = 1.8351...
\ee

We will first discuss the region of small $\xi_{\sigma}$. Here we are facing the same kind of problem with the ansatz as in the disordered region of the Ising model 
that we discussed in section \ref{Ising_sec_disorder}. Namely, for  $\xi_{\sigma}< \xi_{c}^{(1)}$ the vacuum selected by the ansatz has zero magnetisation. 
While in the Ising case to fix this problem we switched on the boundary magnetic field perturbation which generated a boundary RG flow, 
 in the case at hand we only have an irrelevant symmetry breaking operator on the boundary: $\psi_{\epsilon''} \equiv \psi_{3,1}$. 
 This suggests to replace the ansatz (\ref{trial_vac}) with  
 \be \label{irrel_ansatz}
 e^{-\tau H_{0}} e^{-\lambda_{b} \int\limits_{0}^{R} \psi_{\epsilon''}(y) dy} |0\rangle \!\rangle 
 \ee
 where $\lambda_{b}$ is the boundary coupling. This is a schematic expression in which the first factor is an explicit operator acting in the bulk CFT state space while  the second exponential stands for the boundary 
 state of the theory perturbed by the boundary operator $\psi_{\epsilon''}$. Note that $\psi_{\epsilon''}$ has a scaling dimension of $3/2$ so that the 
 integrated operator scales like $1/\sqrt{R}$ that means that at large $R$ and at low weights this term will dominate over the $H_{0}$ term while the latter dominates at large conformal weights. Although the perturbation by a $\psi_{(3,1)}$ boundary field, provided one can handle all of the divergences,  may well be integrable we will not try to find any exact analytic expression for the above boundary state. Rather we will rewrite it, change to a simplified ansatz and then analyse the leading perturbative terms in $\lambda_{b}$.

 For the perturbation at hand we can  write the perturbed 
 boundary state as a result of the action of a certain loop operator on the conformal boundary state $|+\rangle\!\rangle$. To that end we follow 
 the discussion of analogous  integrable boundary conditions   in \cite{BachasGaberdiel}, \cite{Ingo_defects}, \cite{Gaiotto_Kondo}. To utilise the 
 construction of those papers we use the topological defects of TIM as well as  supersymmetry. As in any A-series minimal model, these defects are labelled by the 
 irreducible Virasoro representations \cite{Petkova_Zuber}. We will denote the corresponding defect lines as ${\cal L}_{1}$, ${\cal L}_{\epsilon}$, etc. and the  defect operators as ${\cal D}_{1}$, ${\cal D}_{\epsilon}$ etc. 
 The defect ${\cal L}_{\epsilon''}$ in TIM is the spin reversal symmetry defect and ${\cal L}_{\sigma'}$ is the duality defect. As was realised in \cite{Friedan_etal} 
 and further developed in \cite{Qiu} there is a fermionic version of TIM CFT that carries a representation of $N=1$ superconformal algebra. The fermionic fields 
 are realised as non-local fields living at the end of the spin reversal defect   (see \cite{KW_defects} and  \cite{TIM4} for a thorough discussion of the disorder fields  and duality in the Ising and TIM).  The fermionic field at the tip of ${\cal L}_{\epsilon''}$ is labelled by a pair $(a,b)$ of holomorphic and ant-holomorphic representations 
 whose fusion must contain $\epsilon''$. This gives us 3 types of fermionic fields: the disorder fields $\phi_{\mu}$ and $\phi_{\mu'}$ labelled by 
 the pairs $(\sigma, \sigma)$ and $(\sigma',\sigma')$ respectively, the chiral spin $\pm 1/2$  fields $G(z)$, $\bar G(\bar z)$ with labels 
 $(\epsilon'',1)$, $(1,\epsilon'')$ and spin $\pm 1/2$ fields $\phi_{\pm}$ with labels $(\epsilon, \epsilon')$ and $(\epsilon', \epsilon)$. The fields $G$ and $\bar G$ are the odd
 superconformal generators and the fields $\phi_{\pm}$  are up to normalisation the descendant fields $G_{-1/2}\phi_{\epsilon}$, $\bar G_{-1/2}\phi_{\epsilon}$.
 The ${\bf 1}, \epsilon, \epsilon', \epsilon''$ sectors furnish a representation of the NS-sector of the superconformal algebra while 
 the sectors $\sigma, \sigma'$ carry the Ramond sector representation. Noting that we  have 
 \be
|0\rangle\!\rangle =  {\cal D}_{\sigma ' } |+\rangle\!\rangle 
 \ee
  and that ${\cal L}_{\sigma'}$ has a chiral defect field $\psi_{\epsilon''}^{d}$ with weights  $(3/2,0)$  we can represent the perturbed boundary state 
  in (\ref{irrel_ansatz}) as 
  \be \label{defect_exp1}
  e^{-\lambda_{b} \int\limits_{0}^{R} \psi_{\epsilon''}(y) dy} |0\rangle \!\rangle  =
   {\rm Pexp} \left(\lambda_{b} \int\limits_{0}^{R}\psi_{\epsilon''}^{d} (y) dy  \right) |+\rangle \!\rangle  
  \ee
where the  path ordered exponential\footnote{Although there are no representation matrices which are present in non-abelian Wilson loops we 
put in path ordering as there are contact terms in the commutator of $\psi_{\epsilon''}(y)$ inserted at different points in the integration contour. Those in principle should be handled by a renormalisation prescription which is the trickiest part in defining the loop operator. See \cite{BachasGaberdiel}, \cite{Gaiotto_Kondo} for a discussion.} acts in the bulk CFT state space. We can further rewrite this loop operator by 
representing $\psi_{\epsilon''}^{d}$ as a composition of the bulk fermion $G(z)$ and a topological junction $a$ of ${\cal L}_{\sigma'}$ 
with ${\cal L}_{\epsilon''}$ as illustrated on Figure \ref{defect_fig}. 
The topological junction $a$ plays the role of a boundary fermion field used in the construction of the boundary magnetic field operator in the Ising model 
\cite{GZ}. The systematics behind such constructions is explained in \cite{Runkel_Watts_fermions}. If we shrink the ${\cal L}_{\epsilon''}$ 
we obtain a representation $\psi_{\epsilon''}^{d}=iaG$ that is similar to the expression of the boundary magnetic field $\psi_{\epsilon} $ in the Ising model  in terms of the bulk and boundary free fermion fields: $\psi_{\epsilon} = ia(\psi + \bar \psi) $. If instead we shrink ${\cal L}_{\sigma'}$ onto 
a bulk field  $\phi_{i}$ inserted at the origin we obtain a pair of fermionic fields: $G$ and a disorder field $\tilde \phi_{i}$. This configuration vanishes 
unless $\phi_{i}$ is one of the spin fields: $\phi_{\sigma}$ and $\phi_{\sigma'}$.

\begin{center}
\begin{figure} [H]
\centering
\begin{tikzpicture} [>=latex,scale=1.1]
\draw[very thick,blue,->] (-5,0) circle (1.5);
\draw[very thick, blue,->] (-4.94,1.5)--(-4.93,1.5); 
\draw (-4.7,0) node {$\phi_{i}$};
\draw (-5,0) node  {$\mathsmaller{\times} $};
\draw (-6.5,0) node  {$\mathsmaller{\times} $};
\draw (-6.9,0) node {$ \psi_{\epsilon''}^{d}$};
\draw (-5,1.8) node {${\cal L}_{\sigma'}$};

\draw (-2.8,0) node {$=$};

\draw[very thick,blue,->] (0,0) circle (1);
\draw[very thick, blue,->] (0.06,1)--(0.07,1) ;
\draw (0.3,0) node {$\phi_{i}$};
\draw (0,0) node  {$\mathsmaller{\times} $};
\draw[very thick, red ,->] (-2,0)--(-1.3,0) ;
\draw[very thick, red ] (-1.4,0)--(-1,0) ;
\draw (-1,0) node {$\bullet$};
\draw (-2.03,0) node {$\mathsmaller{\times}$};
\draw (-2,0.3) node {$G$};
\draw (-0.8,0) node {$a$};
\draw (0,1.3) node {${\cal L}_{\sigma'}$};
\draw (-1.5,-0.3) node {${\cal L}_{\epsilon''}$};

\draw (2,0) node {$=$};

\draw[very thick, red] (3.4,0)--(4,0);
\draw[very thick, red ,->] (3,0)--(3.6,0) ;
\draw (2.96,0) node  {$\mathsmaller{\times} $};
\draw (4.04,0) node  {$\mathsmaller{\times} $};
\draw (4.1,0.34) node {$\tilde \phi_{i}$};
\draw (3,0.3) node {$G$};
\draw (3.65,-0.33) node {${\cal L}_{\epsilon''}$};

\end{tikzpicture}
\caption{On the left side the defect  ${\cal L}_{\sigma'}$  with an insertion of the defect field $\psi_{\epsilon''}^{d}$ 
is wrapped around  an insertion of  a bulk operator $\phi_{i}$. This configuration is deformed by shrinking ${\cal L}_{\sigma'}$ slightly and extending a segment of the 
 spin reversal defect ${\cal L}_{\epsilon''}$. Shrinking ${\cal L}_{\sigma'}$ further onto the bulk insertion we obtain a pair of fermionic fields. 
}
\label{defect_fig}
\end{figure}
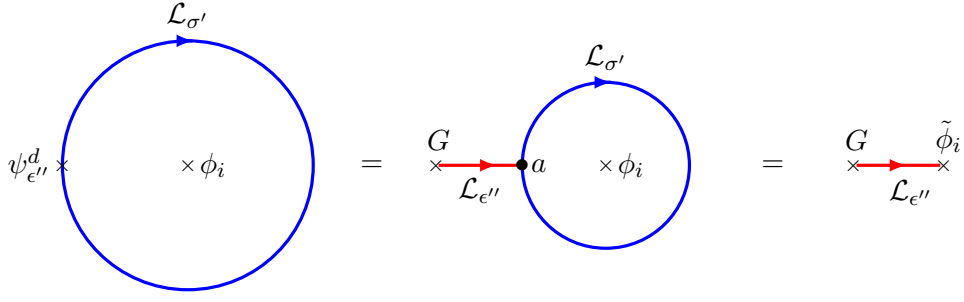
\end{center}

It is clear that when we expand the exponential in (\ref{defect_exp1}) only the configurations with an   odd number of insertions will 
contribute in the Ramond sectors (which contain the spin fields) and only those with an even number of insertions contribute in the NS sector 
(containing the thermal fields). The leading order perturbation of the boundary state can be described as follows. Consider a descendant vector in the Ramond sector 
\be
|n, \bar n; i\rangle = L_{-n_{1}} L_{-n_{2}} \dots L_{-n_{k}} \bar L_{-\bar n_{1}}  \bar L_{-\bar n_{2}} \dots \bar L_{-\bar n_{l}} |i\rangle  \, , \quad 
n_{i}\in {\mathbb N} \, , \enspace n_{j}\in {\mathbb N} \, 
\ee
where $i\in \{\sigma, \sigma'\}$. 
Then at the leading order  up to a normalisation constant 
\be \label{leading_act}
 {\rm Pexp} \left(\lambda_{b} \int\limits_{0}^{R}\psi_{\epsilon''}^{d} (y) dy  \right) |n, \bar n; i\rangle \sim  \left(\frac{2\pi}{R}\right)^{1/2} \lambda_{b} \hat L_{G} |n, \bar n; i\rangle 
\ee
where $\hat L_{G}$ is a linear operator defined by 
\be
\hat L_{G}  |n, \bar n; i\rangle=G_{0} L_{-n_{1}} \dots L_{-n_{k}} \bar L_{-\bar n_{1}}   \dots \bar L_{-\bar n_{l}} G_{0}  |i\rangle \, .
\ee
This operator is  representing the pair of defect fields depicted on the right hand side of Figure \ref{defect_fig} integrated over the position of $G(z)$. 
Its action  amounts to a linear combination of states of the form $|n, \bar n\rangle$ at the same holomorphic and anti-holomorphic level. Since $G_{0}$ 
commutes with $\bar L_{-\bar n_{i}}$ the anti-holomorphic content of the state remains the same. 
Furthermore the relation
\be
G_{0}^{2} = L_{0} - \frac{7}{240} 
\ee
implies that we can find eigenvectors at each holomorphic level $N$ for which (\ref{leading_act}) amounts to multiplication by 
\be
\pm \sqrt{(h_{i} + N - \frac{7}{240}  ) ( h_{i} - \frac{7}{240})  }   \, . 
\ee
We find by explicit calculation\footnote{In these calculations one needs to use the explicit form of supersymmetric null vectors.}  the expressions for levels 0,1 and 2:
\be \label{calc1}
G_{0}^2 |\sigma\rangle = \frac{1}{120} |\sigma\rangle \, , \qquad G_{0}^2 |\sigma'\rangle = \frac{49}{120} |\sigma'\rangle \, , 
\ee
\be \label{calc2}
G_{0}L_{-1}G_{0}|\sigma\rangle = \frac{1}{6} L_{-1}|\sigma\rangle \, , \qquad 
G_{0}L_{-1}G_{0}|\sigma'\rangle = \frac{91}{120} L_{-1}|\sigma'\rangle \, , 
\ee
\be \label{calc3} 
G_{0} {\cal O}_{\pm }^{(2)} G_{0}|\sigma\rangle = \pm \frac{\sqrt{241}}{120}  {\cal O}_{\pm }^{(2)} |\sigma\rangle \, , \qquad 
G_{0}(L_{-1})^2 G_{0}|\sigma'\rangle = \frac{119}{120} (L_{-1})^2 |\sigma' \rangle 
\ee
where 
\be
 {\cal O}_{\pm }^{(2)} = \frac{1}{40}(11\pm \sqrt{241}) L_{-1}^2 + L_{-2} \, .
\ee
In general it is clear that there is at each level a basis of (non-orthogonal)  vectors which under the action of $\hat L_{G}$ are multiplied by a positive or negative number 
which asymptotically grows like the  square root of the holomorphic descendant level.

\begin{center}
\begin{figure}[H]
\begin{minipage}[b]{0.5\linewidth}
\centering
\includegraphics[scale=0.82]{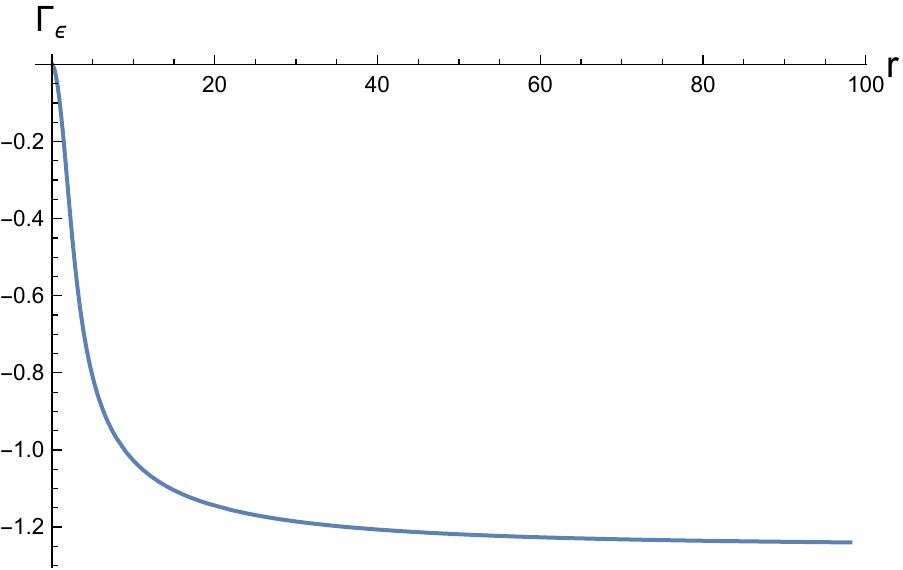}
\end{minipage}%
\begin{minipage}[b]{0.5\linewidth}
\centering
\includegraphics[scale=0.82]{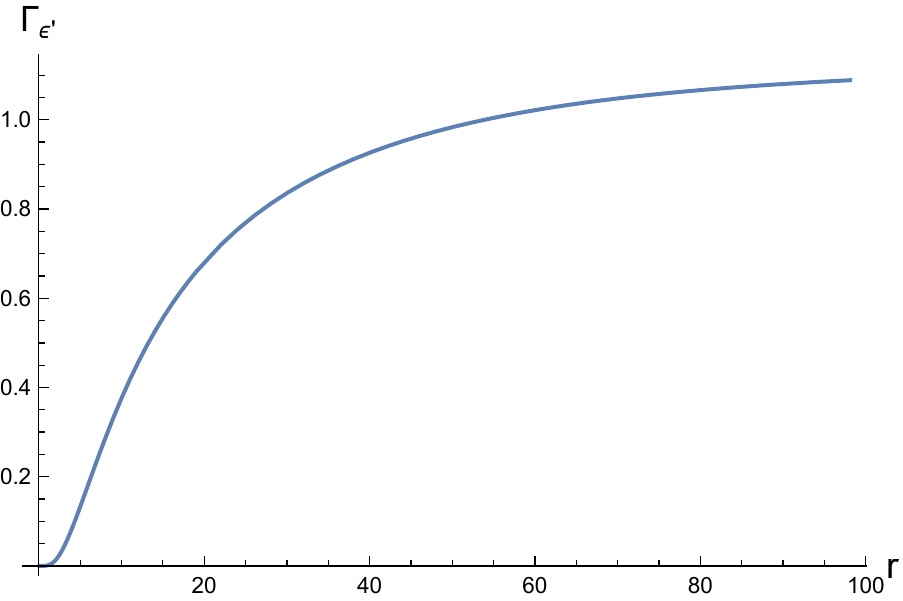}
\end{minipage}
\caption{The component ratios $\Gamma_{\epsilon}$ (left), $\Gamma_{\epsilon'}$  (right) obtained via TCSA for 
$\xi_{\sigma}=0.1$, $\lambda_{\epsilon}>0$ at $n_{c}=11$.  }
\label{gammas_pic1}
\end{figure}
\end{center}

\begin{center}
\begin{figure}[H]
\begin{minipage}[b]{0.5\linewidth}
\centering
\includegraphics[scale=0.8]{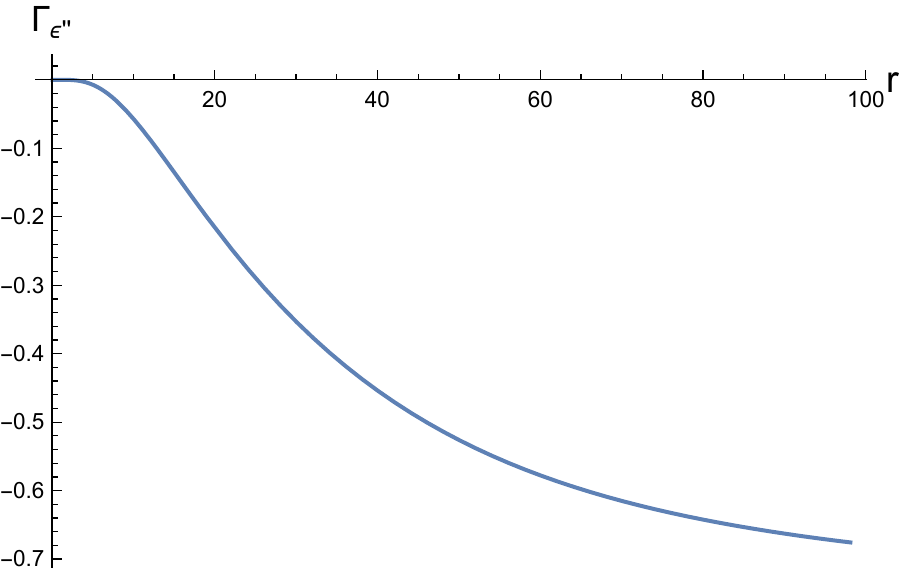}
\end{minipage}%
\begin{minipage}[b]{0.5\linewidth}
\centering
\includegraphics[scale=0.8]{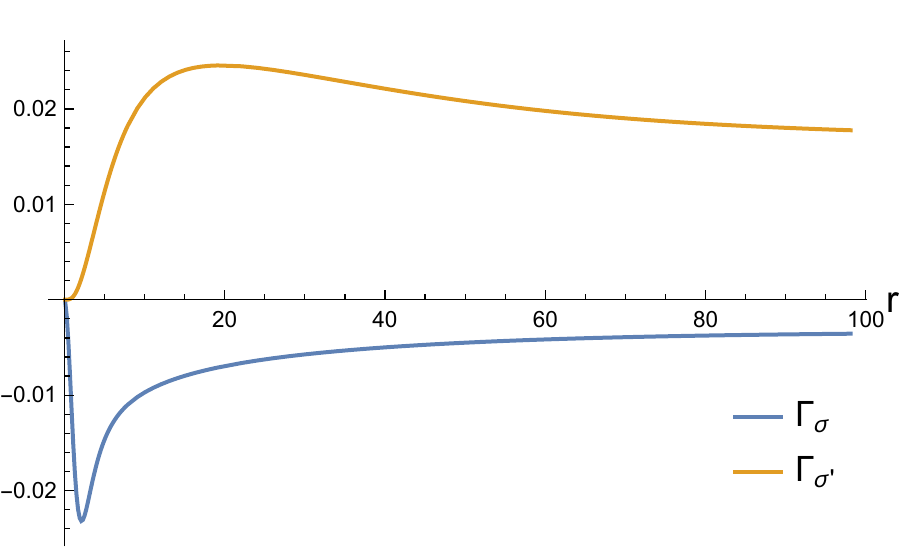}
\end{minipage}
\caption{The component ratios $\Gamma_{\epsilon''}$ (left), $\Gamma_{\sigma}$ and $\Gamma_{\sigma'}$ (right) obtained via TCSA for 
$\xi_{\sigma}=0.1$, $\lambda_{\epsilon}>0$ at $n_{c}=11$.  }
\label{gammas_pic2}
\end{figure}
\end{center}

To compare (\ref{leading_act}) with TCSA data we first look at the lowest weight vacuum vector ratios 
\be \label{gamma_i}
\Gamma_{i} = \frac{\langle i|0\rangle_{\lambda}}{\langle 0|0\rangle_{\lambda}}
\ee
where $|0\rangle_{\lambda}$ stands for the vacuum of perturbed theory (\ref{TIFT}) and $|i\rangle$ are the primary states in the TIM CFT. 
On Figures \ref{gammas_pic1} -  \ref{gammas_pic2} we see the plots for these ratios for $\xi_{\sigma}=0.1$. 
We see that the thermal sector ratios all asymptotically approach\footnote{This approach is the slower the higher is the weight of the 
primary as in Cardy's ansatz.}  constant values monotonically growing in magnitude.  The magnetic ratios: $\Gamma_{\sigma}$ and $\Gamma_{\sigma'}$ 
behave differently: while initially gaining some magnitude, at large $r$ they approach zero. This is consistent with an ansaz like (\ref{irrel_ansatz}) for which at large $r$ we get the conformal boundary $|0\rangle\!\rangle$.   

\begin{center}
\begin{figure}[H]
\begin{minipage}[b]{0.5\linewidth}
\centering
\includegraphics[scale=0.82]{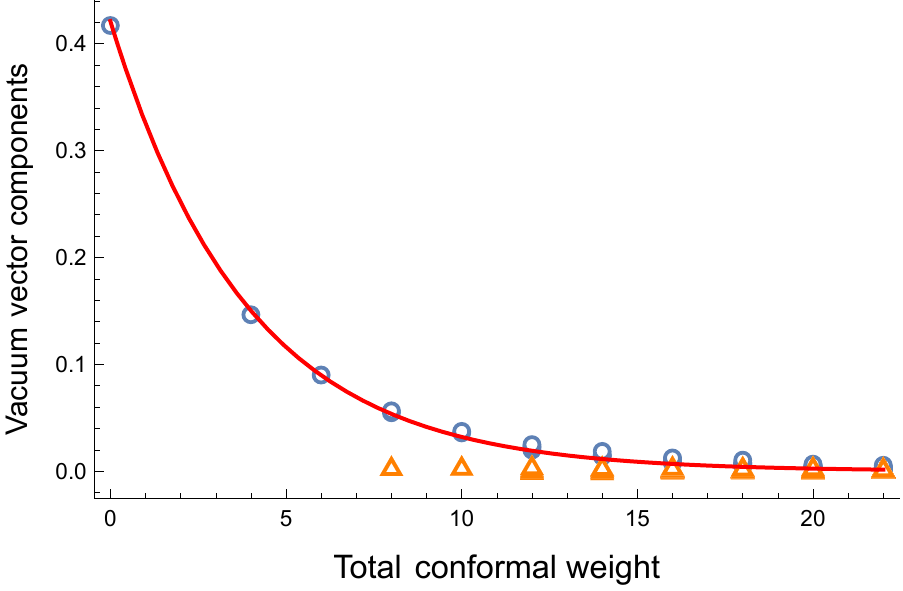}
\end{minipage}%
\begin{minipage}[b]{0.5\linewidth}
\centering
\includegraphics[scale=0.82]{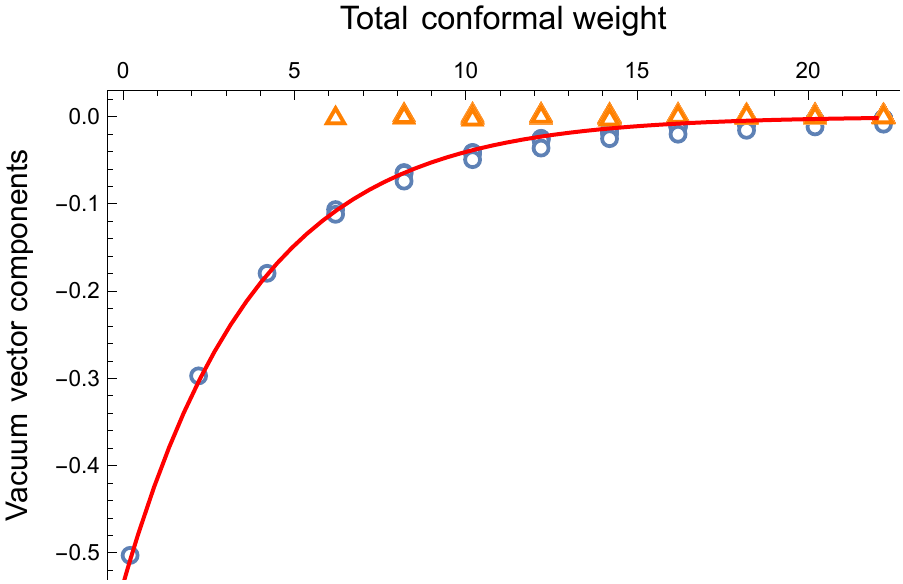}
\end{minipage}
\caption{The TCSA vacuum vector components $C_{i}$ against the total conformal weight of the basis vectors taken for the TIM with $\xi_{\sigma}=0.1$, $\lambda_{\epsilon}>0$ and at  $r=40$. The left plot 
represents the vacuum sector and the right plot -- the $\epsilon$-sector. The red line coresponds to the distribution in the Cardy's ansatz built on $|0\rangle\!\rangle$  with 
$\tau^{*}m=1.64$ found by fitting  from TCSA data with $n_c=11$. The orange triangles mark the non-diagonal components.}
\label{spec_pic1}
\end{figure}
\end{center}

\begin{center}
\begin{figure}[H]
\begin{minipage}[b]{0.5\linewidth}
\centering
\includegraphics[scale=0.82]{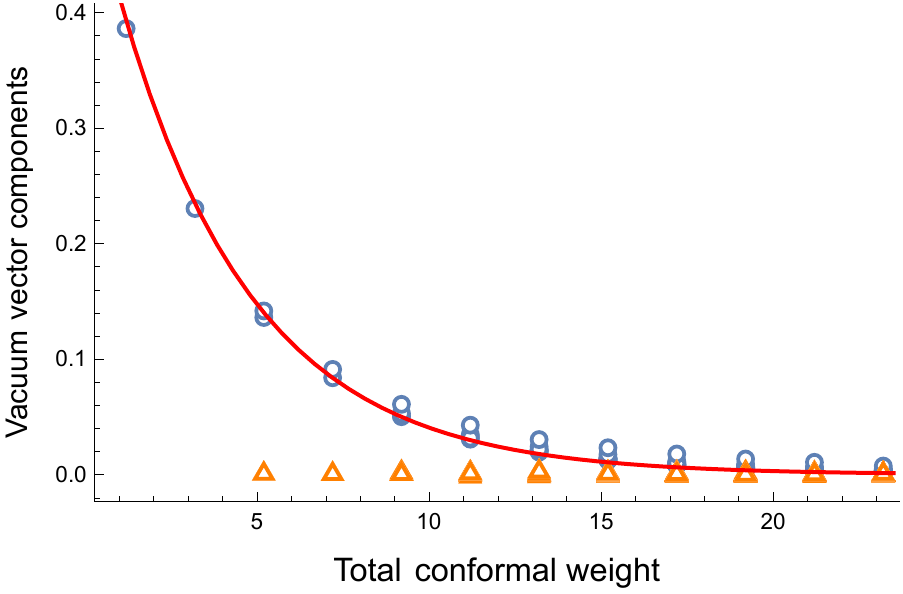}
\end{minipage}%
\begin{minipage}[b]{0.5\linewidth}
\centering
\includegraphics[scale=0.82]{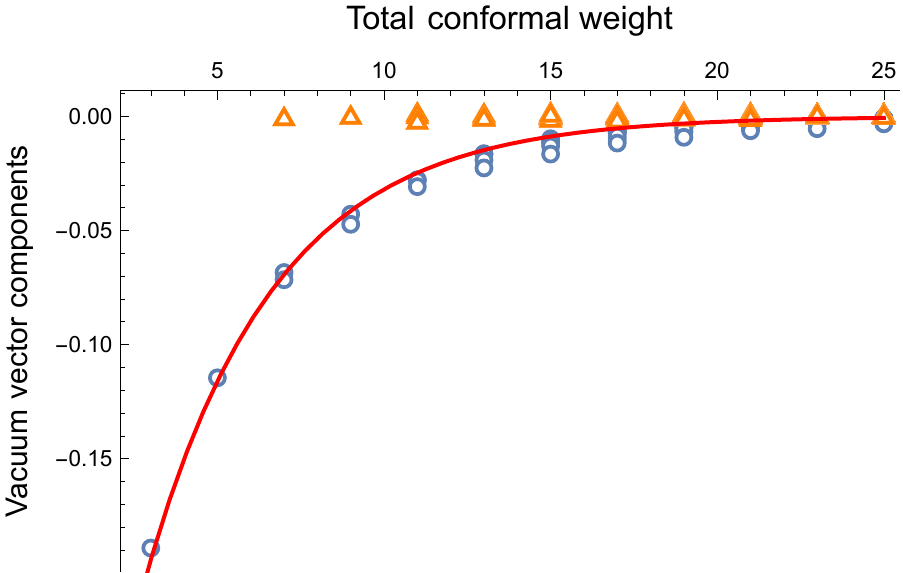}
\end{minipage}
\caption{The TCSA vacuum vector components in the  $\epsilon'$-sector (left) and  the $\epsilon''$-sector (right). See the caption to Fig. \ref{spec_pic1} for 
more detail.}

\label{spec_pic2}
\end{figure}
\end{center}

\begin{center}
\begin{figure}[H]
\begin{minipage}[b]{0.5\linewidth}
\centering
\includegraphics[scale=0.82]{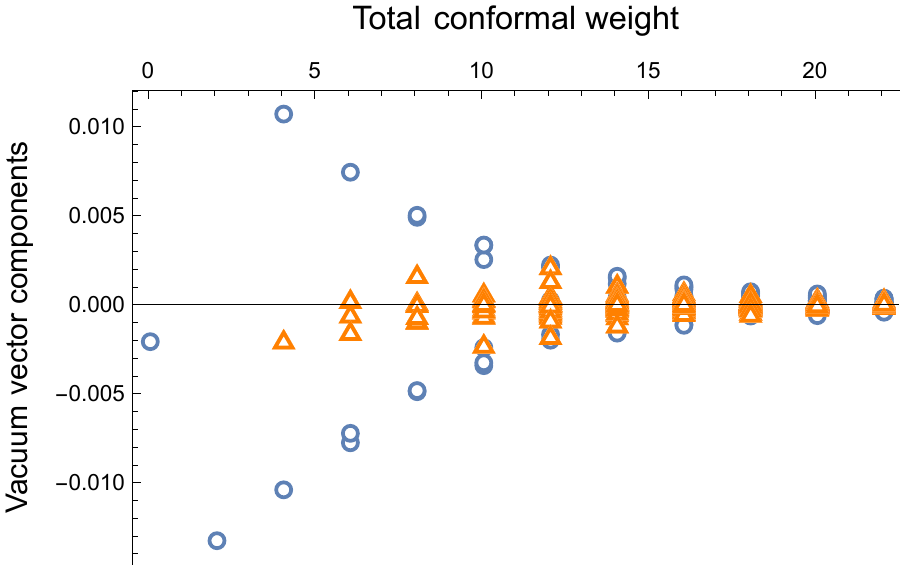}
\end{minipage}%
\begin{minipage}[b]{0.5\linewidth}
\centering
\includegraphics[scale=0.82]{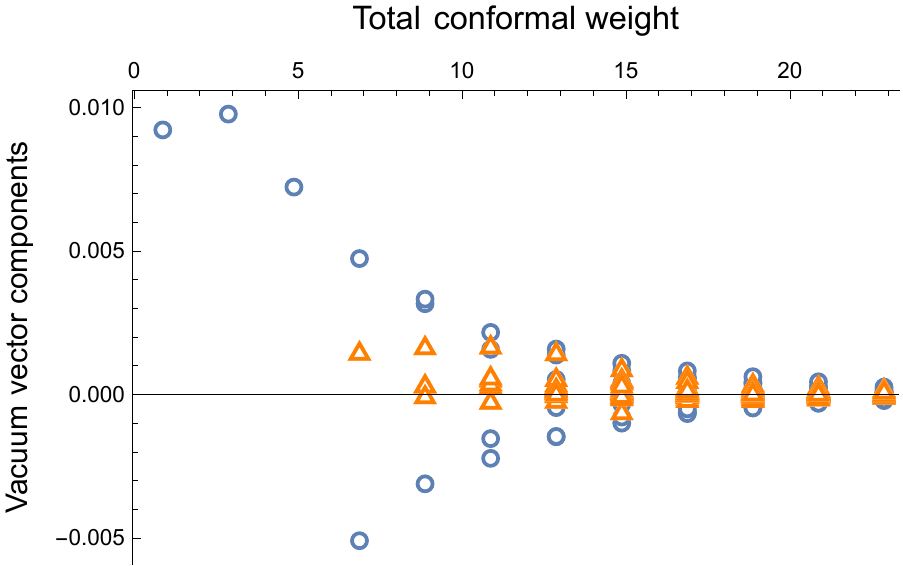}
\end{minipage}
\caption{The TCSA vacuum vector components in the  $\sigma$-sector (left) and  the $\sigma'$-sector (right). See the caption to Fig. \ref{spec_pic1} for 
more detail.}

\label{spec_pic3}
\end{figure}
\end{center}

We get a more detailed information from the weight spectrum of the components of the vacuum vector. We present 
this spectrum on Figures \ref{spec_pic1} -- \ref{spec_pic3} for the TCSA vacuum vector taken for $\xi_{\sigma}=0.1$ and\footnote{The value of $r$ is chosen close to the end of the physical  window.}  $r=40$. On the plots in the thermal sectors we show the distribution corresponding to 
$e^{-\tau_{\rm fit}H_{0}}|0\rangle\!\rangle$ with $\tau_{\rm fit}m=1.64$ 
found by fitting the lowest weight components in the vacuum sector of TCSA data. 
This value is different from $\tau^{*}m\approx 1.484...$  
found by minimising $e_{0}$. The $\sigma$ and $\sigma'$ sectors reveal a more complicated structure which qualitatively is similar to  the one that corresponds to (\ref{leading_act}). The lowest components we worked out have the same 
relative signs as the ones obtained via TCSA. We also observe that while most of the diagonal components fall onto some curve with decreasing magnitude,  the lowest component in each sector curiously seems to have a position with a smaller magnitude away from those curves\footnote{We checked that this feature percists for a range of $r$ and $\xi_{\sigma}$.}. It is tempting to connect this to the small values of the 
coefficients at level zero that we found in (\ref{calc1}). 

While (\ref{leading_act}) was conjectured to hold at the leading order only in the magnetic sectors, looking at the TCSA  data we can come up with a 
tentative  ansatz for the whole vacuum vector:
\be \label{ansatz2} 
|0\rangle_{\lambda}= {\cal N}( e^{-\tau H_{0}} |0\rangle\!\rangle + \lambda_{b}  e^{-\tau H_{0}} \hat L_{G} (\sqrt{\varphi}|\sigma\rangle\!\rangle + |\sigma'\rangle\!\rangle) 
\ee
where $\tau$ and $\lambda_{b}$ are free variables and ${\cal N}$ is a normalisation constant. Given the value of $\tau_{\rm fit}$ 
and our calculations (\ref{calc1}), (\ref{calc2}), (\ref{calc3}) the ansatz (\ref{ansatz2}) predicts certain ratios among the lowest components in the $\sigma$ and $\sigma'$ sectors: 
\be \label{gamma1}
\gamma_{1} \equiv \frac{\lVert L_{-1}\bar L_{-1}|\sigma\rangle\rVert \, \langle \sigma | 0\rangle_{\lambda}}{\langle \sigma |L_{1}\bar L_{1}|0\rangle_{\lambda} } = 
\frac{1}{20} e^{4\pi \frac{\tau m}{r}} \, , 
\ee 
\be \label{gamma2} 
\gamma_{2} \equiv \frac{\lVert L_{-1}\bar L_{-1}|\sigma'\rangle\rVert \, \langle \sigma' | 0\rangle_{\lambda}}{\langle \sigma' |L_{1}\bar L_{1}|0\rangle_{\lambda} } = 
\frac{49}{91} e^{4\pi \frac{\tau m}{r}} \, , 
\ee
\be \label{gamma3} 
\gamma_{3} \equiv \frac{\lVert L_{-1}^2\bar L_{-1}^2|\sigma'\rangle\rVert \, \langle \sigma' | 0\rangle_{\lambda}}{\langle \sigma' |L_{1}^2 \bar L_{1}^2|0\rangle_{\lambda} } = 
\frac{49}{119} e^{8\pi \frac{\tau m}{r}} \, .
\ee
In Table \ref{gamma:table} we compare the values measured directly via TCSA for $\xi_{\sigma}=0.1$, $r=40$, $n_{c}=11$ and found using the right hand sides of (\ref{gamma1}), (\ref{gamma2}), (\ref{gamma3}) 
with $\tau m=\tau_{\rm fit} m = 1.64$. 
We refer to the latter as theoretical values. 

\begin{center}
\begin{table}[H]
\centering
\begin{tabular}{|c|c|c|}
\hline 
Component Ratio& TCSA value & Theoretical value \\
\hline 

$\gamma_{1}$ & 0.1569...& 0.0837... \\
\hline
$\gamma_{2}$ & 0.9434...& 0.9013...\\
\hline
$\gamma_{3}$ &1.2756... & 1.1538...\\
\hline 
\end{tabular}
\caption{The TCSA values of component ratios $\gamma_{i}$ against the theoretical values predicted by (\ref{ansatz2}) }
\label{gamma:table}
\end{table}
\end{center}
We see that while the theoretical values  of  $\gamma_{2}$, $\gamma_{3}$ are roughly  in the ballpark of the numerical values, 
the theoretical value of $\gamma_{1}$ is about twice smaller than the TCSA value. It would be nice to understand how our conjectured
ansatz  (\ref{ansatz2}) can be improved to describe the magnetic sector quantitatively better.

\subsection{Phase transition in the $\lambda_{\epsilon}>0$, $\lambda_{\sigma} \ne 0$  region} \label{transition_subsec} 

As we discussed at the beginning of section \ref{small_sigma_sec} when we increase  $\xi_{\sigma}$ from small values Cardy's ansatz predicts that 
the variational energies $e_{+0}$ and $e_{+}$ go down so that the ansatz   states based on $|+0\rangle\!\rangle $ and $|+\rangle\!\rangle$ conceivably can 
take over from the deformed $|0\rangle\!\rangle$. On the other hand the three boundary states at hand are connected by 
the boundary RG flows triggered by $\psi_{\epsilon'}^{(+0)}$   \cite{Chim}, \cite{Affleck}. These flows can be represented by 
the following diagram 
\be \label{bd_flows}
(0) \longleftarrow (+0) \longrightarrow (+)
\ee
They are induced by inserting on the boundary the exponential 
\be
\exp\left( -\alpha_{\epsilon'} \int\limits_{0}^{R}  \psi_{\epsilon'}^{(+0)}(y) dy \right) \, .
\ee
When the boundary coupling  constant $\alpha_{\epsilon'}$ is negative the flow ends up in $(0)$ while when it is positive it ends up in $(+)$. 
It is known \cite{RRS,LSS,GrahamW,KM} 
that the leading irrelevant boundary operator along which these flows arrive to the IR fixed points is the stress-energy tensor $T$ for the flow to $(+)$ and $\psi_{\epsilon''}^{(0)}$ for the flow to $(0)$. Thus, the irrelevant boundary perturbation  (\ref{irrel_ansatz}) we adopted to describe the vacuum state for small $\xi_{\sigma}$ corresponds to moving against the  left arrow  in (\ref{bd_flows}) that is towards the $(+0)$ boundary condition. 

Both of the above observations suggest that at some critical value $\xi_{\sigma}=\xi_{\rm c}$ the vacuum will be described by a smeared 
$|\!+\!0\rangle\!\rangle$ boundary condition. Increasing $\xi_{\sigma}$ past the critical value we may expect generating the right flow in (\ref{bd_flows}) 
and getting into the IR phase labelled by $(+)$. We can further substantiate this scenario and find a tentative value for $\xi_{\rm c}$ using our 
general analysis of stability of the variational ansatz in section \ref{Cardy_gen_section}. 

Let us consider the Cardy's variational energy for the smeared $|\!+\!0\rangle\!\rangle$. To analyse the stability of this ansatz against the boundary relevant perturbation $\psi_{\epsilon'}^{(+0)}$  we note that both of our bulk perturbing fields: $\phi_{\epsilon}$ and $\phi_{\sigma}$ couple to $\psi_{\epsilon'}^{(+0)}$ 
via the bulk-boundary OPE coefficients: $ {}^{(+0)}\!B_{\sigma}^{\epsilon'}$,  ${}^{(+0)}\!B_{\epsilon}^{\epsilon'}$. These coefficients are calculated in Appendix  \ref{appendix_OPE}, where the reader can also find more details on our conventions,  and are given by formulae (\ref{btb1}), (\ref{btb2}). The two coefficients have the opposite signs so that the $\phi_{\epsilon}$ perturbation with $\lambda_{\epsilon}>0$ drives the system (by lowering the energy) towards 
the $(0)$-phase while $\phi_{\sigma}$ drives the system in the opposite direction towards the $(+)$-phase. Both effects cancel each other at the leading order when 
 condition (\ref{equilibrium_gen}) is satisfied. Using the scaling variables (\ref{xi_def}), (\ref{tt_def}) and assuming $\lambda_{\epsilon}>0$ this condition can be written as 
\be \label{equilibrium_cond1}
\lambda_{\epsilon}^{10/9} \Bigl(  \xi_{\sigma} \frac{ {}^{(+0)}\!B_{\sigma}^{\epsilon'} }{(\tilde \tau^{*})^{3/40}} + \frac{{}^{(+0)}\!B_{\epsilon}^{\epsilon'}}{(\tilde \tau^{*})^{1/5}}  \Bigr) = 0  
\ee
where $\tilde \tau^{*} =  \tilde \tau^{*}(\xi_{\sigma})$ is the value of $\tilde \tau$ that minimises the variational energy  $e_{+0}$ given in (\ref{var_e}).
Using (\ref{B_ratio}) the last equation can be written as 
\be \label{equilibrium_cond2}
 \tilde \tau^{*}(\xi_{\sigma}) = \frac{2^{6}}{(\xi_{\sigma})^{8}} \, .
\ee
This is an equation on $\xi_{\sigma}$. Finding the function $ \tilde \tau^{*}(\xi_{\sigma})$ numerically we obtain that (\ref{equilibrium_cond2}) 
is satisfied for 
\be \label{crit_value}
\xi_{\sigma} = \xi_{\rm c}^{\rm (3)} \equiv 1.78138...
\ee
We will present now some TCSA numerical evidence that indicates the presence of a transition between different RG boundaries at a value of $\xi_{\sigma}$ close to $\xi_{\rm c}^{\rm (3)}$.

\begin{center}
\begin{figure}[H]
\centering
\includegraphics[scale=1]{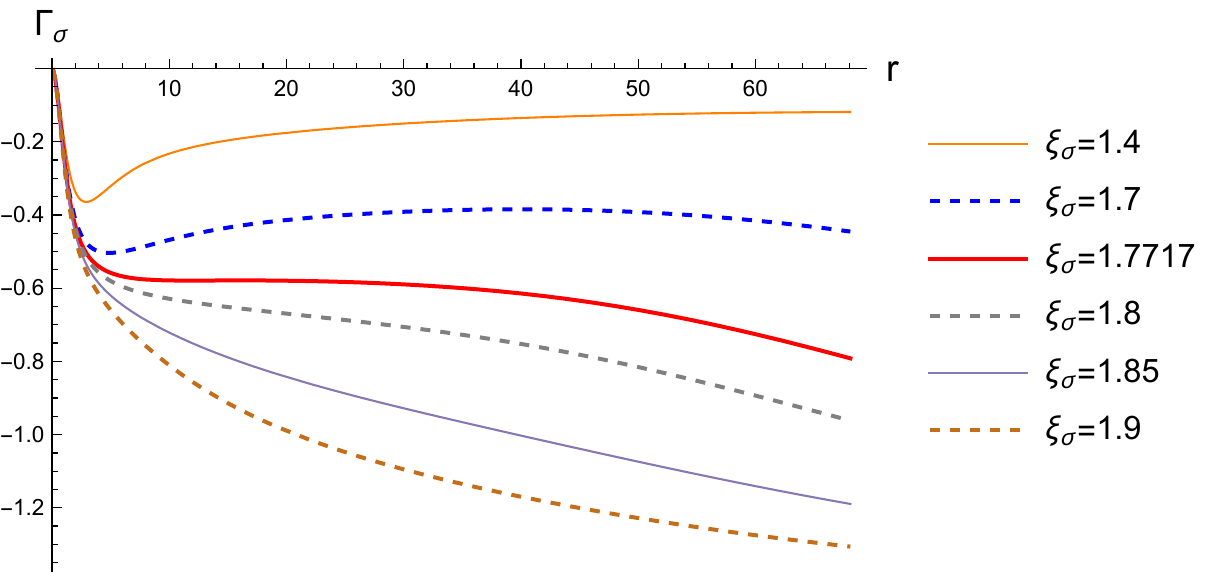}
\caption{The component ratio $\Gamma_{\sigma}$ for the perturbed TIM with $\lambda_{\epsilon}>0$ and at different values of $\xi_{\sigma}$, obtained using TCSA with $n_{c}=11$.  }
\label{fig_gammas_crit}
\end{figure}
\end{center}

We will start by looking at  the component ratios $\Gamma_{i}$ (\ref{gamma_i}) which serve as a good rough probe into the behaviour  of the system. 
The finite size effects are the higher the higher is the weight of the primary $i$ so that $\Gamma_{\sigma}$ should approach its infinite volume value faster than the rest of the ratios. This is particularly important in TCSA calculations where the value of $R$ should't be too high to ensure the smallness of the truncation errors (see Appendix  \ref{appendix_TCSA} for more technical details on TCSA calculations). Luckily $\Gamma_{\sigma}$ is also 
the ratio that demonstrates the most interesting behaviour in the vicinity of $\xi_{\rm c}^{\rm (3)}$. We find that at a certain critical value 
$\xi_{\sigma}= \xi_{c}^{\rm T} \approx 1.77$ 
the curve for  $\Gamma_{\sigma}(r)$ flattens. For $\xi_{\sigma} <  \xi_{c}^{\rm T}$ the ratio after growing in magnitude goes into a decaying regime 
as was previously observed for small $\xi_{\sigma}$, see Figure \ref{gammas_pic2}. For $\xi_{\sigma} > \xi_{c}^{\rm T}$ the ratio monotonically increases in magnitude. On Figure \ref{fig_gammas_crit} we present a sample of plots illustrating the transition. 

The precise value of $ \xi_{c}^{\rm T}$ depends on the truncation parameter $n_{c}$. The values for four different truncation parameters 
are shown in Table \ref{xicrit:table}. We see that the values increase with the dimension of the truncated space getting closer to the value of $\xi_{\rm c}^{\rm (3)}$. 

\begin{center}
\begin{table}[H]
\centering
\begin{tabular}{|c|c|}
\hline 
\rule{0pt}{3ex} \rule{-3ex}{0pt}
$n_{c}$ &$ \xi_{c}^{\rm T}$   \rule{0pt}{3ex} \rule{-3ex}{0pt}   \\

\hline

9 & 1.7710... \\
\hline
10 & 1.7714...\\
\hline
11& 1.7717...\\
\hline
12 & 1.7719...\\
\hline 
\end{tabular}
\caption{The TCSA values of  $ \xi_{c}^{\rm T}$ at which $\Gamma_{\sigma}$ has a flat interval at different values of truncation parameter $n_{c}$.}
\label{xicrit:table}
\end{table}
\end{center}

We note that the value of $\Gamma_{\sigma}$ 
on the flat interval at $\xi_{\sigma}= \xi_{c}^{\rm T}$ is $\Gamma_{\sigma} = -0.579...$ for $n_{c}=11$. This is close to the value of $\Gamma_{\sigma}$ for 
the $(+0)$ boundary condition that equals $-2^{1/4}\varphi^{-3/2}=-0.577...$. 

For the other component ratios at $\xi_{\sigma}= \xi_{c}^{\rm T}$  we observe a variety of behaviour: $\Gamma_{\epsilon}$, $\Gamma_{\sigma'}$
 $\Gamma_{\epsilon''}$ each  have a local minimum or maximum where the values are of the same sign as the corresponding ratios in  $(+0)$ 
 but are quite different in magnitude (see  Figures \ref{gammas_appendixpic1}, \ref{gammas_appendixpic2} in Appendix \ref{xicrit_appendix}). The ratio $\Gamma_{\epsilon'}$ becomes negative and continues to asymptotically decrease reaching a local 
 minimum at very large values of $r$. Certainly this does not look like the vacuum state has reached the asymptotic regime in which 
 the component ratios are  close to the conformal ones. 
 
 Looking next at the weight distribution of the vacuum vector components 
 we find that when $r$ increases from the onset of the flat $\Gamma_{\sigma}$ region at $r\approx 10$ all components either have the same sign\footnote{We note that in the TIM CFT the signs of the coefficients at Ishibashi states determine the conformal boundary states unambiguously. Thus to identify a TCSA vacuum with a smeared conformal boundary state it seems to be a good strategy to first find the region where all components in each primary representation have the same sign and then check that they follow a smeared profile.  } as 
 in $(+0)$ (as is the case in sectors: $1$, $\epsilon$, $\epsilon''$, $\sigma'$) or are moving in that direction (in sectors $\epsilon'$ and $\sigma$). The decrease of components with weight becomes more and more exponential-like. However past some point 
 at about $r\approx 30$ while some sectors continue to improve  others start moving away from $(+0)$. We should note that the physical window where we have confidence in TCSA results also ends at about $r=50$. We remark that  it is known that TCSA truncation errors are in general sector dependent \cite{GW}. 
 To summarise, we find the vacuum state to be close to a smeared $|\!+\!0\rangle\!\rangle$ state but the fit is not as impressive as the ones we previously found in other ranges of parameters e.g. the fits presented on Figures  \ref{fig_TIM_spec1} - \ref{fig_TIM_spec3}.  A curious reader can see the weight distribution for $r=40$ on Figures \ref{fig_TIM_critspec1} - \ref{fig_TIM_critspec3} in Appendix \ref{xicrit_appendix}.  
 
 Of course the component  ratio $\Gamma_{\sigma}$ is not a physical quantity so one can use the flattening observed above only as a rough indicator of 
 a possible transition. To determine whether some physical transition actually takes place and to analyse its nature we turn to the vacuum expectation values of the scaling fields. Following  \cite{GM1} we will use TCSA to calculate the one-point functions on the plane: 
  $\langle \phi_{i}\rangle_{\rm pl}$ for the perturbed theory (\ref{TIFT}). We have 
  \be
  \langle \phi_{i}\rangle_{\rm pl} = m^{\Delta_{i}} f_{i}(\xi_{\sigma})
  \ee
 where the dimensionless scaling functions $f_{i}$ can be calculated as 
 \be
  f_{i}(\xi_{\sigma})=  \lim_{r\to \infty} \left( \frac{2\pi}{r}\right)^{\Delta_{i}} {}_{\lambda}\langle 0|\phi_{i}^{\rm pl} |0\rangle_{\lambda} 
 \ee
 where $\phi_{i}^{\rm pl}$ is the conformal field on the plane and $ |0\rangle_{\lambda} $ is the vacuum of the perturbed theory on the cylinder.
 
 In the region of large $\xi_{\sigma}$ it is interesting to compare the TCSA one-point functions with those predicted by Cardy's ansatz. 
 For the ansatz based on a boundary condition $|a\rangle\!\rangle$ the predicted scaled one-point functions  are 
 \be
 f_{i}^{a} = A_{i}^{a} \frac{\kappa_{\epsilon}^{5\Delta_{i}/9}}{(\tilde \tau^{*}_{a})^{\Delta_{i}}}
 \ee  
where $A_{i}^{a}$ are given in (\ref{ai_1pt}) and $\tilde \tau^{*}_{a}$ gives the minimum of the variational energy $e_{a}(\tilde \tau)$.

\begin{center}
\begin{figure}[H]
\begin{minipage}[b]{0.5\linewidth}
\centering
\includegraphics[scale=0.82]{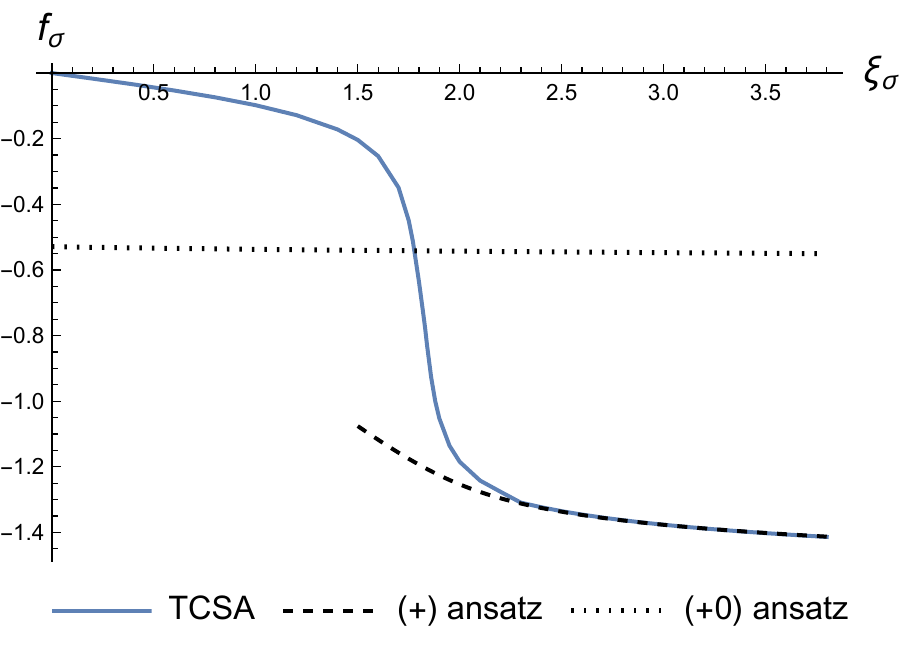}
\end{minipage}%
\begin{minipage}[b]{0.5\linewidth}
\centering
\includegraphics[scale=0.82]{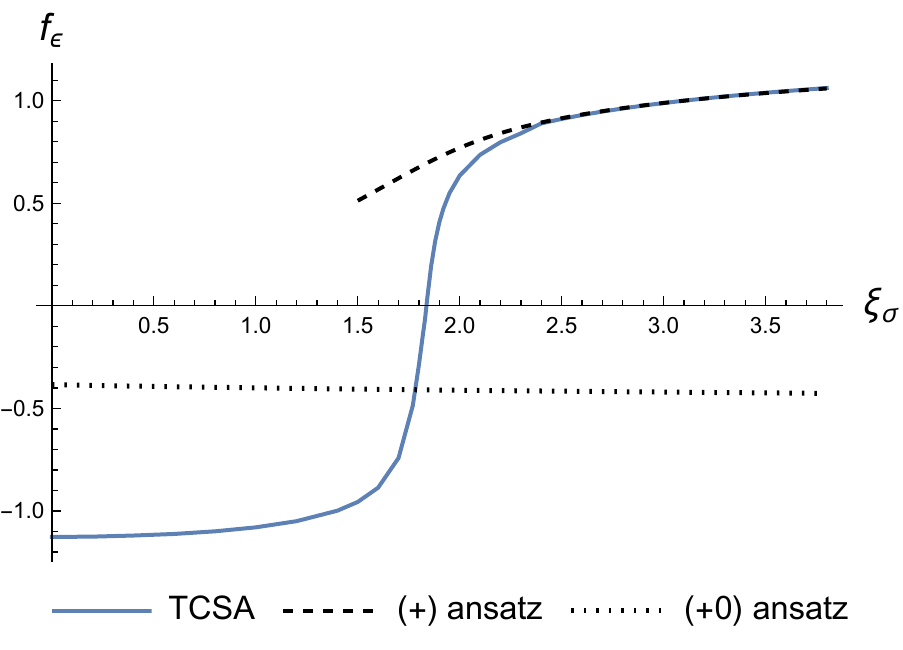} 
\end{minipage}
\caption{ Scaled vacuum expectation values on the plane of $\phi_{\sigma}$ (left) and $\phi_{\epsilon}$ (right) plotted against $\xi_{\sigma}$. 
The solid blue line is from TCSA with $n_{c}=11$, the dashed  line is from Cardy's ansatz for the $(+)$ boundary condition and the dotted line is 
from Cardy's ansatz for  the $(+0)$ boundary condition.}
\label{vev_pic1}
\end{figure}
\end{center}

\begin{center}
\begin{figure}[H]
\begin{minipage}[b]{0.5\linewidth}
\centering
\includegraphics[scale=0.82]{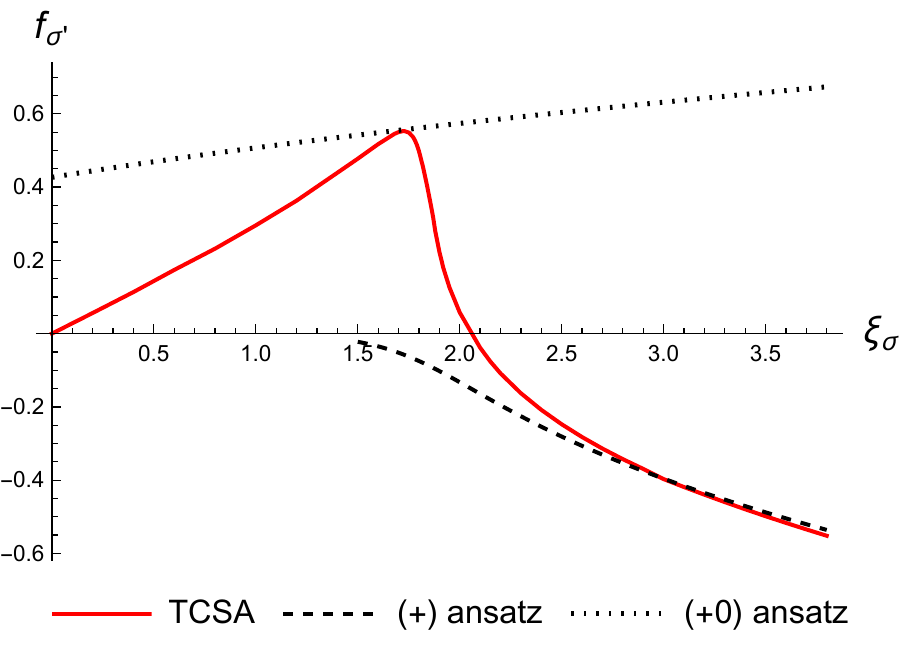}
\end{minipage}%
\begin{minipage}[b]{0.5\linewidth}
\centering
\includegraphics[scale=0.82]{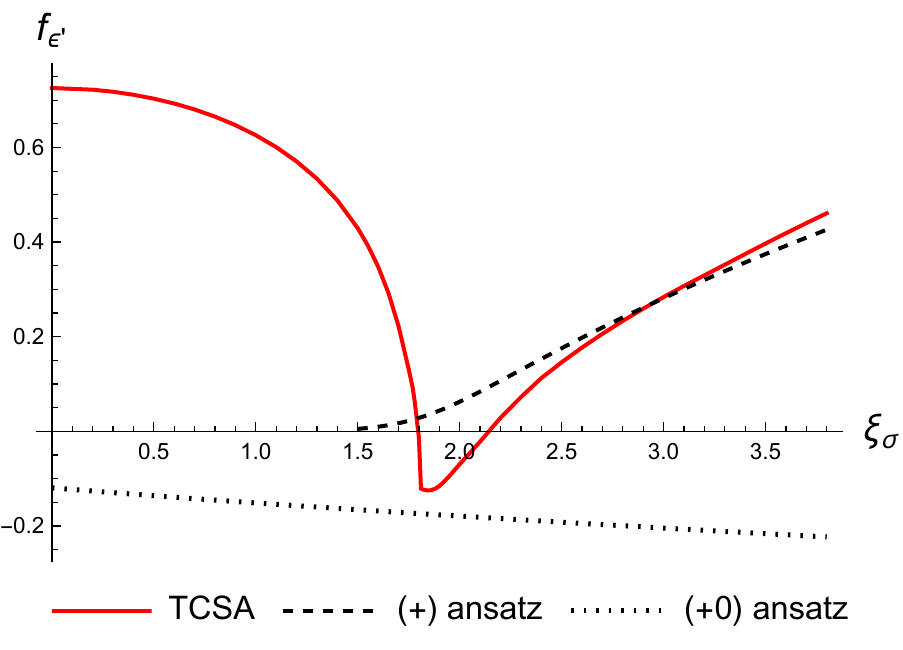} 
\end{minipage}
\caption{ Scaled vacuum expectation values on the plane of $\phi_{\sigma'}$ (left) and $\phi_{\epsilon'}$ (right) plotted against $\xi_{\sigma}$. 
The solid line is from TCSA at $n_{c}=11$, the dashed  line is from Cardy's ansatz for the $(+)$ boundary condition and the dotted line is 
from Cardy's ansatz for  the $(+0)$ boundary condition.}
\label{vev_pic2}
\end{figure}
\end{center}

On Figures \ref{vev_pic1} and \ref{vev_pic2} we present the scaled expectation values of $\phi_{\sigma}$, $\phi_{\epsilon}$ , 
$\phi_{\sigma'}$, and $\phi_{\epsilon'}$ obtained using TCSA.

The details of the plots on    Figures  \ref{vev_pic1} and \ref{vev_pic2} are as follows. The  dashed lines give portions of plots of $f_{i}^{+}$ and we see that for large enough $\xi_{\sigma}$ 
they approximate the TCSA data reasonably well with some visible discrepancies for $f_{\sigma'}$ and $f_{\epsilon'}$. 
The dotted lines give the values of  $f_{i}^{+0}$ which in the range of $\xi_{\sigma}$ chosen  vary little. 
The functions $f_{\sigma}^{+0}$, $f_{\epsilon}^{+0}$ cross the TCSA plots for $f_{\sigma}$ and $f_{\epsilon}$ at $\xi_{\sigma} \approx 1.78$. 
For $f_{\sigma'}$ the plot of  $f_{\sigma'}^{+0}$ touches the TCSA plot at the tip of the smoothed-out $\lambda$-like singularity that appears at approximately 
$\xi_{\sigma}=1.72$. For the TCSA plot of $f_{\epsilon'}$ we observe that it has what appears to be a smoothed-out singularity at about $\xi_{\sigma}=1.8$. Close to that point the left branch 
of the function dips down and gets close to the graph of $f_{\epsilon'}^{+0}$ before it goes into the increasing right branch.  
It is empirically known that the truncation errors for the one-point functions grow with  the scaling dimension of the operator.  
Among our TCSA calculations the   $f_{\epsilon'}$ data is then most susceptible to truncation errors. 
It is possible that in the continuum limit  at the transition point the $f_{\epsilon'}$ plot touches the $f_{\epsilon'}^{+0}$ function similar to 
how that happens for the other three plots.  We give a detailed discussion of how the one-point functions were calculated in Appendix \ref{appendix_TCSA}.

We see from the above plots  that there is undoubtedly some   transition that takes place near 
$\xi_{\sigma}=1.78$ as was predicted by our theoretical calculation  that lead to (\ref{crit_value}). The $f_{\sigma}$ and $f_{\epsilon}$ profiles 
look like a smoothed-out step function, or a kink,  that resembles a smoothed-out first order transition. 
Moreover, we find that the mass gap in the region of the transition  is small  but finite (see Figure \ref{b_pic}) so it does not look like we are dealing with a thermodynamic phase  transition here. Rather it looks like the $\xi_{\sigma}=\xi_{c}$ line is similar to the disorder line in the Ising model. We also found such lines (or rather a whole surface of RG trajectories) in TIM using Landau theory by imposing the condition (\ref{third_der}) of vanishing third order derivative of Landau free energy with respect to the order parameter. Near such a surface the magnetisation profile looks like a kink (see Figure \ref{Landau_kinks}) that is similar to the magnetisation plot on Figure \ref{vev_pic1}. To use the terminology we introduced in section \ref{Landau_sec}, moving across the 
$\xi_{\sigma}=\xi_{c}$ line we move from the weakly magnetised phase to the strongly magnetised one. 
In this paper we refer to this transition as a ``kink transition''. 
Although our phase labels (RG boundaries) change when going across the transition line we avoid calling it a "phase transition" (except for in the title of this subsection) not to confuse it with thermodynamic phase transitions.
We will see in the next subsection that the analogy with  Landau theory surface (\ref{kink_Landau}) is even closer as we will find a whole surface 
of kink transitions  which has the same features as the one we analysed in section \ref{Landau_sec}.

We note that the precise value of  $\xi_{\rm c}$  needs to be better determined. A working definition for it could be the position of the minimum of 
$f_{\epsilon'}$ or of the maximum $f_{\sigma'}$. Unfortunately both of those quantities are subject  to  corrections that shows in the difference of the numerical 
values we obtained for them: $1.72$ and $1.8$. 

 For $\xi_{\sigma}>\xi_{\rm c}$  in addition to comparing the TCSA one-point functions to the Cardy's ansatz based on $(+)$ we 
 also looked at the conformal weight spectrum of the vacuum vector. Starting from $\xi_{\sigma}\approx 2.2 $ we find a  good fit 
 with the $(+)$-ansatz with plots like the ones shown on Figures \ref{fig_TIM_spec1} -  \ref{fig_TIM_spec3}. 
 The closer we are to the critical value $\xi_{\sigma}\approx 1.78$ the harder it becomes to find a physical window and 
 the larger are deviations from Cardy's ansatz.

\subsection{Switching on $\phi_{\sigma'}$ and symmetry breaking critical lines} \label{sigmaprime_section} 

\begin{center}
\begin{figure}[H]
\centering
\includegraphics[scale=0.8]{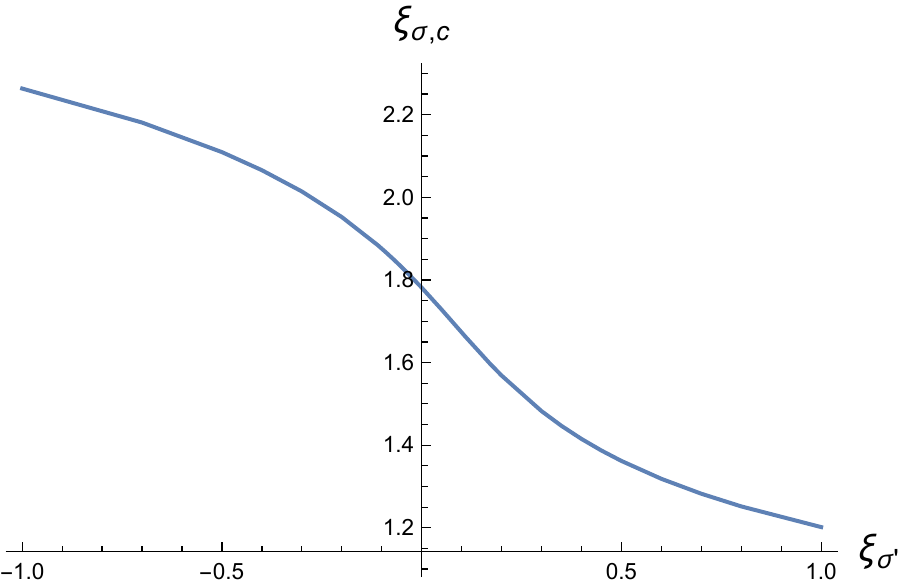}
\caption{The curve $ \xi_{\sigma} = \xi_{\sigma,c}(\xi_{\sigma'})$ that specifies a
surface  of kink phase transitions in the $\lambda_{\sigma}$, $\lambda_{\epsilon}$, $\lambda_{\sigma'}$ space.
   }
\label{fig_xixi}
\end{figure}
\end{center}

The tentative location (\ref{crit_value})  of the kink transition discussed in the previous section was obtained by requiring that the first order 
perturbation of  $e_{+0}$  in the boundary coupling vanishes. This happens due to a mutual cancellation 
of the contributions from $\phi_{\epsilon}$ and $\phi_{\sigma}$.   The subleading magnetisation field $\phi_{\sigma'}$ has vanishing bulk-boundary 
coefficient: ${}^{(+0)}\!B_{\sigma'}^{\epsilon'} =0$, and thus does not couple to the relevant boundary field.  This means that the transition line we 
observed on the $(\lambda_{\sigma}, \lambda_{\epsilon})$- plane extends to a surface in the three dimensional space of triple perturbations 
parameterised by  $\lambda_{\sigma}$, $\lambda_{\epsilon}$, $\lambda_{\sigma'}$. To label the RG trajectories in this space we introduce 
in addition to $\xi_{\sigma}$ a new parameter 
\be
\xi_{\sigma'} =  \frac{\lambda_{\sigma'}}{|\lambda_{\epsilon}|^{5/8}} \, .
\ee
The scaled variational energy for the $(+0)$ boundary condition is now given by 
\be
e_{+ 0}(\tilde \tau) = r \Bigl[[\frac{7}{60 \pi \tilde \tau^2} - a \frac{\varphi^{-3/2}}{(\tilde \tau)^{1/5}}  - \xi_{\sigma}  \frac{2^{1/4} \varphi^{-3/2}}{(\tilde \tau)^{3/40}}  
+ \xi_{\sigma'}  \frac{2^{1/4} }{(\tilde \tau)^{7/8}}   \Bigr] \, . 
\ee
Minimising this function numerically we find a function $\tilde \tau^{*}=\tilde \tau^{*}(\xi_{\sigma},\xi_{\sigma'})$. Substituting it 
into the equilibrium condition (\ref{equilibrium_cond2}) and solving for $\xi_{\sigma}$ we obtain a curve 
$\xi_{\sigma} = \xi_{\sigma,c}(\xi_{\sigma'})$ containing  the kink transition points. A portion of this curve is presented on Figure \ref{fig_xixi}. 

Since a pair $(\xi_{\sigma}, \xi_{\sigma'})$ specifies an RG trajectory in the three-coupling space: $\lambda_{\sigma}$, $\lambda_{\epsilon}$, $\lambda_{\sigma'}$, the curve $\xi_{\sigma,c}$ specifies a two dimensional surface in the coupling space. By a slight abuse of notation we also denote this surface as $\xi_{\sigma,c}$.

TCSA numerics shows that as we move along the curve $\xi_{\sigma, c}$ from $\xi_{\sigma'}=0$ in the positive $\xi_{\sigma'}$ direction the mass gap shrinks 
and the system gets into the critical regime when the energy gaps scale as 
\be
E_{i} -E_{0} \sim \frac{1}{R} \, .
\ee
To locate the critical line with precision we use the first scaled gap $r(E_{1}-E_{0})$ and minimise its variation with $r$. At $n_{c}=11$ 
we obtain the critical line at
\be \label{crit_xixi}
\xi_{\sigma}^{*}=1.682... \qquad  \xi_{\sigma'}^{*}= 0.096...
\ee
This is close to the curve  $\xi_{\sigma, c}$ which contains for example the point $(\xi_{\sigma}, \xi_{\sigma'}) = ( 1.677...,0.096) $. 
It is tempting to conjecture that the discrepancy is due to truncation errors and that in the $n_{c}\to \infty$ limit the critical curve actually lies on the 
$\xi_{\sigma, c}$ surface.

\begin{center}
\begin{figure}[H]
\begin{minipage}[b]{0.5\linewidth}
\centering
\includegraphics[scale=0.82]{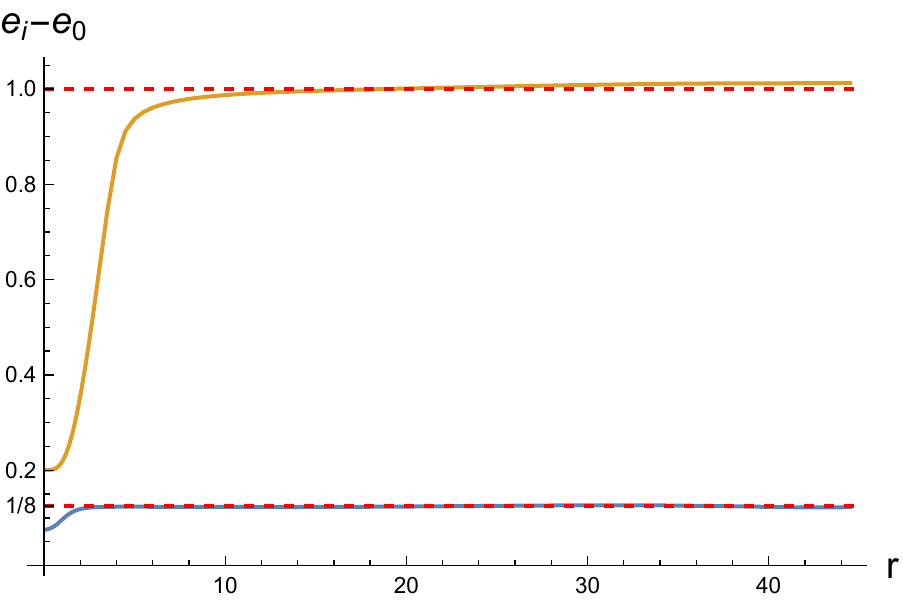}
\end{minipage}%
\begin{minipage}[b]{0.5\linewidth}
\centering
\includegraphics[scale=0.82]{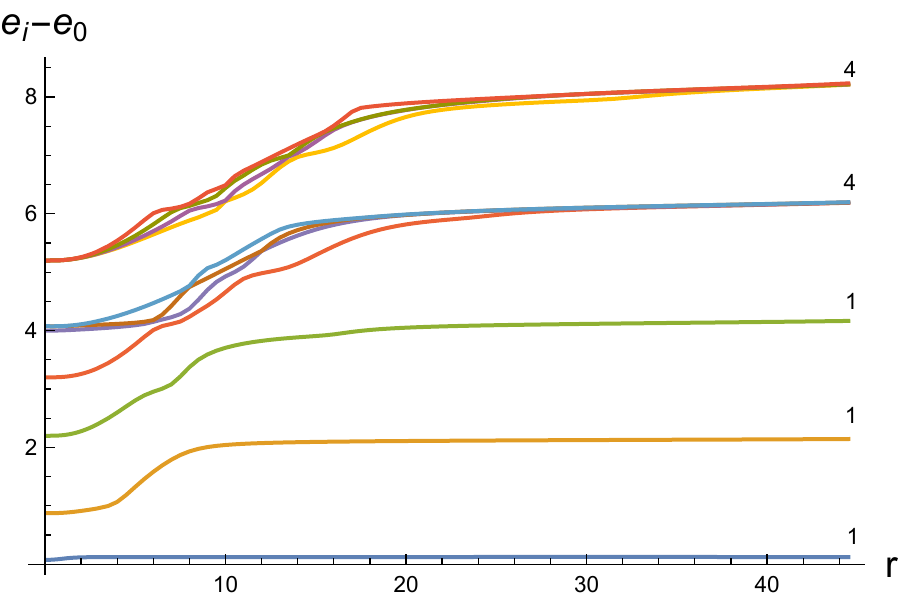}
\end{minipage}
\caption{On the left plot, the lowest two scaled gaps $e_{i}-e_{0}= (E_{i}-E_{0})R/2\pi$ are shown together with  the expected values: $1/8$ and $1$. 
On the right plot, scaled gaps for  the first 11 eigenstates in the $\sigma$-sector of the Ising model. The plots are obtained using TCSA with $n_{c}=11$ for 
$\xi_{\sigma}^{*}, \xi_{\sigma'}^{*}$ given in (\ref{crit_xixi}).}
\label{gap_pic1}
\end{figure}
\end{center}

\begin{center}
\begin{figure}[H]
\begin{minipage}[b]{0.5\linewidth}
\centering
\includegraphics[scale=0.82]{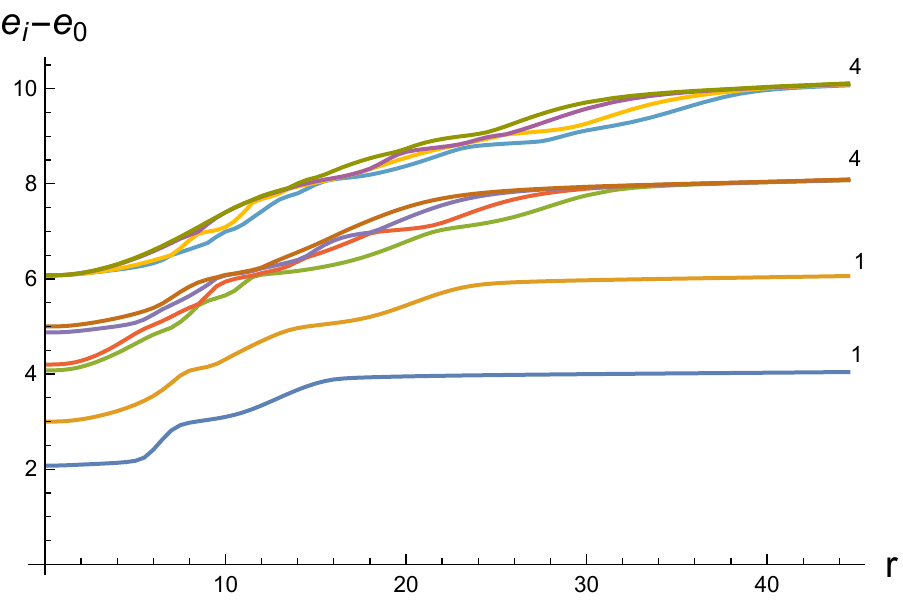}
\end{minipage}%
\begin{minipage}[b]{0.5\linewidth}
\centering
\includegraphics[scale=0.82]{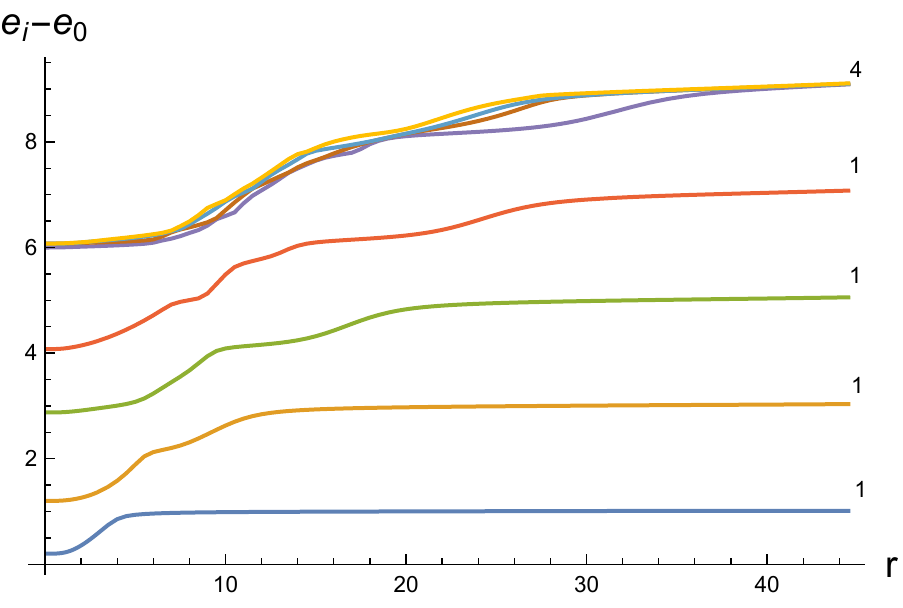}
\end{minipage}
\caption{Scaled gaps $e_{i}-e_{0}= (E_{i}-E_{0})R/2\pi$ for the first 10 states in the vacuum sector (left plot) and for 
the first 8 states in the $\epsilon$-sector of the Ising model (right plot). The plots are obtained using TCSA with $n_{c}=11$ for 
$\xi_{\sigma}^{*}, \xi_{\sigma'}^{*}$ given in (\ref{crit_xixi}).}
\label{gap_pic2}
\end{figure}
\end{center}

The detailed study of scaled energy gaps $e_{i}-e_{0} = (R/2\pi)(E_{i}-E_{0})$ shows that the system approaches the universality class of the 
critical Ising model. On the left plot on Figure \ref{gap_pic1} we show the numerical comparison of the first two scaled gaps with the conformal 
dimensions of the $\sigma$ and $\epsilon$ primary fields in the Ising model. On the right plot on Figure \ref{gap_pic1} and on two plots 
on Figure \ref{gap_pic2} we show the first 30 scaled gaps which we separated manually into the 3 conformal towers. We observe that 
the gaps level out and merge to give the expected multiplicities.   Similar symmetry breaking critical lines were recently identified using 
TCSA for the perturbed TIM 
with imaginary magnetisation couplings \cite{multicrit1}, \cite{multicrit2} as well as  in  other models  
 \cite{FZ}, \cite{Toth_crit},  \cite{Toth_crit2}.

By the spin reversal symmetry there is another symmetry breaking critical line located at $(\xi_{\sigma},\xi_{\sigma'})=(-\xi_{\sigma}^{*},-\xi_{\sigma'}^{*})$. 
Together with the well known symmetry preserving line given by the $\phi_{\epsilon'}$ perturbation with a positive coupling we get three 
critical lines as predicted by the classical theory \cite{LS}. (The existence of three critical lines ending on the UV CFT gives the tricritical Ising model its name.)

\begin{center}
\begin{figure}[H]
\begin{minipage}[b]{0.5\linewidth}
\centering
\includegraphics[scale=0.82]{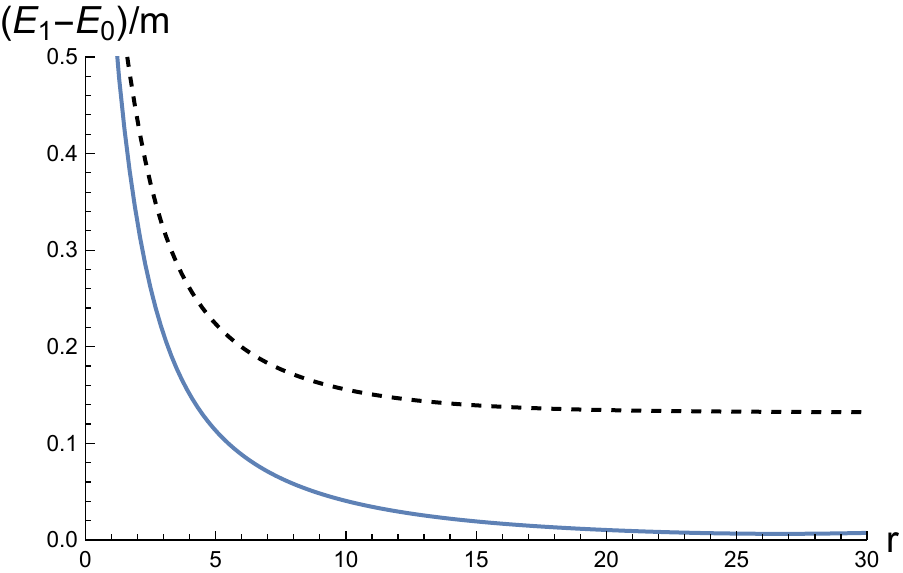}
\end{minipage}%
\begin{minipage}[b]{0.5\linewidth}
\centering
\includegraphics[scale=0.82]{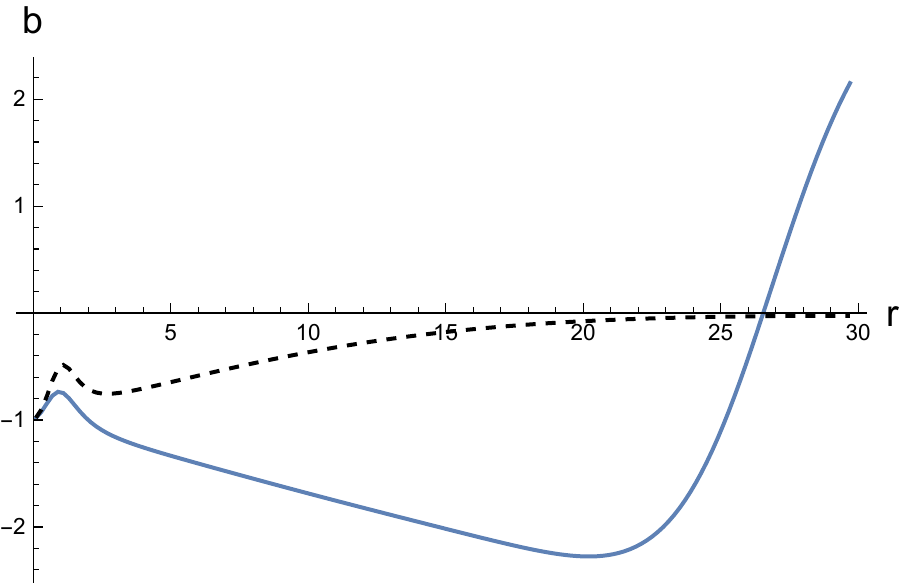}
\end{minipage}
\caption{The first gap $E_{1}-E_{0}$ (left) and its effective scaling exponent $b$ (right). The solid line corresponds to  
$\xi_{\sigma}=1.5983$, $\xi_{\sigma'}=0.17$, which is a point on the curve $\xi_{\sigma,c}$. The dashed line corresponds to  the kink transition studied in section \ref{transition_subsec}: $\xi_{\sigma}=1.78$, $\xi_{\sigma'}=0$.}
\label{b_pic}
\end{figure}
\end{center}

Next we investigate what happens when we move along $\xi_{\sigma,c}$ beyond the critical line. We do find that for the values 
$\xi_{\sigma'}>\xi_{\sigma'}^{*}$ on the surface $\xi_{\sigma,c}$, which are not too large the system develops a doubly degenerate vacuum state. The best way to detect this in TCSA numerics 
is by looking at the first  effective scaling exponent
\be
b=R\frac{d}{dR} \ln (E_{1} - E_{0}) \, .
\ee 
In a massive regime $b$ is close to zero, on a critical line it is close to -1 and when we have vacuum degeneracy the gap decreases exponentially with $R$ so that $b$ is a linear function with a negative slope. On Figure \ref{b_pic} we show the first gap and its effective scaling exponent  for a 
sample point on the curve $\xi_{\sigma,c}$ with $\xi_{\sigma}'>\xi_{\sigma'}^{*}$ that shows an exponentially closing gap in an interval of $r$ which is 
approximately $(3,21)$. For comparison on the same figure we show the same quantities at the kink transition we discussed in the previous subsection. 



 The critical lines then appear on the boundary of the two-phase coexistence region. This is in accord with Landau theory 
 as discussed in section \ref{Landau_sec} where we showed that the two phase coexistence sheets $\Gamma_{+}$, $\Gamma_{-}$ each end 
 on the critical lines $\gamma_{+}$, $\gamma_{-}$. What appears to be new and to the best of our knowledge not previously discussed is that the coexistence
 surfaces continue past the critical lines in a smoothed out form that corresponds to the kink transitions. In Landau theory we 
 defined such surfaces by equation (\ref{kink_Landau}) for which we found an explicit solution (\ref{Lkink_eq}) in the three coupling model.
 It seems to us that the surface $\xi_{\sigma, c}$ with $\xi_{\sigma'}<\xi_{\sigma'}^{*}$ is the field theory counterpart of the surface (\ref{kink_Landau}), (\ref{Lkink_eq}).
 
 A number of questions remain about the phase structure of the triply perturbed TIM. The curve $\xi_{\sigma, c}$ was obtained using the stability analysis of the ansatz built on 
 the $(+0)$ boundary condition. The phases that coexist for  $\xi_{\sigma}'>\xi_{\sigma'}^{*}$ should be labelled by $(0)$ and $(+)$ 
 so that on the coexistence surface we would expect to have a superposition $(0)\oplus (+)$. The latter should replace the $(+0)$ boundary 
 condition that sits on the kink transition lines. As discussed in section \ref{Landau_sec}  we expect the surface of kink transitions 
 to deviate from the phase coexistence surface sufficiently away from the critical line. It would be desirable to develop some analytic methods to describe the whole phase coexistence surface and also to study it in detail using TCSA. 
  We leave these questions to future work. 
 
\subsection{Summary and open questions} \label{summary2_sec}

To summarise our results regarding the perturbed TIM, we have mostly focused on the doubly perturbed model \ref{TIFT} 
for which the phase structure emerges as on the diagram shown in the introduction which for reader's convenience we 
show again on Figure \ref{TIM_diagram2} below. On that diagram we label the two-dimensional regions as well as the boundaries between them by 
conformal boundary conditions. For the two-dimensional regions we have also put a physical interpretation of the phases as strongly and weakly magnetised.
Next to the boundary conditions we put the operators specifying the leading perturbation 
of the conformal boundary state.  Our TCSA results indicate that this operator is $T$, as in the Cardy's ansatz, except for 
 the region $\lambda_{\epsilon}>0$, $|\xi_{\sigma}|<\xi_{c}$  where, as we argued in section \ref{small_sigma_sec}, the leading perturbation should be 
replaced by the boundary irrelevant field $\psi_{\epsilon''}$. It would be desirable to understand this more general ansatz better analytically. 
We proposed  to replace the perturbation by  $\psi_{\epsilon''}$  by a simpler supersymmetry related ansatz (\ref{ansatz2} ) but more work is needed in that direction.

\begin{center}
\begin{figure}[H]  
\centering
\begin{tikzpicture}[>=latex]
\draw[->] (-3,0) --(5.3,0);
\draw[very thick, dashed] (0,0)--(4.7,0);
\draw[->] (0,-3)--(0,3);
\draw[blue,very thick] (-3.5,0) --(0,0);
\draw (5.6,0  ) node {$\lambda_{\epsilon}$};
\draw (0,3.3) node {$\lambda_{\sigma}$};
\draw (-2, 2.5) node {\small strongly magnetised};
\draw (-2.5,2) node {\small  phase:      $|+\rangle\!\rangle , T$};
\draw (-2, -2) node {\small strongly magnetised};
\draw (-2.5,-2.5) node {\small  phase:      $|-\rangle\!\rangle , T$};
\draw (3.9, 2.5) node {\small weakly magnetised};
\draw (3.65,2) node {\small  phase:   $|0\rangle\!\rangle, \psi_{\epsilon''}$  };
\draw (3.9, -2) node {\small weakly magnetised};
\draw (3.8,-2.5) node {\small  phase:   $|0\rangle\!\rangle, -\psi_{\epsilon''}$  };
\draw (-4.8,0) node {\small $|+\rangle\!\rangle \oplus |-\rangle\!\rangle, T$};
\draw (2.9,0.3) node {\small $|0\rangle\!\rangle, T$}; 
\draw[scale=0.5, domain=0:3.2, smooth, variable=\x, red,very thick] plot ({\x}, {1.3*(\x)^(1.4)});
\draw[scale=0.5, domain=0:3.2, smooth, variable=\x, red,very thick] plot ({\x}, {-1.3*(\x)^(1.4)});
\draw (1.9,3.65) node {\small $|\!+\!0\rangle\!\rangle , T$} ;
\draw (1.9,-3.65) node {\small $|\!-\!0\rangle\!\rangle , T$} ;
\end{tikzpicture}
\caption{The phase diagram  of TIM perturbed by $\phi_{\sigma}$ and $\phi_{\epsilon}$. The phases are labeled 
 by conformal boundary states together with the leading irrelevant operator defining their perturbation at finite volume. 
 The red lines correspond to  $\xi_{\sigma}=\pm \xi_{c}$. }
\label{TIM_diagram2}
\end{figure}
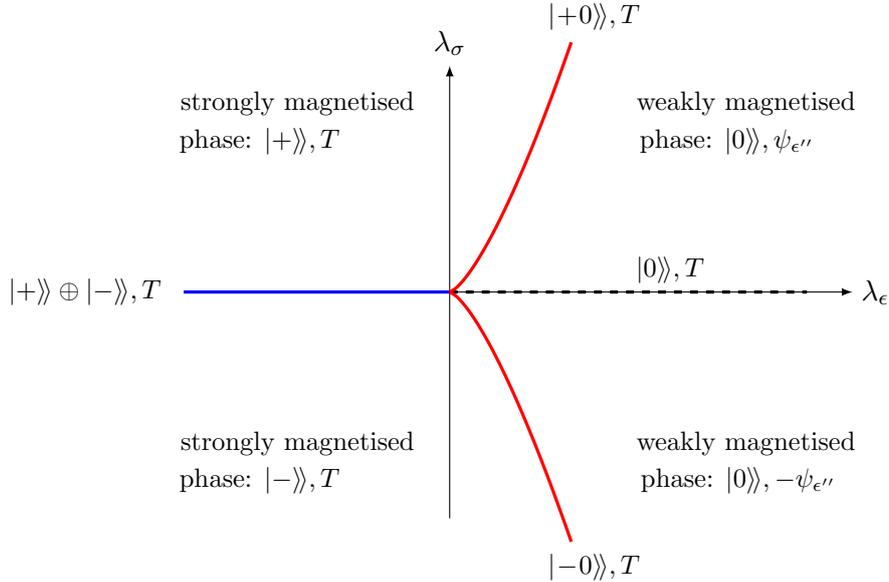
\end{center}

There are several boundaries between the two-dimensional regions  on the phase diagram. The dashed line just separates the regions with 
different leading irrelevant perturbations of the same conformal boundary state. This differs from the disorder line in the Ising model where moving 
away from the line triggers a boundary RG flow and a change of the conformal boundary condition. At the moment we don't know how this difference can 
be characterised in more physical terms. 

The blue line is the first order phase transition line. As in the Ising model it can be associated with the boundary RG flows triggered by the 
identity operators on the components: 
\be
(-) \longleftarrow (+)\oplus (-)  \longrightarrow (+) \, . 
\ee

The red lines correspond to the kink transitions: $\xi_{\sigma} = \pm \xi_{c}$ which we discussed in section \ref{transition_subsec}. 
The kink line in $\lambda_{\sigma}>0$ region   is associated with the pair of boundary RG flows that start from $(+0)$ and are triggered by the relevant boundary operator $\psi_{\epsilon'}$
\be
(0) \longleftarrow (+0) \longrightarrow (+) \, .
\ee
Similarly the red line in the $\lambda_{\sigma}<0$ region is associated with the flows
\be
(0) \longleftarrow (-0) \longrightarrow (-) \, .
\ee
Unlike the Ising model case where we found a region of parameters where we directly observed the boundary magnetic field flow, in the case at hand our 
evidence is less direct in the sense that we could not  see the flow as the evolution of the vacuum vector with $R$. 
This is because it  is  hard to locate numerically the region  where the vacuum is described by a smeared $|\!+\!0\rangle\!\rangle$ boundary state. 
Nevertheless, by measuring the one-point functions we could see that there is a physical transition and its position is well approximated using 
the stability of $|\!+\!0\rangle\!\rangle$ with respect to the boundary flows generated by  $\psi_{\epsilon'}$. We found that the $\langle \phi_{\sigma}\rangle$ 
and $\langle \phi_{\epsilon}\rangle$ one-point functions have a kink shape while $\langle \phi_{\sigma'}\rangle$, $\langle \phi_{\epsilon'}\rangle$ 
have a smoothed-out cusp singularity. We know that TCSA results are affected by  both finite size and truncation corrections which always smooth out 
singularities. Looking however at the behaviour of the effective scaling exponent of the mass gap we see that the mass gap is finite and there is no 
signature of either the second or the first order phase transition. This makes us believe that there is no actual thermodynamic phase transition 
on the $\xi=\pm\xi_{c}$ lines. Instead we interpret them as field theory counterparts of regions of phase space where 
 the third derivative of the Landau free energy vanishes and magnetisation has an inflection point. We discussed in section \ref{Landau_sec} 
 that in Landau theory there is a two-dimensional surface of such points which leads to the symmetry breaking critical lines and continues past them 
 passing close to the two-phase coexistence surface (containing first order phase transitions). On the field theory side we also found that the kink lines: $\xi=\pm \xi_{c}$ extend to a two-dimensional surface, denoted $\xi_{\sigma, c}$ which lies in the 3D space of couplings: $\lambda_{\epsilon}, \lambda_{\sigma}, \lambda_{\sigma'}$. 
 The surface $\xi_{\sigma, c}$ is easily found numerically from the CFT data and Cardy's ansatz variational energy. 
Close to this theoretical surface, using TCSA,  we located two symmetry breaking critical lines leading to the critical Ising model as well as 
 two-phase coexistence points. As TCSA results are affected by truncation errors the position of those features depends on $n_{c}$. We conjecture that in the continuum limit the critical lines  lie on $\xi_{\sigma, c}$ (and its ${\mathbb Z}_{2}$-reflected counterpart) while the phase coexistence surface 
 is only approximated by $\xi_{\sigma, c}$ close to the critical lines. 
  More work is needed to check this conjecture.

 It would be interesting to study the global phase structure of TIM   more systematically in the three- and four-coupling spaces. We feel like we have only scratched the surface here (pun intended). 
 Our stability analysis may prove useful in that task. One interesting question is whether there is a phase labelled by the $(d)$ boundary condition. 
 This boundary condition has two relevant boundary operators: $\psi^{(d)}_{\epsilon}$ and $\psi^{(d)}_{\epsilon'}$, that gives rise to two stability 
 conditions. Using the selection rules from Appendix \ref{appendix_OPE} and the fusion rules 
 \be
 \epsilon\times \epsilon = 1 + \epsilon' \, , \qquad \epsilon'\times \epsilon' = 1 + \epsilon' \, , 
 \ee
 \be
 \sigma' \times \sigma' = 1 + \epsilon'' \, , \qquad \sigma \times \sigma = 1 + \epsilon + \epsilon' + \epsilon''
 \ee
 we see that the stability with respect to $\psi^{(d)}_{\epsilon}$ boundary perturbation requires  $\lambda_{\sigma}=0$ while 
   the stability with respect to $\psi^{(d)}_{\epsilon'}$ requires a condition of the form (\ref{equilibrium_gen}). Using (\ref{btb_general}) we 
    find that 
    \be
    {}^{(d)}\!B_{\epsilon}^{\epsilon'} =   {}^{(d)}\!B_{\epsilon'}^{\epsilon'} \, .
    \ee
 Assuming for simplicity that $\lambda_{\sigma'}=0$ the condition  (\ref{equilibrium_gen}) boils down to 
 \be \label{d_stab}
 \xi_{\epsilon'} = -{\rm sign}(\lambda_{\epsilon}) \tilde \tau^{*}(\xi_{\epsilon'}) 
 \ee
 where 
 \be
 \xi_{\epsilon'} = \frac{\lambda_{\epsilon'}}{|\lambda_{\epsilon}|^{4/9}} 
 \ee
 and $\tilde \tau^{*}$ corresponds to a minimum of $e_{d}$. For a local minimum of $e_{d}$ to exist we need to have $\lambda_{\epsilon}<0$ 
 so that (\ref{d_stab}) implies that $\lambda_{\epsilon'}>0$ when $(d)$ is stable. We do find a solution to (\ref{d_stab}) at the value 
 \be
 \xi_{\epsilon'} = 0.29218...
 \ee
  This suggests that there may be a two-dimensional region\footnote{One can add the deformation by $\lambda_{\sigma'}$ as it neither couples to any boundary fields nor affects the variational energy $e_{d}$.} of phase space labelled by $(d)$.  
  Although it is difficult to imagine any new feature in that region and how it may mesh with the Kramers-Wannier duality, it may be worth  checking  using TCSA.
 The main challenge here is to tackle the large truncation errors related to $\phi_{\sigma'}$ and $\phi_{\epsilon'}$ which have a large dimension\footnote{ Also the resonance condition $\Delta_{\epsilon'}=1 + \Delta_{\epsilon}$ may need to be taken into account.}. 
 We leave these questions to future work.

\section{Outlook and future directions} \label{outlook_sec}
In this section we attempt to give an outlook at the use of Cardy's ansatz and how it could be further developed.  
We also try to clarify the conceptual framework of RG boundaries and interfaces to which it belongs. 
We have already mentioned in sections \ref{summary1_sec} and   \ref{summary2_sec} various loose ends  related to the two models we considered
so here we won't mention these loose ends again focussing  instead on general issues.  

When it comes to phases and phase diagrams Landau  theory has proven to be the easiest and most commonly used way to get a first draft 
version of the phase diagram. Of course it is known that quantitatively it does not work well close to criticality and furthermore sometimes its 
qualitative predictions (like the order of transition) can also  be  wrong. But practitioners usually know  where one should be careful and the 
disadvantages are outweighed  by the theory's simplicity and possibility of improvement via renormalisation group corrections. 
The theory assumes the existence of the order parameters which distinguish the phases and are described by bosonic fields with a free energy functional and in a sense are akin to a Lagrangian QFT with fundamental fields. 

On the other hand, in particular in two dimensions, we have a plethora of CFTs which describe critical statistical mechanical systems and 
for which a Lagrangian description is either not known or is not very useful. Many of these theories are solved exactly by algebraic methods. 
To describe the phase diagram of the perturbed CFTs in their vicinity it seems more natural to use the intrinsic algebraic language both for 
describing the phases and for putting them together. Labelling the gapped phases by conformal boundary states and using Cardy's ansatz to 
construct variational energies that allow the phases to compete seem to be the steps in that direction. The input is the exact CFT data describing the critical point and the variational energies are linear combinations of power functions. These are not as simple as polynomials of Landau theory but 
by modern standards finding their numerical minima is just as easy and transparent. Moreover, it was  found in \cite{LVT} as well as in  this paper 
that in many cases (essentially when the theory is well behaved in the UV) Cardy's ansatz gives quantitatively accurate answers for various quantities. 
 As we discussed earlier in the case of TIM in the disordered region with a small magnetic perturbation the ansatz needs to be modified and we could 
 not get a good analytic control over it. Hopefully some progress can be done in the future in generalising the ansatz for some models e.g. by using supersymmetry or other extended algebra generators.

Labelling the phases by  conformal 
boundaries (the RG boundaries) is more abstract than the Landau theory description but at the same time it is  less phenomenological and may be more precise. 
For example the conformal boundary condition $(+0)$ which appears in TIM does not have a precise analogue as a phase label in the classical Landau theory. We should be careful and remember that our new "phases" may be different from the conventional ones as well as the transitions between them. Nevertheless, for the sake of brevity we will continue to refer to RG boundaries as phases  in this section.

Starting with a given 2D CFT it is not immediately clear which conformal boundary conditions appear as phases and which can coexist. 
For example we don't know whether the boundary condition $(d)$ appears as a phase  in the full four-coupling space of the perturbed TIM (see our discussion of that in section \ref{summary2_sec}).  
Relating the boundaries between the phase regions to  boundary RG flows, that proved to be quite concrete and useful in the two models we studied in this paper, suggests an organising principle here. Given a CFT with its list of conformal boundary conditions we could start by identifying as ``elementary'' or ``prime'' phases with the stable boundary conditions which do not have any relevant boundary operators. We note that for the Ising and TI models such boundary conditions are in one-to-one correspondence with the fundamental  lattice degrees of freedom. The ``composite'' phases then could be described by boundary conditions with higher boundary entropy that may be realisable on  sub-manifolds of smaller dimension (perhaps the smaller the more relevant boundary operators it has) in the coupling space. A tentative rule for 
constructing the phase boundaries   can be the existence of boundary RG flows from the boundary conditions belonging to the boundary to the ones this boundary separates. The simplest boundary in this language  is the one that corresponds to the first order transitions and that is labelled by the superposition 
of the boundary states describing the coexisting phases. The boundary RG flows in this case are just the flows triggered by the boundary identity operators.
We note that for the boundary RG flows there is also a version of Cardy's ansatz developed in \cite{AKB}  which can be put to work when mapping the manifold of boundary RG flows.

At first glance the description using Cardy's ansatz only allows to accommodate for the first order phase transitions but the Ising model disorder line together with our example of the kink transitions in this paper adds more features. These lines do not describe a thermodynamic phase transition. There are 
two angles one can take looking at their significance. On the one hand they refine the picture of the conventional phase space adding more features. 
On the other hand, from the purely utilitarian point of view, they are useful in locating the  second order phase transition lines as we demonstrated 
in the case of TIM where we found the symmetry breaking critical lines close to the kink surface $\xi_{\sigma,c}$.

A  description of critical lines in the algebraic way remains a  challenge in that it requires knowing conformal interfaces between the UV and IR CFTs. 
While for many models, e.g. for the Virasoro minimal models,  we have a complete knowledge of conformal boundary conditions,  we 
have few explicit constructions of conformal interfaces between different CFTs. However, the construction of Gaiotto \cite{Gaiotto} and its generalisations 
\cite{Pog1, Stanishkov, Pog2, Pog3} give examples of RG interfaces which can be described both exactly algebraically and compared to perturbative 
RG calculations. In particular the  construction of  \cite{Gaiotto} describes the RG interface between the TIM and the critical Ising models that corresponds 
to the symmetry preserving critical line\footnote{We have checked some predicted eigenvectors' components by TCSA calculations and found a good match. The details will be reported elsewhere.}. 
Hopefully more progress will be made in the future which will lead to a better description of critical lines within the general paradigm of RG boundaries and interfaces.

Lastly we want to mention  the quantity called chiral (or left-right) entanglement entropy \cite{ch1,ch2,ch3} which was studied in \cite{LVT} in conjunction with  both Cardy's ansatz and TCSA numerics. 
Unlike the low weight components of the vacuum vector, on which we focused for the  most part in  this paper, the chiral entanglement carries 
information on the high weight tail of the vacuum vector. It is an interesting quantity that looks promising as a tool to  distinguish different  phases and should in our opinion be studied    further. We plan to report some findings about it in \cite{Kon_wip}.

\begin{center}
{\bf \large Acknowledgements} 
\end{center}
I would like to thank Gabor Tak\' acs and Des Johnston for comments on the draft version of the paper.
This work was supported by EPSRC grant ``Renormalisation group interfaces in Tricritical Ising Model'', grant reference number: EP/W010283/1.
All numerical calculations were done using Wolfram Mathematica 13.1.

\appendix
\renewcommand{\theequation}{\Alph{section}.\arabic{equation}}
\counterwithin{figure}{section}
\setcounter{equation}{0}

\section{Phase coexistence surface of TIM in Landau theory} \label{appendix_Landau}
\setcounter{equation}{0}
In this appendix we give an explicit description of the two-phase coexistence surface of the three-coupling theory with Landau 
potential (\ref{LF2}). We will parameterise this surface by the smallest and largest real roots of $\frac{d{\cal F}}{d\mu}$ : $\mu_{1}$ and $\mu_{3}$. 
The absence of the quartic and quintic terms in ${\cal F}$ can be used to express the complex conjugate pair of roots in terms of the three real roots: $\mu_{1}$, $\mu_{2}$ and 
$\mu_{3}$. We can then express $h$, $t_{2}$ and $t_{3}$ in terms of $\mu_{i}$, $i=1,2,3$. Substituting these expressions into the phase coexistence condition (\ref{coex_eq}) we obtain that $\mu_{2}$ must be a root of the following cubic polynomial 
\be
P(x) = x^3 + x^2 (\mu_{1} + \mu_{3}) + (x-\mu_{1} - \mu_{3}) (\mu_{1}^2 + \mu_1\mu_3 + \mu_{3}^2)  
\ee
In the region of interest this polynomial has only one real root that specifies a function $M(\mu_{1} , \mu_{3}) $ which can be either calculated numerically or expressed via cubic roots.  Hence the middle root is 
\be
\mu_{2} = M(\mu_{1} , \mu_{3}) \, .
\ee
We find the following expressions for the three couplings on the coexistence surface 
\bea \label{3par}
&& h = M\mu_{1} \mu_{3} (M^2 +  \mu_{1}^2+ \mu_{1}\mu_{3} + \mu_{3}^2 + M(\mu_1 + \mu_3)      ) \, , \nonumber  \\
&& t_{2} = M^2(\mu_{1} + \mu_3)(M+\mu_1+\mu_3) + (\mu_{1}^2+ \mu_{1}\mu_{3} + \mu_{3}^2)(\mu_{1}\mu_{3} + M(\mu_{1} + \mu_{3})) \, , 
\nonumber \\
&& t_{3} = M^3 +  (\mu_1 + \mu_3) (M^2 + \mu_1^2+\mu_3^2) +  M (\mu_{1}^2+ \mu_{1}\mu_{3} + \mu_{3}^2) \, 
\eea
where $M$ stands for $M(\mu_{1},\mu_{3})$. Formulae (\ref{3par}) give a parametric representation of the coexistence surface which can be used to plot the surface numerically. 

We remark that in a similar way one can describe the metastability surface  where the second phase first appears as a metastable state. 
We can parameterise such free energies by two root values one of which gives a coincident root: $x_{1}=x_{2}$, $x_{3} \ne x_{1}$ or $x_{3}=x_{2}$, $x_{1} \ne x_{2}$.

\section{Bulk-boundary OPE coefficients} \label{appendix_OPE}
\setcounter{equation}{0}
A general formula for the bulk-boundary OPE coefficients in the A-series minimal models was derived 
in \cite{Runkel}. To set the conventions we consider a minimal model on the upper half plane with 
a Cardy boundary condition labelled  by $a$. The bulk primary operators $\phi_{i}(z,\bar z)$ have conformal weights 
$(h_{i},h_{i})$ and are canonically normalised 
to have an OPE of the form 
\be
\phi_{i}(z,\bar z) \phi_{i}(z',\bar z') = \frac{1}{ |z-z'|^{4h_{i}}} + \mbox{  less singular terms } 
\ee
The boundary primary fields are denoted $\psi_{k}^{(a)}$. They 
have dimension $h_{k}$ and are normalised to have OPE 
\be
\psi_{k}^{(a)}(x)\psi_{k}^{(a)}(x') = \frac{1}{ |x-x'|^{2h_{i}}} + \mbox{  less singular terms } \, . 
\ee
The bulk-boundary OPE coefficients are defined as 
\be
\phi_{i}(x+iy,x-iy) = \sum_{k} {}^{(a)}\! B_{i}^{k}\psi_{k}^{(a)}(x) (2y)^{h_{k}-2h_{i}} + \mbox{  descendants } 
\ee

With these conventions the formula for ${}^{(a)}\! B_{i}^{k}$ derived in \cite{Runkel} reads 
\bea \label{btb_general}
{}^{(a)}\! B_{i}^{k} = &&\left(\frac{S_{11}}{S_{1a}}\right)\frac{\left(F_{11}\left[ \begin{array}{cc} i&i\\
i & i
\end{array} \right]F_{a1}\left[ \begin{array}{cc} a&a\\
k& k
\end{array} \right]\right)^{1/2}}{F_{11}\left[ \begin{array}{cc} k&k\\
k & k
\end{array} \right]F_{k1}\left[ \begin{array}{cc} i&i\\
i & i
\end{array} \right]F_{k1}\left[ \begin{array}{cc} a&a\\
a & a
\end{array} \right]} \nonumber \\
&& \qquad \qquad \times \sum_{r} e^{i\pi(\frac{1}{2}h_{k}-2h_{a}-2h_{i}+2h_{r})}F_{kr}\left[ \begin{array}{cc} a&i\\
a & i
\end{array} \right]F_{r1}\left[ \begin{array}{cc} i&i\\
a & a
\end{array} \right] \, . 
\eea
Here $S_{ij}$ are the $S$-matrix entries and $F_{pq}\left[ \begin{array}{cc} i&j\\
k & l
\end{array} \right]$ are the fusion matrices which can be calculated using the recurrence relations as 
explained in \cite{Runkel}, \cite{Runkel_PhD}. The summation in (\ref{btb_general}) runs over the primary labels 
for which both fusion matrices in the sum exist. This is ensured by two necessary conditions: $k$ is contained in the fusion $a\times a$
and $k$ is contained in the fusion $i\times i$. The first condition holds for all boundary fields with the boundary condition labelled by $a$ 
while the second condition can be considered as a useful superselection rule which says that certain bulk-boundary coefficients are zero. 

For the TIM the $(+0)$ boundary condition corresponds to $a=\epsilon'=(1,3)$ and there is a relevant boundary field $\psi_{\epsilon'}^{(+0)}$ which couples 
to $\phi_{\sigma}$, $\phi_{\epsilon}$ and $\phi_{\epsilon'}$ but not to $\phi_{\sigma'}$. We calculate the corresponding bulk-boundary OPE coefficients for  $\phi_{\sigma}$ and  $\phi_{\epsilon}$ using (\ref{btb_general}) and the expressions for the $F$-matrices found in \cite{Runkel},  \cite{Runkel_PhD}. We  obtain 
  \be \label{btb1}
 {}^{(+0)}\!B_{\sigma}^{\epsilon'} = b_{\epsilon'}\frac{\left( 5(7-3\sqrt{5})\right)^{1/4}}{2} \, ,
 \ee
  \be \label{btb2}
 {}^{(+0)}\!B_{\epsilon}^{\epsilon'} = -b_{\epsilon'} \frac{\sqrt{3\sqrt{5}-5}}{\sqrt{2}} 
 \ee
 where 
 \be
 b_{\epsilon'}=\left(F_{\epsilon'1}\left[ \begin{array}{cc} \epsilon'&\epsilon'\\
\epsilon'& \epsilon'
\end{array} \right]\right)^{-1/2} =\frac{2\pi\sqrt{2}}{\Gamma(1/5)\Gamma(8/5)}\sqrt{\frac{\Gamma(2/5)}{(5-\sqrt{5})\Gamma(-6/5)}} 
\approx 0.881\, .
 \ee
This gives the ratio 
\be \label{B_ratio}
\frac{{}^{(+0)}\!B_{\epsilon}^{\epsilon'}}{{}^{(+0)}\!B_{\sigma}^{\epsilon'}}=-2^{3/4} \, .
\ee

\section{TCSA computations: technical issues}  \label{appendix_TCSA}
\setcounter{equation}{0}
In this appendix we discuss how one finds a physical window and give details on how the one-point functions plotted on Figures \ref{vev_pic1}, \ref{vev_pic2}  
  were calculated. 

\begin{center}
\begin{figure}[H]
\begin{minipage}[b]{0.5\linewidth}
\centering
\includegraphics[scale=0.82]{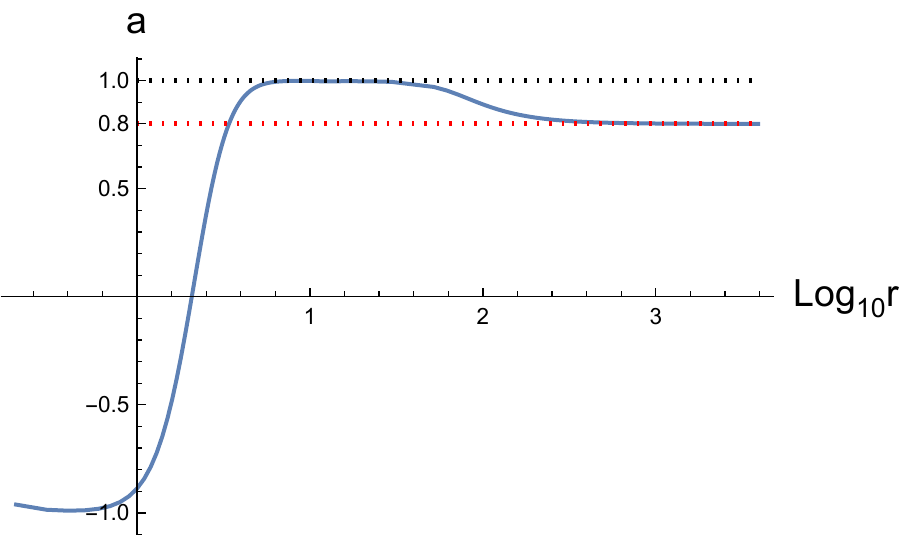}
\end{minipage}%
\begin{minipage}[b]{0.5\linewidth}
\centering
\includegraphics[scale=0.82]{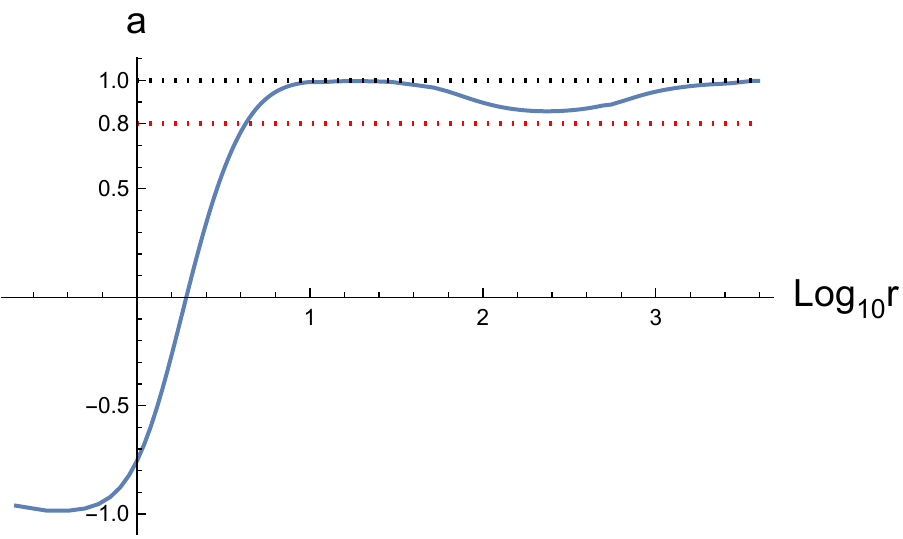}
\end{minipage}
\caption{The effective scaling exponent of the vacuum energy $a$ for a pure $\phi_{\epsilon}$ perturbation (left) and $\xi_{\sigma}=1.4$ mixed perturbation (right) of TIM. Both perturbations are taken with $\lambda_{\epsilon}>0$ 
and calculated with $n_{c}=11$.  }
\label{fig_aa}
\end{figure}
\end{center}

In TCSA in order to calculate  quantities describing a QFT on a plane we need to take sufficiently large $R$ to make finite size corrections small. 
On the other hand, as  was noticed  in \cite{Cardy_etal}, the truncation errors increase with $R$. Thus, to measure anything with reasonable accuracy we need to be within the so called physical window. To quantify this it was proposed in  \cite{Cardy_etal} to use an effective scaling exponent of 
the vacuum energy:
\be
a = \frac{R}{E_{0} }\frac{d E_{0}  }{dR} \, .
\ee
Close to the UV fixed point, at small $R$, this quantity is approximately equal to -1. As $R$ increases and we enter the physical window 
characterising a massive trivially gapped QFT $a$ becomes close to 1. As we increase $R$ further, since the CFT Hamiltonian $H_{0}$ and the 
perturbing operators are represented by finite size matrices, the truncated perturbing operators start dominating  the Hamiltonian and the exponent $a$ starts decreasing. For a single perturbing operator with dimension $\Delta$ the exponent $a$ decreases to a plateau at the value $1-\Delta$, see the left plot on Figure \ref{fig_aa} for illustration.

For multiple perturbations the more relevant operators become more important for larger values of $R$ so that more different regimes can appear as we change $R$. Sometimes we can discern more than one plateau where $a\approx 1$.  See Figure  \ref{fig_aa} for an illustration, the right plot there 
gives $a$ for the doubly perturbed TIM (\ref{TIFT}) with $\xi_{\sigma}=1.4$, $\lambda_{\epsilon}>0$. 

\begin{center}
\begin{figure}[H]
\begin{minipage}[b]{0.5\linewidth}
\centering
\includegraphics[scale=0.82]{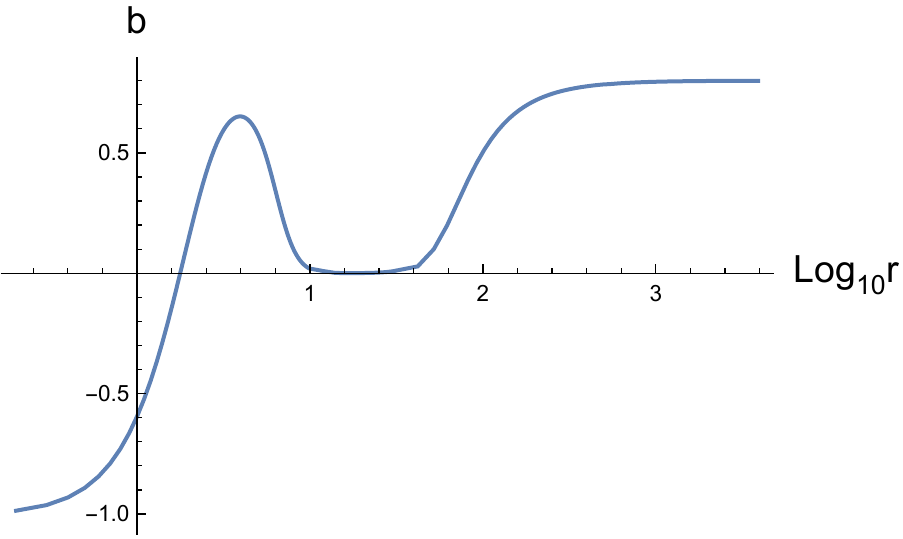}
\end{minipage}%
\begin{minipage}[b]{0.5\linewidth}
\centering
\includegraphics[scale=0.82]{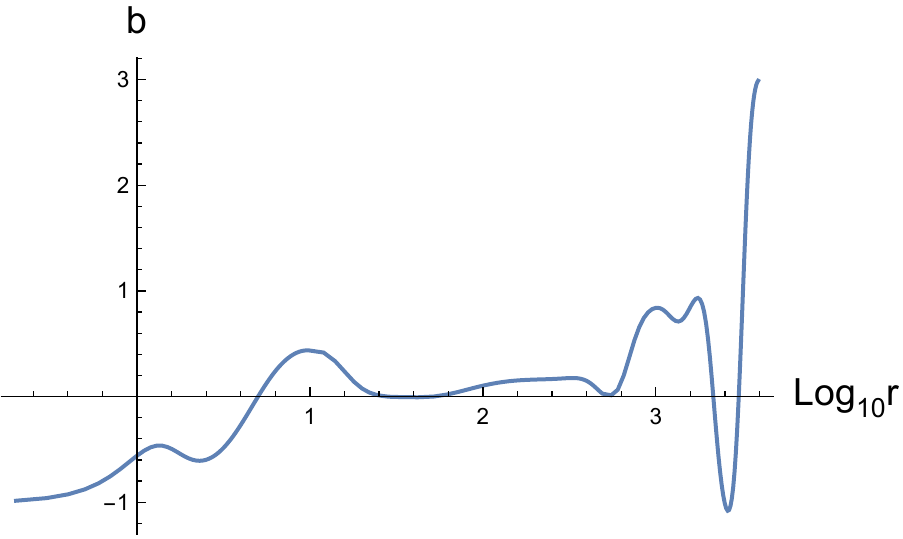}
\end{minipage}
\caption{The effective scaling exponent of the  energy gap $b$ for a pure $\phi_{\epsilon}$ perturbation (left) and $\xi_{\sigma}=1.4$ mixed perturbation (right) of TIM. Both perturbations are taken with $\lambda_{\epsilon}>0$ 
and calculated with $n_{c}=11$. }
\label{fig_bb}
\end{figure}
\end{center}

An additional quantity which helps to isolate the physical window and which in particular helps to distinguish between multiple  plateaus of $a\approx 1$ is 
the effective scaling exponent of the mass gap:
\be
b = R\frac{d  \ln(E_{1}-E_{0})}{dR} \, .
\ee
In the physical window it should be close to zero. From the plots on Figure \ref{fig_bb} we see that  for the pure $\phi_{\epsilon}$ perturbation the exponent   $b$  refines 
the physical window in comparison to the one singled out by $a$ only. For the mixed perturbation the $b$ exponent is very large at the second plateau 
where  $a\approx 1$, and   that effectively rules it out\footnote{We also note that the second plateau is located at very large values of $r$ where the $H_{0}$ part of the Hamiltonian can be neglected that also looks unphysical.}.

The second plateau with $a\approx 1$ is always present for mixed perturbations and is located at large values of $r$. 
As we increase  $\xi_{\sigma}$ this plateau moves towards the first one located at smaller values of $r$. Near the kink transition $\xi_{\sigma}=\xi_{c} \approx 1.78$ both plateaus collide and eventually, as we keep increasing $\xi_{\sigma}$, they merge. 

\begin{center}
\begin{figure}[H]
\begin{minipage}[b]{0.5\linewidth}
\centering
\includegraphics[scale=0.82]{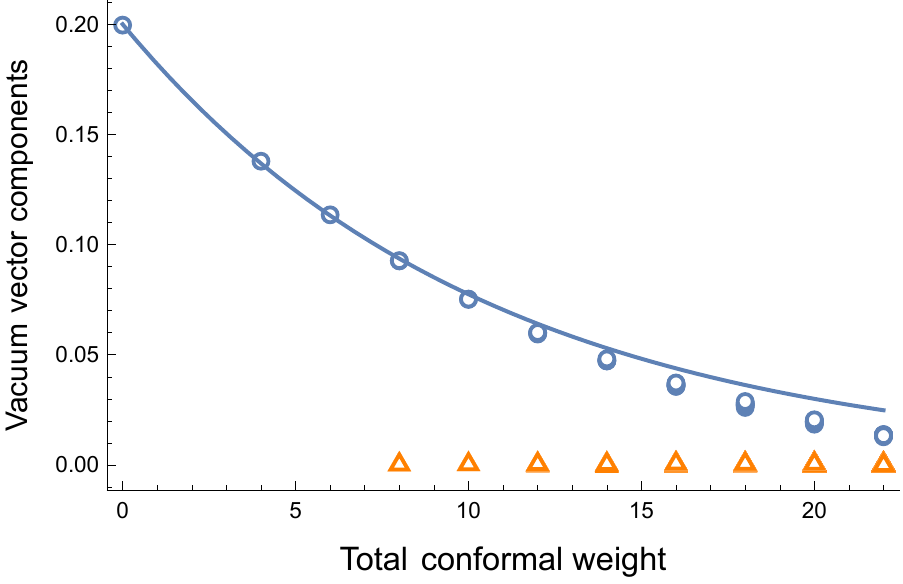}
\end{minipage}%
\begin{minipage}[b]{0.5\linewidth}
\centering
\includegraphics[scale=0.82]{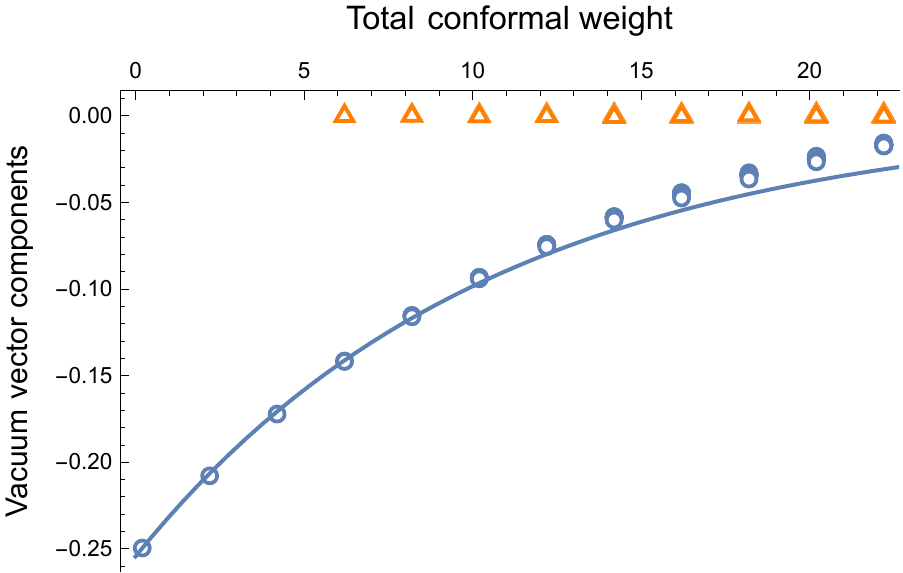}
\end{minipage}
\caption{The conformal weight spectrum of the lowest eigenvalue eigenvector of the integrated $\phi_{\epsilon}$ operator in TIM. The left plot is for the identity sector and the right -- for the $\epsilon$-sector. The solid line corresponds to a smeared $|0\rangle\!\rangle$ boundary state.}
\label{fig_V1}
\end{figure}
\end{center}

\begin{center}
\begin{figure}[H]
\begin{minipage}[b]{0.5\linewidth}
\centering
\includegraphics[scale=0.82]{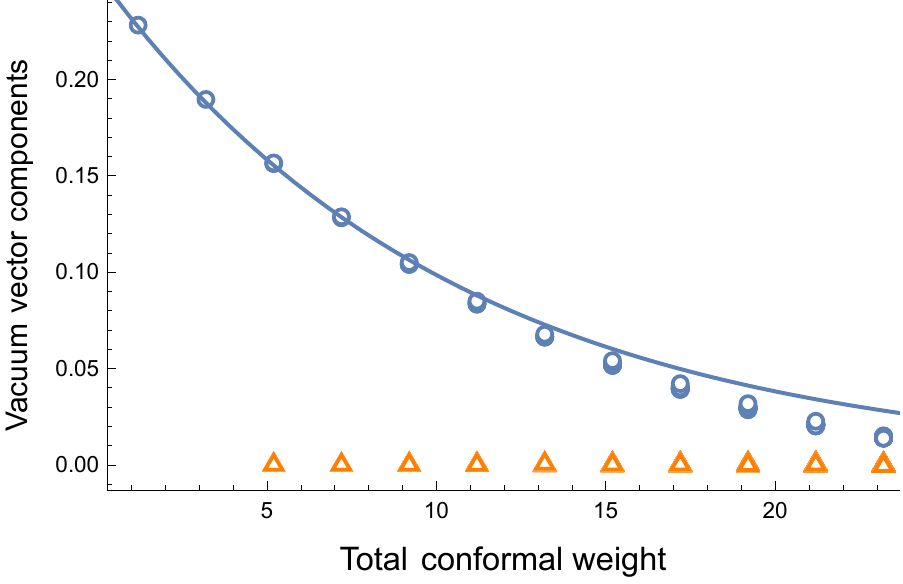}
\end{minipage}%
\begin{minipage}[b]{0.5\linewidth}
\centering
\includegraphics[scale=0.82]{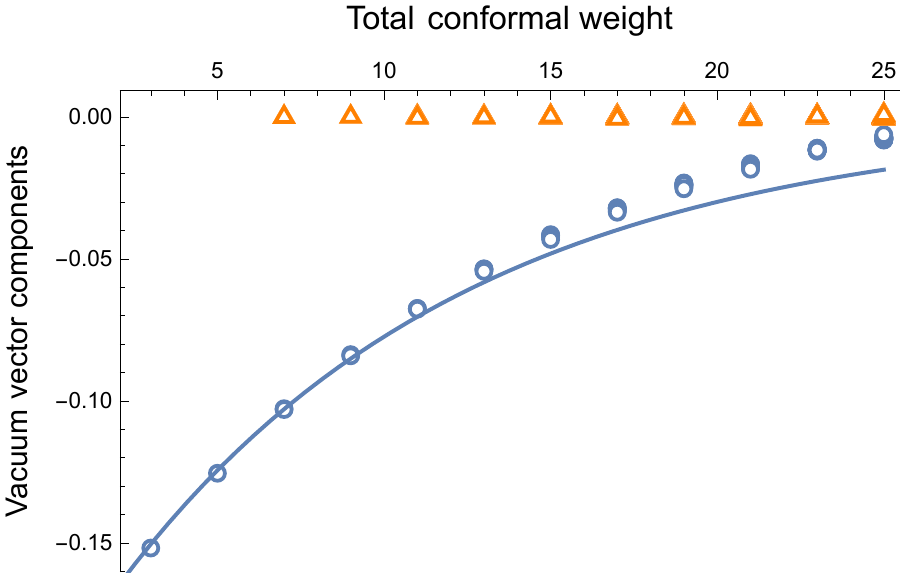}
\end{minipage}
\caption{The conformal weight spectrum of the lowest eigenvalue eigenvector of the integrated $\phi_{\epsilon}$ operator in TIM. The left plot is for the $\epsilon'$ sector and the right -- for the $\epsilon''$-sector. The solid line corresponds to a smeared $|0\rangle\!\rangle$ boundary state.}
\label{fig_V2}
\end{figure}
\end{center}

One should be especially careful dealing with very large values 
of $r$ when comparing the vacuum vector with smeared conformal boundary states because the eigenvectors of the perturbing (integrated) operators 
also have a similar form. For illustration we show\footnote{Only four energy sectors are shown. The components in the magnetic sectors are very small.} on Figures \ref{fig_V1}, \ref{fig_V2} the conformal weight spectrum of the integrated $\phi_{\epsilon}$ operator in TIM 
found by TCSA with $n_{c}=11$ 
compared to a smeared state 
\be
e^{-\eta (L_{0} + \bar L_{0})} |0\rangle\!\rangle 
\ee
with $\eta\approx 0.095$ found by numerical fitting.  

We see that this fits  with  Cardy's ansatz quite well. Clearly $\eta$ is a dimensionless parameter that replaces $2\pi \tau/R$ and depends only on the truncation parameter. We find similar fitting for the integrated $\phi_{\sigma}$ operator and smeared $|\pm\rangle\!\rangle$ states (the plus and minus correspond to the choice of the lowest or highest eigenvalue).  For  the integrated $\phi_{\sigma'}$ we get smeared superpositions of $|+\rangle\!\rangle$ and $|\!-\!0\rangle\!\rangle$ (or $|-\rangle\!\rangle$ and $|\!+\!0\rangle\!\rangle$, again depending  on whether we choose the two highest or the two lowest eigenvalues) 
same as the ones discussed in \cite{LVT}. 

It follows from these observations that if we take $R$ too large, in the region where one perturbing operator dominates, we are risking misidentifying the conformal boundary state characterising the true vacuum. 

Next we would like to discuss the calculation of the one-point functions  which lead to the plots on Figures \ref{vev_pic1}, \ref{vev_pic2}. 
The scaled one-point functions are formally defined as a large $r$ limit
\be
  f_{i} =  \lim_{r\to \infty} \left( \frac{2\pi}{r}\right)^{\Delta_{i}} {}_{\lambda}\langle 0|\phi_{i}^{\rm pl} |0\rangle_{\lambda} \, .
 \ee 
In TCSA we calculate numerically the quantities 
\be
F_{i}(r)=\left( \frac{2\pi}{r}\right)^{\Delta_{i}} {}_{\lambda}\langle 0|\phi_{i}^{\rm pl} |0\rangle_{\lambda}
\ee
and then we choose the value  $r=r^{*}$ at which we take $F_{i}(r^{*})$ as an approximation of $f_{i}$.
To choose $r^{*}$ we calculate the derivative 
\be
\frac{dF_{i}(r)}{dr}
\ee
and take either the point where it vanishes, that corresponds to a local minimum or maximum of $F_{i}$, or where it has a local minimum 
that looks like a kink (the function keeps decreasing or increasing but has a more flattened region). Similarly to our  discussion of the physical window above, 
for multiple perturbations we typically have several candidate values of $r^{*}$ present  in the regions (plateaus) where the function levels out. As a general rule we give preference to smaller values of $r^{*}$ unless they are too small and appear before the start of the physical window. For $f_{\sigma}$ and $f_{\epsilon}$ this strategy is implemented in a straightforward way as there is always a single value of $r^{*}$ selected  in the physical window by the rules stated. The second plateau for small $\xi_{\sigma}$ is at very large values of $r$. As $\xi_{\sigma}$ increases the second plateau moves towards the first one and merges with it at about $\xi_{\sigma}=2.3$. 

\begin{center}
\begin{figure}[H]
\centering
\includegraphics[scale=0.8]{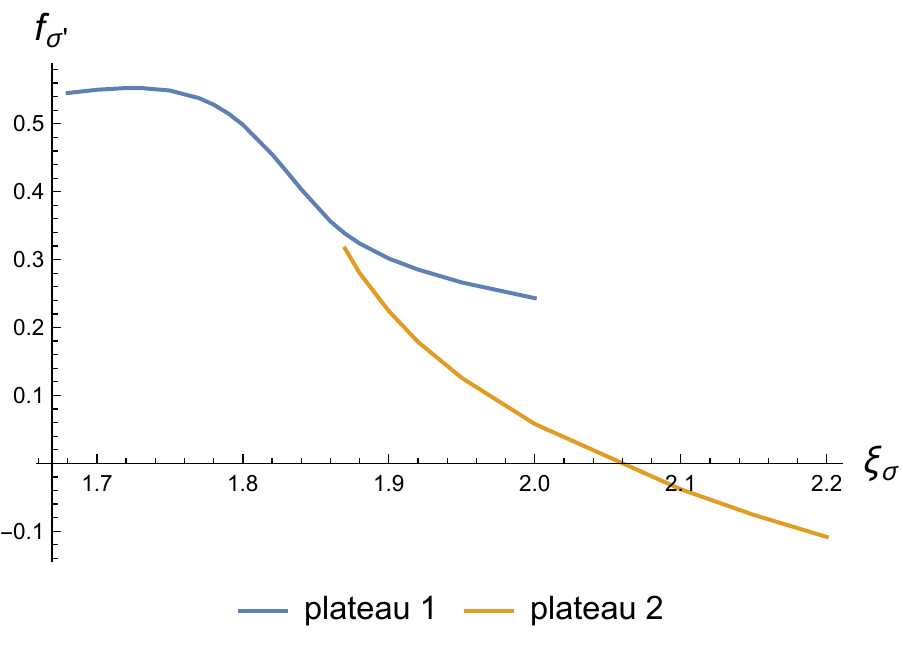}
\caption{Two branches of functions $f_{\sigma'}$ corresponding to values of $F_{\sigma'}(r^{*})$ with $r^{*}$ located on different plateaus.
   }
\label{fig_splice}
\end{figure}
\end{center}

For $f_{\sigma'}$ 
and $f_{\epsilon'}$ the situation is a bit more involved. For $f_{\sigma'}$ when $\xi_{\sigma}<1.86$ we always take $r^{*}$  on the first (small $r$) plateau. 
However  this plateau, which contains local maxima, moves to small values of $r$ and eventually gets outside of the physical window. At the same time the second plateau, which contains local minima of derivative, and located at larger values of $r$ emerges in the physical window. To splice the two branches we plot $f_{\sigma'}$ on a segment near $\xi_{\sigma}=1.86$ (see Figure \ref{fig_splice}) and determine the switching point  by continuity.
 
At about $\xi_{\sigma}=2.15$ the  plateau in the physical window disappears altogether  with the first plateau appearing at $r\approx 216$ that is outside the physical window. From that value of $\xi_{\sigma}$ onwards we read off our approximation to $f_{\sigma'}$ at a fixed point $r^{*}=31.5$ that is located roughly where the values of $r^{*}$ were for  $\xi_{\sigma}\le 2.1$. As we increase $\xi_{\sigma}$ from that value the unphysical plateau moves to smaller $r$ and the 
corresponding value of $r^{*}$, that gives a local minimum of $F_{\sigma'}$, hits our chosen point $r=31.5$ at about $\xi_{\sigma}=3.3$ from which point on 
we start choosing $r^{*}$ at a local minimum. To summarise, in four different intervals of $\xi_{\sigma}$ we used four different ways to choose $r^{*}$ which we spliced in a continuous way to 
obtain the plot on Figure  \ref{vev_pic2}.

We encounter  similar problems when determining $f_{\epsilon'}$. The plateau in the physical window disappears at about $\xi_{\sigma}=2.4$. 
Between that value and $\xi_{\sigma}=3.1$ we read off  $f_{\epsilon'}$ at a fixed point $r^{*}=32$. Close to  $\xi_{\sigma}\approx 3.1$ a plateau, 
which contains a local maximum, enters the physical region from the region of large $r$ and we switch to reading off at local maxima for  $\xi_{\sigma}\ge 3.1$. The right plot 
on  Figure  \ref{vev_pic2} is thus obtained using three different intervals with different ways for choosing $r^{*}$.

Next we comment on truncation corrections. In \cite{LVT} a two-step procedure developed in \cite{LT_Potts} was used.  First RG corrections 
by local $n_{c}$-dependent counterterms were used as developed in  \cite{GW,HRvR,RV,LT_RG},  followed by  extrapolation in $n_{c}$. 
For the extrapolation a sum of power functions of the form 
\be \label{interpolation}
\sum_{i} C_{i} \left( \frac{1}{n_{c}}\right )^{x_{i}}
\ee is used where the powers $x_{i}$ are determined from 
perturbation theory using OPE of the perturbing operator with itself. The values of physical quantities are calculated by TCSA at different values of $n_{c}$ which 
are then fitted numerically to an ansatz \ref{interpolation} contaning a number of powers $x_{i}$ which is extrapolated to $n_{c}\to \infty$. 
The method was applied in \cite{LVT} to single operator perturbations in the Ising and TI models. It has demonstrated a certain consistency by checking 
it against the exact answers known for integrable perturbations and seeing how it behaves for quantities whose functional dependence on $R$
 is known on physical grounds. For example the ground state energy and the chiral entanglement entropy are expected to be linear functions of $R$ in the physical window. 
Performing extrapolation in $n_{c}$ separately at different values of $R$ one finds that the  variance of the function at hand decreases versus the one obtained from  the raw data, it thus gets better approximated by a straight line. 

For mixed perturbations that we studied in this paper  we first checked numerically that the 
corrections due to the running couplings are small. We then looked at interpolation of vacuum energy and chiral entanglement entropy based on keeping 
a few smallest exponents in functions like (\ref{interpolation}). We found the results less consistent than in the single perturbation cases. The variance in some 
intervals of $R$ (in the physical window) would go down after extraploation while in others it would go up. Perhaps this difficulty is  due to the fact that for 
 two perturbing operators the set of exponents $x_{i}$ is more dense with some values coming from different OPE terms being quite close to each other.  
 This may mean that more values of $n_c$ would be needed to fit such functions efficiently, but with our computing capabilities the truncation 
 parameter  is limited so that we can only collect data at no more than 10 different  values of $n_{c}$. On the other hand in TIM, which is our main focus in this paper, 
 both of the perturbing operator's dimensions are quite small so we expect the truncation errors to be small. 
 Thus, throughout the paper we do not attempt to implement any corrections always using the raw TCSA data. 
 
 A few comments about our TCSA code. In our algorithm we  first construct a non-orthonormal basis $|j\rangle$ of Virasoro descendants of the form (\ref{tvec1})
 iteratively level by level, removing null states. This gives us a Gram matrix $G=\langle i|j\rangle$ of inner products of the basis states which is real symmetric. 
 The perturbed Hamiltonian in this basis has matrix elements 
 \be
 H_{ij} = \langle i|H|j\rangle \, .
 \ee
 We further construct an orthonormal basis at each level using the Cholesky decomposition  of the Gram's matrix. The latter represents $G$ 
 as a product 
 \be
 G=( \eta \eta^{T})^{-1} 
 \ee
 where $\eta$ is a lower triangular matrix and $ \eta^{T}$ is its transposition. The associated orthonormal basis is given by 
 \be
 |i\rangle_{\rm ort} = \sum_{j} \eta_{j, i} |j\rangle 
 \ee
 and the  Hamiltonian in the orthonormal basis has matrix elements 
 \be
{}_{\rm ort} \langle i|H|j\rangle_{\rm ort} = \sum_{k,l} \eta^{T}_{i,k} H_{kl} \eta_{l, j} \, . 
 \ee
We use Wolfram's Mathematica which has a built in algorithm for finding the Cholesky decomposition. To find the lowest eigenvalues and eigenvectors  of the truncated Hamiltonian 
we use Arnoldi's method which for large matrices works a lot faster than the built in default method. For the Ising model, with a few exceptions stated in the main text, we used the mode truncation with $n_{c}=17$ that gives the truncated space of dimension 8774. For the perturbed TIM we used the truncation 
by the level of Virasoro descendants as in (\ref{tvec2}) with $n_{c}=11$ that gives dimension 8810.

\section{TCSA computations: additional plots}  \label{appendix_TCSA_plots}
\setcounter{equation}{0}

\subsection{Ising field theory  with $\xi=10$ and $t>0$ } \label{xi10_appendix}

Here we present some plots for $\xi=10$ and $t>0$  -- an example with a large magnetic field in the disordered region.
 \begin{center}
\begin{figure}[H]
\begin{minipage}[b]{0.5\linewidth}
\centering
\includegraphics[scale=0.82]{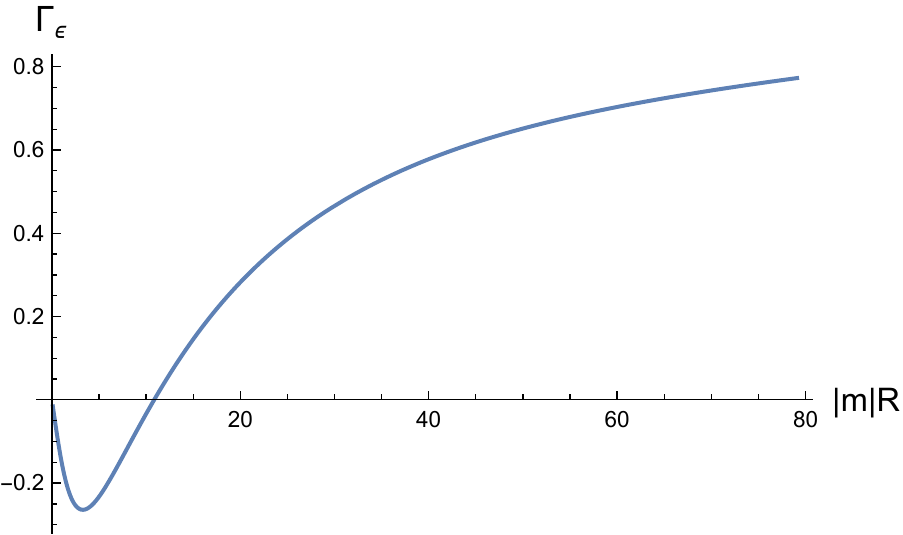}
\end{minipage}%
\begin{minipage}[b]{0.5\linewidth}
\centering
\includegraphics[scale=0.82]{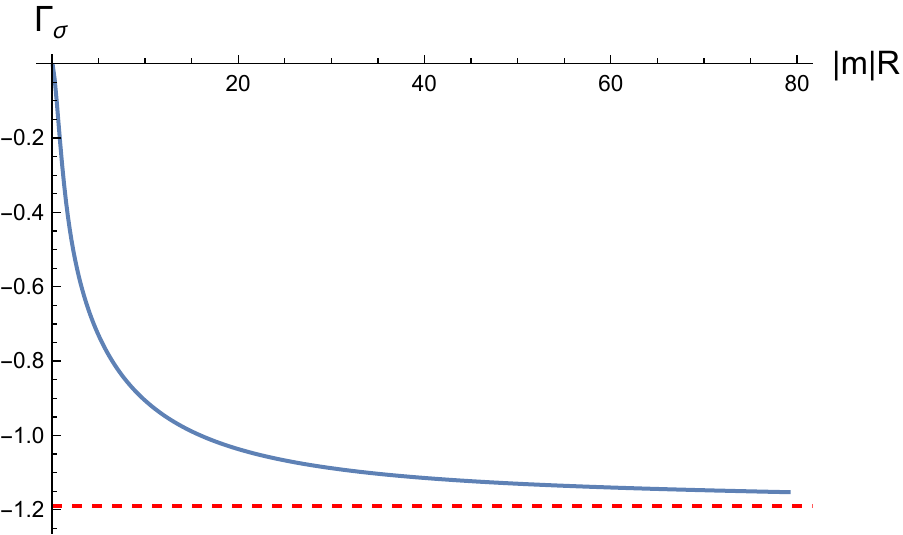}
\end{minipage}
\caption{Ratios of vacuum components obtained using TCSA with oscillator basis truncated at level 17 (dimension 8774) for 
$\xi=10$, $t>0$.  }
\label{overlap_plots_xi10}
\end{figure}

\end{center}

\begin{center}
\begin{figure}[H]
\begin{minipage}[b]{0.5\linewidth}
\centering
\includegraphics[scale=0.82]{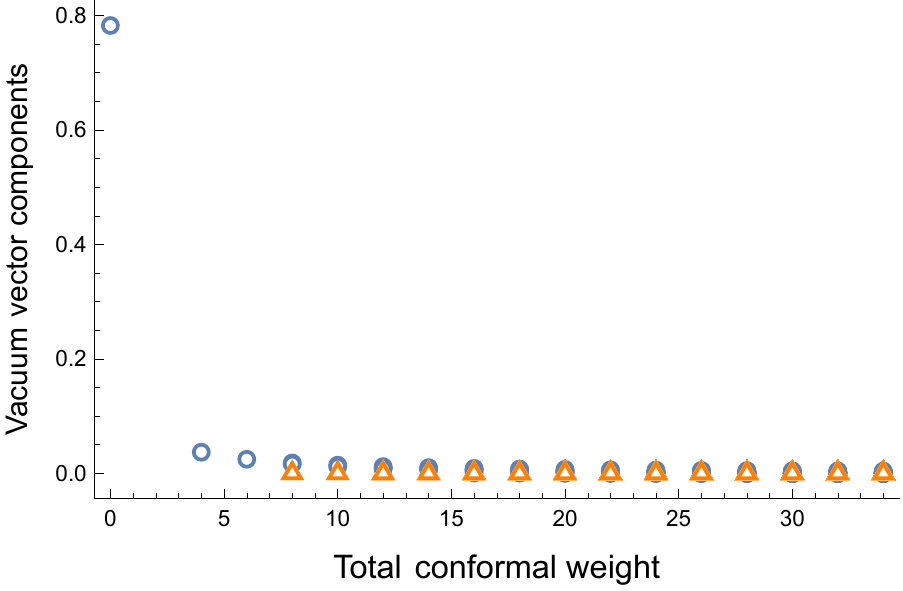}
\end{minipage}%
\begin{minipage}[b]{0.5\linewidth}
\centering
\includegraphics[scale=0.82]{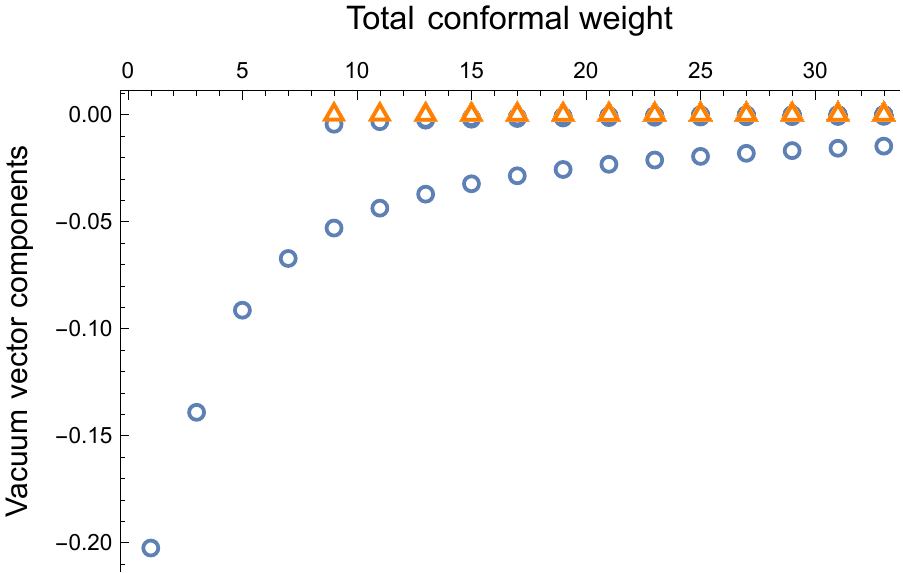}
\end{minipage}
\caption{The TCSA vacuum vector components $C_{i}$ against the total conformal weight of the basis vectors taken for the mixed perturbation of the Ising model with $\xi=10$, $t>0$ and at  $|m|R=4$. The left plot 
represents the vacuum sector and the right plot -- the $\epsilon$-sector. The blue circles mark the diagonal components and the orange triangles -- the non-diagonal ones.}
\label{fig_bmfspec1b}
\end{figure}

\end{center}

\begin{center}
\begin{figure}[H]
\centering
\includegraphics[scale=0.85]{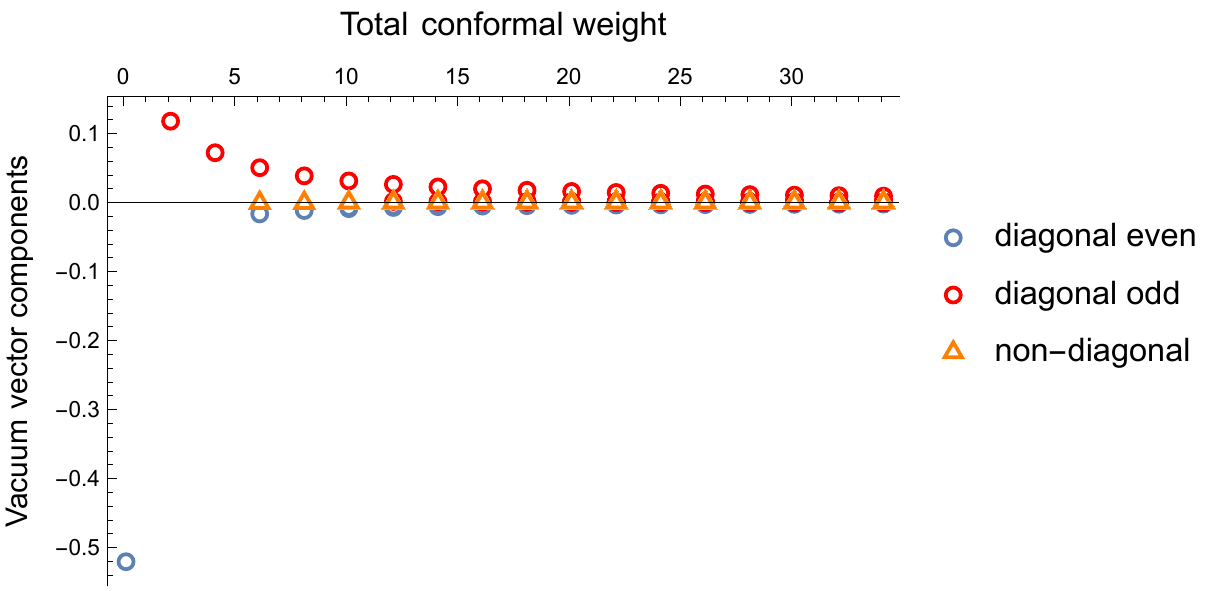}
\caption{The TCSA vacuum vector components $C_{i}$ in the $\sigma$-sector plotted against the total conformal weight of the basis vectors taken for the mixed perturbation of the Ising model with $\xi=10$, $t>0$ and at  $|m|R=4$.  }
\label{fig_bmfspec2b}
\end{figure}
\end{center}

 \begin{center}
\begin{figure}[H]
\centering
\includegraphics[scale=1]{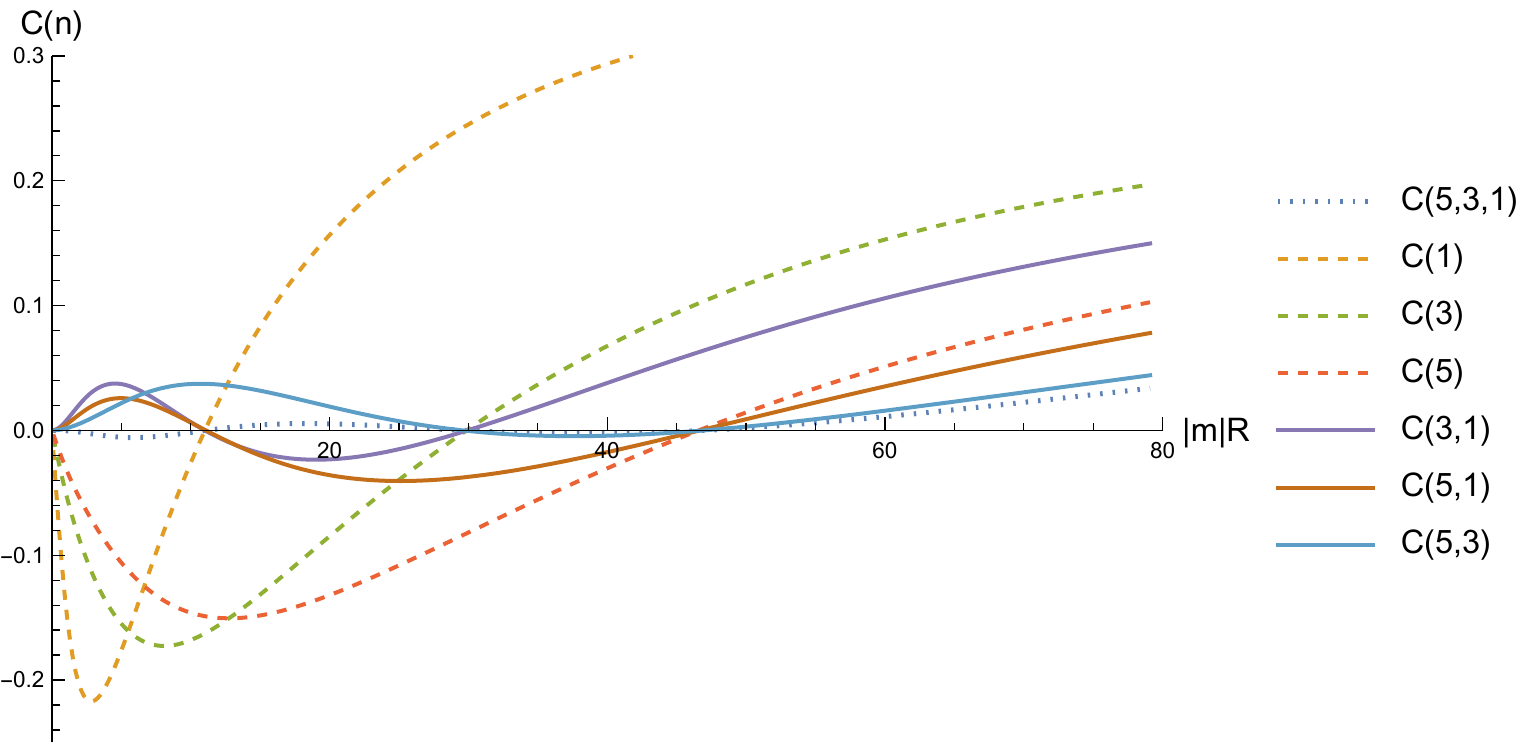}
\caption{ The seven low weight components in the NS sector of the vacuum as a function of $R$ for $\xi=10$. The TCSA numerics is  obtained with truncation at level $n_{c}=17$.   }
\label{fig_oscNS_xi10}
\end{figure}

\end{center}

  \begin{center}
\begin{figure}[H]
\centering
\includegraphics[scale=1]{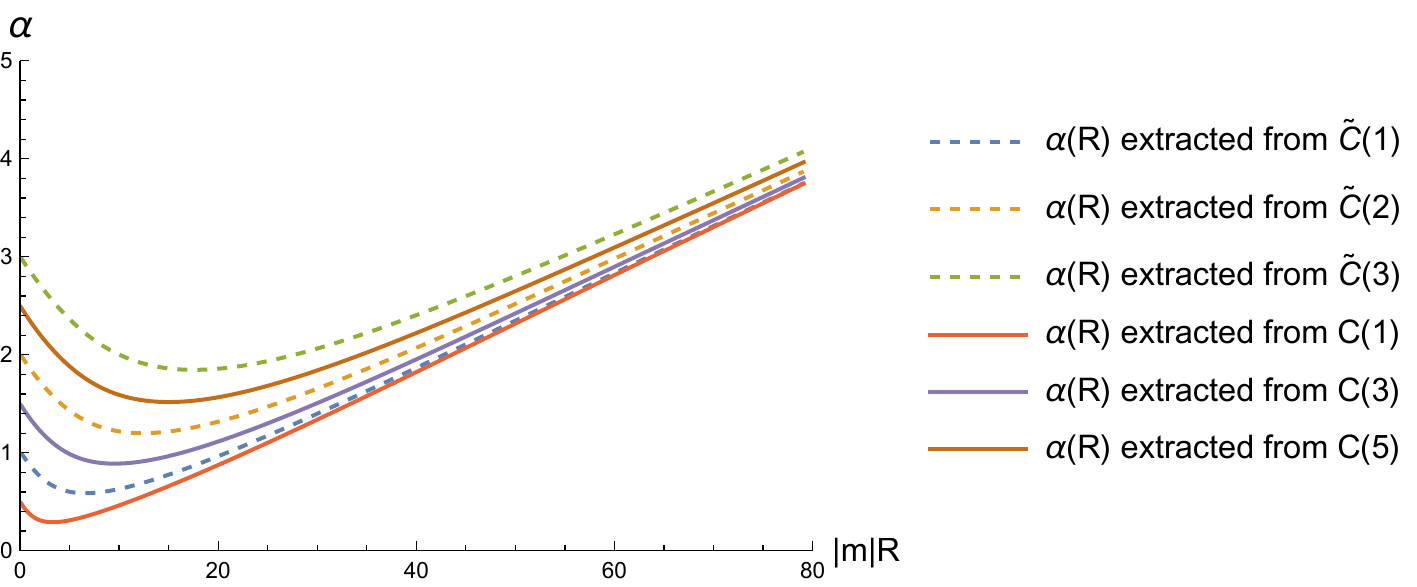}
\caption{The running boundary coupling $\alpha(R)$ for $\xi=10$, $t>0$ extracted from the one-particle states in the NS and Ramond sector and the value   $\tau_{\rm uv}=0$. 
The plots are obtained using TCSA  with oscillator basis truncated at level 17. }
\label{fig_alpha_NSR_xi10}
\end{figure}

\end{center}

\subsection{Perturbed TIM   with $\xi_{\sigma}=1.7717$ and $\lambda_{\epsilon}>0$ } \label{xicrit_appendix} 

 \begin{center}
\begin{figure}[H]
\begin{minipage}[b]{0.5\linewidth}
\centering
\includegraphics[scale=0.82]{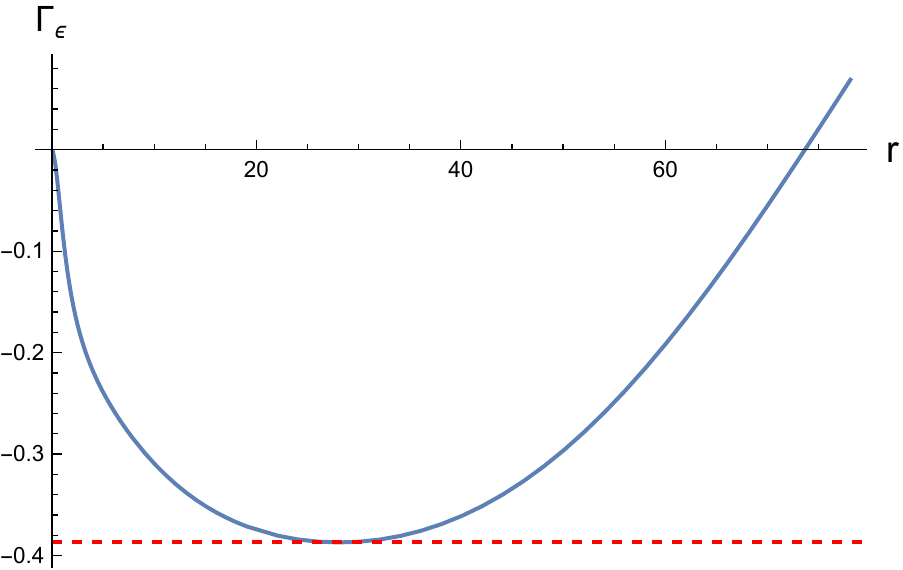}
\end{minipage}%
\begin{minipage}[b]{0.5\linewidth}
\centering
\includegraphics[scale=0.82]{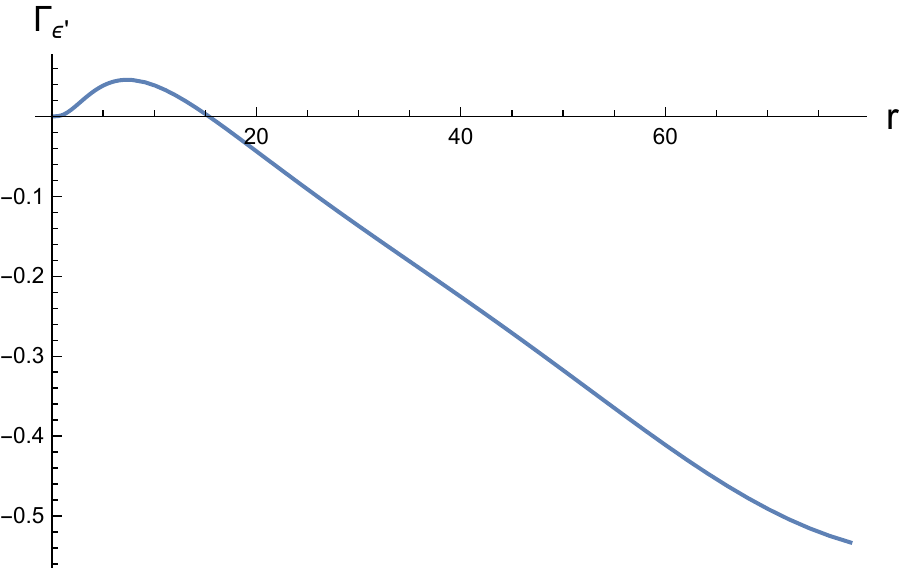}
\end{minipage}
\caption{The component ratios $\Gamma_{\epsilon}$ (left), $\Gamma_{\epsilon'}$  (right) obtained via TCSA for 
$\xi_{\sigma}=1.7717$, $\lambda_{\epsilon}>0$ at $n_{c}=11$.  }
\label{gammas_appendixpic1}
\end{figure}
\end{center}

 \begin{center}
\begin{figure}[H]
\begin{minipage}[b]{0.5\linewidth}
\centering
\includegraphics[scale=0.82]{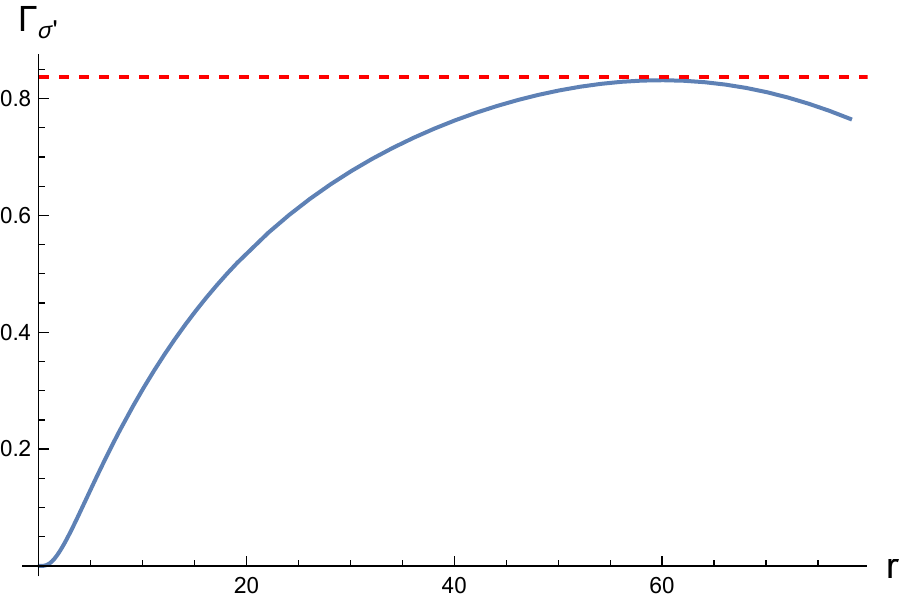}
\end{minipage}%
\begin{minipage}[b]{0.5\linewidth}
\centering
\includegraphics[scale=0.82]{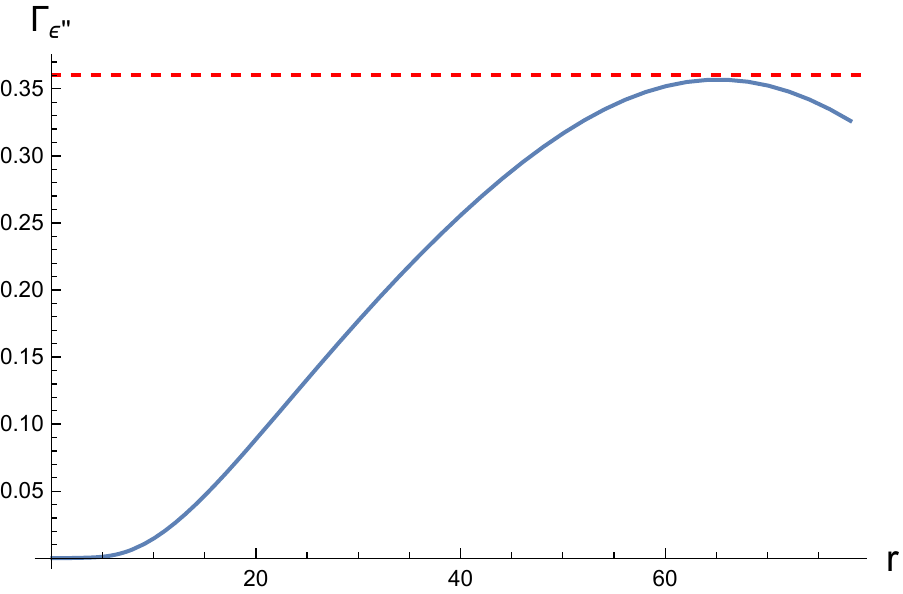}
\end{minipage}
\caption{The component ratios $\Gamma_{\sigma'}$ (left), $\Gamma_{\epsilon''}$  (right) obtained via TCSA for 
$\xi_{\sigma}=1.7717$, $\lambda_{\epsilon}>0$ at $n_{c}=11$.  }
\label{gammas_appendixpic2}
\end{figure}
\end{center}


 \begin{center}
\begin{figure}[H]
\begin{minipage}[b]{0.5\linewidth}
\centering
\includegraphics[scale=0.82]{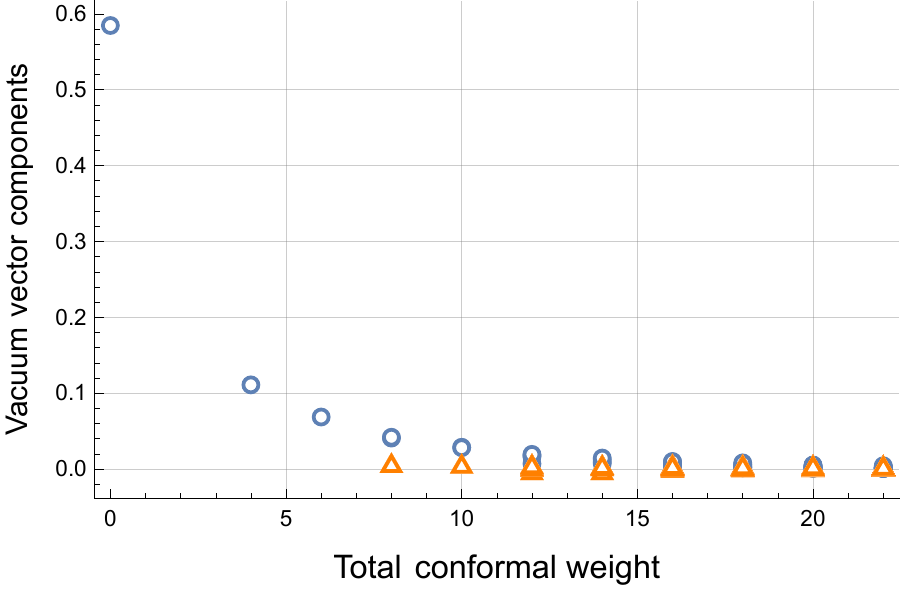}
\end{minipage}%
\begin{minipage}[b]{0.5\linewidth}
\centering
\includegraphics[scale=0.82]{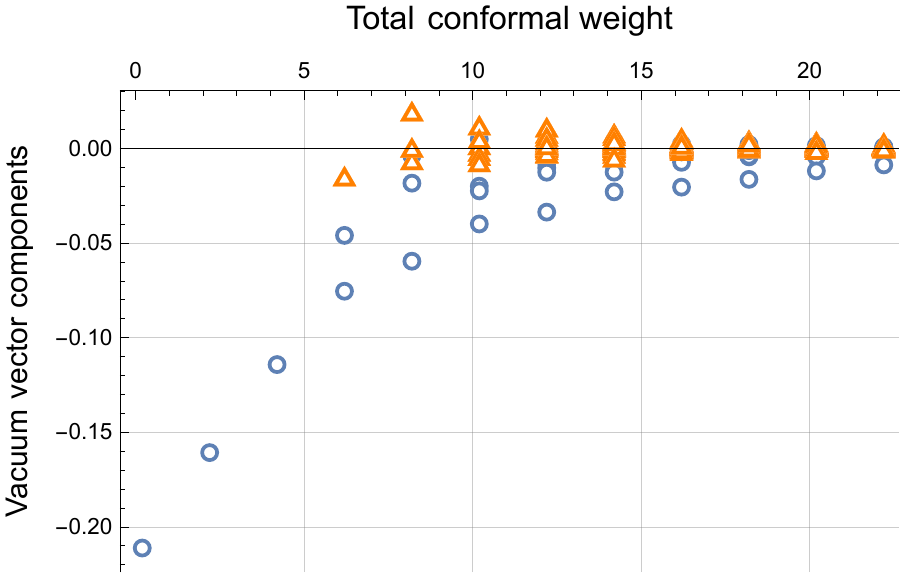}
\end{minipage}
\caption{The TCSA vacuum vector components  against the total conformal weight of the basis vectors taken for the TIM with $\xi_{\sigma}=1.7717$, $\lambda_{\epsilon}>0$ and at  $r=40$. The left plot 
represents the identity sector and the right plot -- the $\epsilon$-sector. 
The orange triangles mark the non-diagonal components. The truncation parameter $n_{c}=11$.}
\label{fig_TIM_critspec1}
\end{figure}

\end{center}

 \begin{center}
\begin{figure}[H]
\begin{minipage}[b]{0.5\linewidth}
\centering
\includegraphics[scale=0.82]{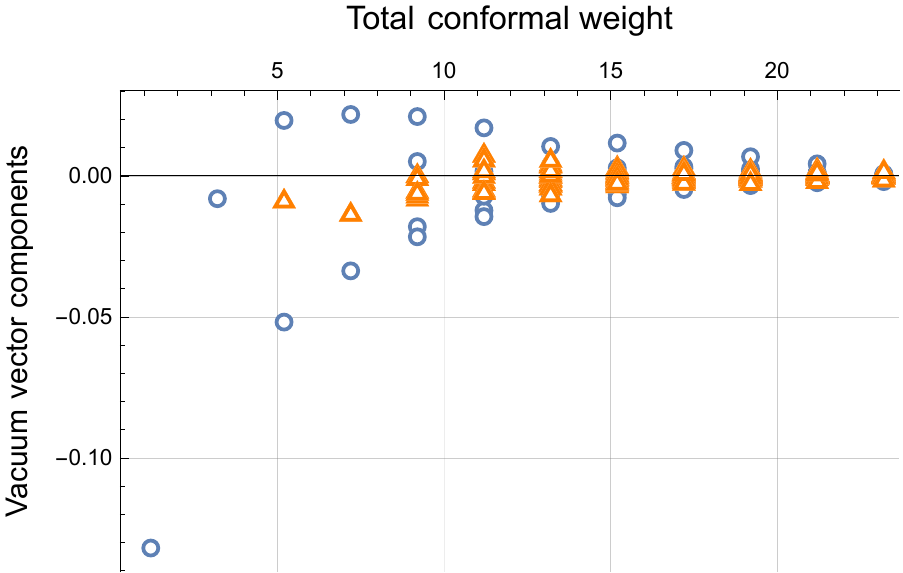}
\end{minipage}%
\begin{minipage}[b]{0.5\linewidth}
\centering
\includegraphics[scale=0.82]{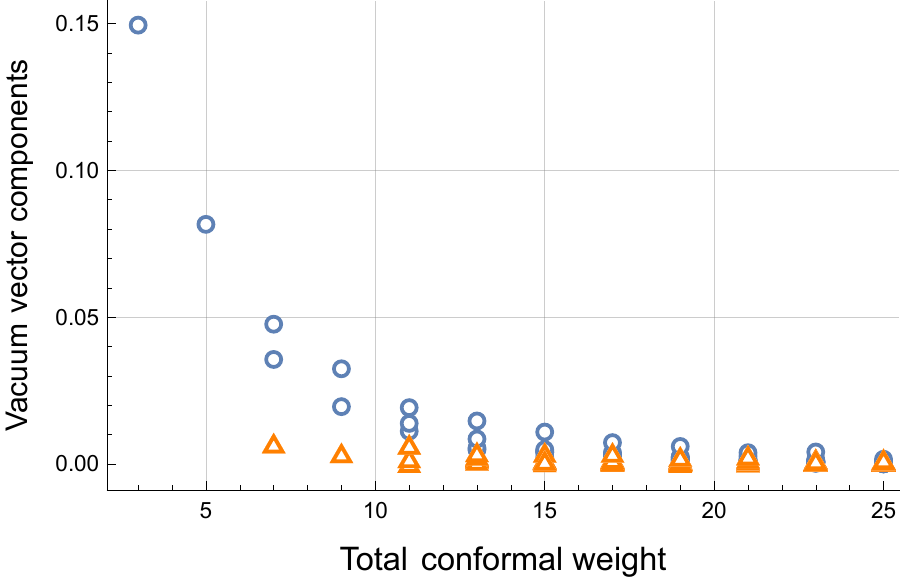}
\end{minipage}
\caption{The TCSA vacuum vector components  in the $\epsilon'$ sector (left) and $\epsilon''$ sector (right). See caption to Figure \ref{fig_TIM_critspec1} 
for more detail. }
\label{fig_TIM_critspec2}
\end{figure}

\end{center}
 \begin{center}
\begin{figure}[H]
\begin{minipage}[b]{0.5\linewidth}
\centering
\includegraphics[scale=0.82]{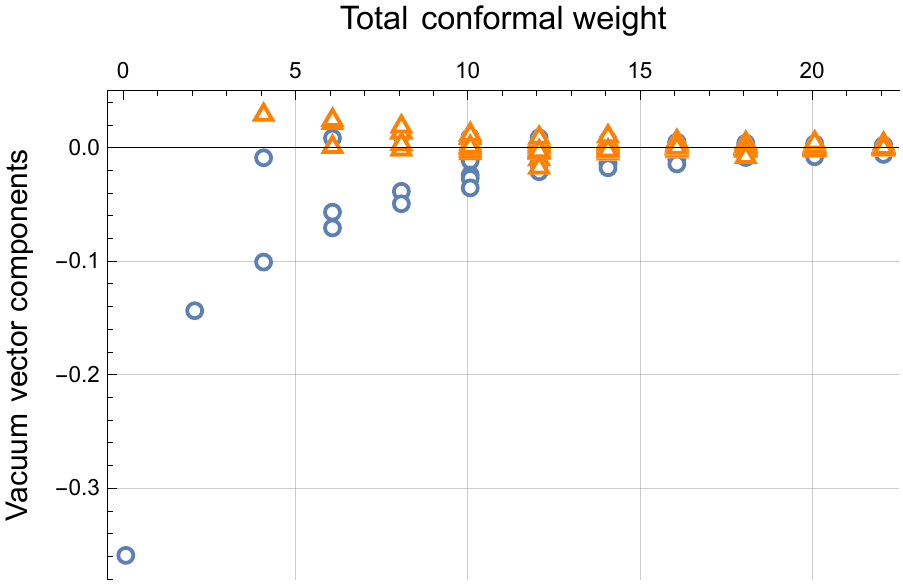}
\end{minipage}%
\begin{minipage}[b]{0.5\linewidth}
\centering
\includegraphics[scale=0.82]{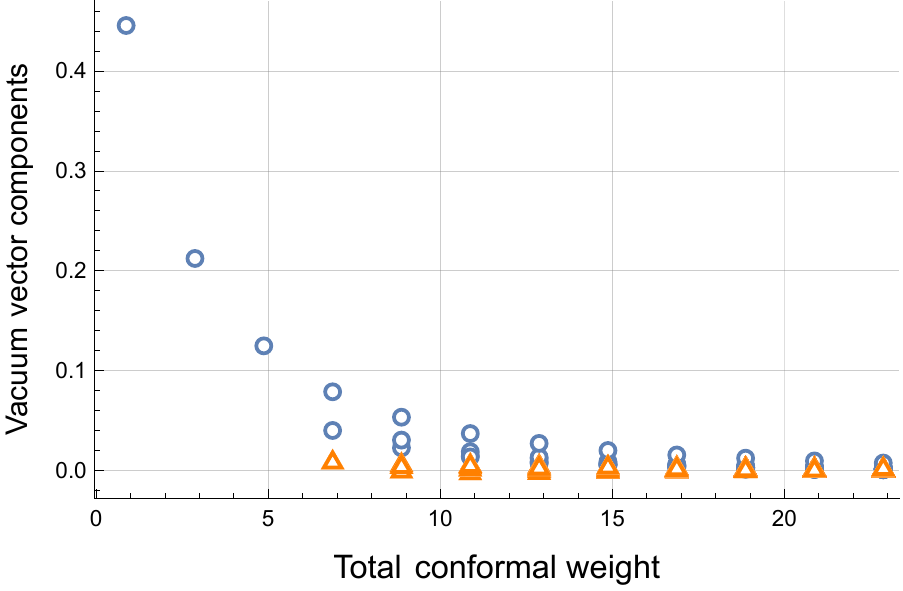}
\end{minipage}
\caption{The TCSA vacuum vector components  in the $\sigma$ sector (left) and $\sigma'$ sector (right). See caption to Figure \ref{fig_TIM_critspec1} 
for more detail. }
\label{fig_TIM_critspec3}
\end{figure}

\end{center}

\end{document}